Research Article

# Density deficit of the Earth's core revealed by a multi-megabar primary pressure scale

Daijo Ikuta[1,*,†], Eiji Ohtani[1,*], Hiroshi Fukui[2,3], Tatsuya Sakamaki[1], Rolf Heid[4], Daisuke Ishikawa[2,3], and Alfred Q. R. Baron[2,3,*]

[*] Corresponding authors. Email: dikuta@okayama-u.ac.jp, eohtani@tohoku.ac.jp, and baron@spring8.or.jp

***Affiliations***

*[1] Department of Earth Science, Tohoku University, Sendai, Miyagi 980-8578, Japan*
*[2] Materials Dynamics Laboratory, RIKEN SPring-8 Center, Sayo, Hyogo 679-5148, Japan*
*[3] Japan Synchrotron Radiation Research Institute, Sayo, Hyogo 679-5198, Japan*
*[4] Inst. Quantum Mater. Technol., Karlsruhe Institute of Technology, D-76021 Karlsruhe, Germany*
*† Present affiliation: Institute for Planetary Materials, Okayama University, Misasa, Tottori 682-0193, Japan*

## Abstract

An accurate pressure scale is a fundamental requirement to understand planetary interiors. Here, we establish a primary pressure scale extending to the multi-megabar pressures of the Earth's core, by combined measurement of the acoustic velocities and the density from a rhenium sample in a diamond anvil cell using inelastic x-ray scattering and x-ray diffraction. Our scale agrees well with previous primary scales and shock Hugoniots in each experimental pressure range, and reveals that previous scales have overestimated laboratory pressures by at least 20% at 230 gigapascals. It suggests that the light element content in the Earth's inner core (the density deficit relative to iron) is likely to be double what was previously estimated, or the Earth's inner core temperature is much higher than expected, or some combination thereof.

## One-sentence summary

A primary pressure scale to multi-megabar pressure may double the light element content of the Earth's inner core.

## Introduction

Precise information about the composition of the Earth's core is critical for understanding planetary evolution (*1–3*) and discussing current important topics in geodynamic behavior, such as core–mantle boundary heat flow (*3, 4*). However, samples from deep in the planetary interior are not available, so our knowledge is based on comparison of laboratory measurements (*5–8*) with seismological observations (*9*), information from meteorite composition (*3*), and indications of the Earth's core temperature (*10–12*). One of the most interesting results of such work has been the suggestion that Earth's core must contain light elements because the density of the core, as determined from seismological observations (*9*), is lower than the density of pure iron, its main constituent, as determined by laboratory measurements (*5–7*) and theoretical work (*10, 11*). However, this conclusion critically relies on having an accurate pressure scale to relate lab-generated pressures to geological pressures.

Establishing an accurate pressure scale has been the subject of intensive research (*13–20*), but present scales still rely on large extrapolation and approximations, especially at high pressures (*21*). Further, a pressure scale to multi-megabar pressures is indispensable for discussing super-Earth planets (*22, 23*). Previously, the compression curve for rhenium has been used as a secondary pressure scale determined on the basis of the pressure scales derived from shock compression measurements of several metals (*24–26*). The shock compression work, which occurs along a non-isothermal Hugoniot curve, is converted to an isothermal scale by the Rankine–Hugoniot equations with the Mie–Grüneisen–Debye (MGD) equation of state (EoS) (*1*). However, these derived scales show discrepancies of ~50% at density of ~33 g cm$^{-3}$ (*26*). Other work in static conditions provides primary pressure scales based on thermodynamic relations that allow the pressure to be determined when the density and both acoustic velocities, longitudinal (or compressional, $v_p$) and transverse (or shear, $v_s$) waves, are measured (*13–20*). However, most of the static experiments have been limited to lower-mantle pressures (up to 55 GPa) (*13–19*) with only one recent result (*20*) extending to ~120 GPa, as is close to the core–mantle boundary pressure: The measurement techniques used in the previous work, Brillouin scattering (BS) measurements [single crystal of periclase (*13, 14*) and polycrystalline sample of sodium chloride (*20*)], ultrasonic (US) measurements [polycrystalline sample of wadsleyite (*15*), periclase (*16*), and tungsten (*19*)], and inelastic x-ray scattering (IXS) measurements [single crystal of platinum (*17*) and sodium chloride (*18*)] become increasingly difficult as pressure increases.

Here, we measure acoustic velocities ($v_p$ and $v_s$) of rhenium in a diamond anvil cell (DAC) under extreme pressure using IXS and *in situ* x-ray diffraction (XRD) at BL43LXU (*27*) of the RIKEN SPring-8 Center. The XRD measurements were performed *in situ*, with the same x-ray beam and probed sample volume used for the IXS measurements. The energy of the longitudinal acoustic (LA) and transverse acoustic (TA) modes was measured using IXS, determining $v_p$ and $v_s$, while *in situ* XRD was used to determine the density, $\rho$. A highly optimized setup with a 5-μm beam size and special optics to reduce backgrounds (*28*) allowed us to extend the range of our work in static conditions in a DAC to the multi-megabar pressures of the Earth's core, 230 GPa in our rhenium scale, or what would be 274–300 GPa based on previous scales (*25, 29–31*) (see also Methods Sects. 1 to 5).

## Results

### Acoustic velocity measurement by IXS

An example of an IXS spectrum measured from rhenium at 230 GPa (the highest pressure: IXS-Re-12) is shown in Fig. 1A and shows clear peaks that we identify as the being due to the TA and LA modes. Fits to the IXS spectra allow us to determine $v_p$ and $v_s$ of rhenium (Fig. S1 and Table S1). We also measured $v_p$, $v_s$, and $\rho$ of rhenium at ambient conditions (in air) using a rhenium foil (Fig. S2 and Table S1) and confirmed that $v_p$, $v_s$, and $\rho$ are consistent with the ultrasonic measurement (*32*) (see also Note S1). The presence of the clear TA peak in the IXS spectrum was unexpected, as generally it should be weak in our small scattering angle geometry. This is discussed in Note S2 and Figs. S3 and S4, and we conclude that it is due to a large defect density that occurs when rhenium is compressed. Figure 1B shows the relations of $\rho$ with $v_p$ and $v_s$ at high pressure and ambient temperature. High-pressure experiments in this study were performed both with and without a periclase (MgO) pressure medium and laser annealing. In the experiments with the MgO pressure medium and laser annealing, the rhenium sample was annealed at temperatures over 1000 K by a double-sided laser heating method (Methods Sects. 1 and 2) before the IXS–XRD measurements, to minimize the deviatoric stress. As shown in Fig. S5A, the observed *c*/*a* ratios of the rhenium sample under non-hydrostatic conditions (direct compression, without pressure medium and laser annealing) were smaller than the calculated model *c*/*a* ratio of rhenium under hydrostatic pressure (*31*). However, both cases showed essentially similar acoustic velocities (Fig. S5B). A detailed and careful analysis of the data, including the impact of the crystal preferred orientation, the lattice strain (LS), and other factors, may be found in Notes S3 to S10 and Figs. S6 to S14. We find that the observed preferred orientations, and LSs, have negligible impact on the acoustic velocity. This is consistent with previous studies for hexagonal close-packed (hcp) iron (*8*, *33*).

The $\rho$–$v_p$ relation is well described by a linear function, Birch's law (*34*) with:

$$v_p = v_{p,0} + (\partial v_p / \partial \rho)(\rho - \rho_0) \,, \quad (1)$$

where we find $\rho_0$ = 20.8(±0.1)×10$^3$ kg m$^{-3}$, $v_{p,0}$ = 5.30(±0.03)×10$^3$ m s$^{-1}$, and $\partial v_p/\partial \rho$ = 0.313(±0.002) m$^4$ kg$^{-1}$ s$^{-1}$ for rhenium (see Table S2). The subscript zero indicates ambient conditions. The $\rho$–$v_s$ relation (red line in Fig. 1B) is derived from Eq. 1 with the MGD EoS (Methods Sect. 5 and Table S2). Comparing our result to previous studies, we find that $v_s$ in our study is consistent with the XRD–LS measurements (*35*) and the first-principles generalized gradient approximation (GGA) calculation of ref. (*31*), but is not consistent with the GGA calculation of ref. (*36*). Meanwhile, $v_p$ in our study is consistent with the GGA calculations of ref. (*36*), but we have a different trend compared with the XRD–LS measurements (*35*) and the GGA calculation of ref. (*31*), especially at multi-megabar pressures.

**Primary pressure scale of rhenium at multi-megabar pressures**

The primary pressure scale can be derived from $v_p$, $v_s$, and $\rho$ following the procedure of previous work (*15, 16, 19, 20*). We used a *K*-primed EoS (*37–39*) to express the relation between density and pressure at multi-megabar pressures. This EoS is based on the finite strain theory with the isothermal bulk modulus, *K*, and density, $\rho$, determined by the present IXS–XRD measurements. *K* and $\rho$ are fitted with finite strain parameters of the bulk modulus at ambient pressure, $K_0$, and its first pressure derivatives ($\partial K/\partial P$) at ambient pressure and infinite pressure, $K'_0$ and $K'_\infty$, respectively. We employed this EoS to keep consistency with the pressure dependence of thermodynamic Grüneisen parameter, $\gamma_{th}$. Details are given in Methods Sect. 5. A good fit was found with $\rho_0$ = 20.8(±0.1) g cm$^{-3}$, $K_0$ = 340(±9) GPa, $K'_0$ = 3.25(±0.12), and $K'_\infty$ = 2.15(±0.11). The obtained EoS parameters and pressures for rhenium are given in Tables S1 and S2. The uncertainty of the present pressure scale was evaluated by careful error propagation (Notes S1 to S9), with the detailed discussion presented in Note S10 and Table S3. Figure 2A shows our primary pressure scale of rhenium, compared with previous pressure scales (*24, 25, 29–31*). Our rhenium scale and the previous pressure scales are reasonably consistent up to ~60 GPa ($\rho$ ~ 24 g cm$^{-3}$). However, differences are observed above 85 GPa ($\rho$ ~ 25 g cm$^{-3}$), and large differences, beyond the uncertainties, appear above 120 GPa ($\rho$ ~ 26.5 g cm$^{-3}$). The previous pressure scales give pressures 20% higher at $\rho$ = 30.24 g cm$^{-3}$, and the overestimation increases with increasing pressure. Our rhenium scale agrees with previous primary scales at lower pressures (*13–20*). Investigation shows the recent primary scale study is consistent with our scale, suggesting that previous secondary pressure scales overestimate pressures by 2–10% at 120 GPa (*21*). Comparing our scale with previous secondary scales, previous scales have overestimated the laboratory pressures by at least 20% at 230 GPa. The discrepancy of the rhenium scale and previous scales originates from the density dependence of the $v_p$ of rhenium determined in this work (Fig. 1B). The experimental uncertainties derived from fitting the phonon dispersion, and $v_p$ and $v_s$ (Fig. S6), preferred orientation and anisotropy (Figs. S7 to S10), LS (Fig. S11), density gradient (Fig. S12), and diamond cupping (Fig. S13) were evaluated in Notes S4 to S10, Fig. S14, and Table S3. Even with the maximum uncertainty, there is still discrepancy of pressure values between present and previous scales (Fig. S14).

**EoSs of rhenium, iron, and MgO**

To understand the impact of our rhenium scale, we performed simultaneous compression experiments of rhenium, iron, and MgO by laser annealing the samples to minimize the deviatoric stress (Methods Sect. 4). Both the *c*/*a* ratios of rhenium and hcp-iron were consistent with those of calculated model *c*/*a* ratios (*31, 36*) even at multi-megabar pressures (Fig. S15), which indicates that annealing of the sample well worked to minimize stress. We obtained *K*-primed EoS parameters of hcp-iron with $\rho_0$ = 8.25(±0.05) g cm$^{-3}$, $K_0$ = 162(±5) GPa, $K'_0$ = 5.12(±0.08), and $K'_\infty$ = 2.55(±0.09) and those of MgO with $\rho_0$ = 3.58(±0.03) g cm$^{-3}$, $K_0$ = 159(±6) GPa, $K'_0$ = 3.79(±0.08), and $K'_\infty$ = 2.29(±0.12) (Methods Sects. 6 and 7, and Tables S2 and S4). Figure 2B and Fig. S16 show the calibrated *K*-primed EoS of hcp-iron and MgO, respectively. The compression curve of hcp-iron based on our rhenium scale is consistent with the curves based on previous scales (*5–7, 23*) up to 100 GPa within the uncertainties, but the differences become greater than 20% in the present maximum experimental pressure range. The compression curve of MgO based on our rhenium scale is consistent with the curves based on previous scales (*14, 40, 41*) in their respective experimental pressure ranges, within the uncertainties of our scale. The detailed comparison between the compression curves of MgO based on present and previous scales is given in Note S11 and Fig. S16.

Previous measurements using shock compression along the Hugoniot curve can be brought into agreement with our scale by careful consideration of the $\rho$-dependence of the Grüneisen parameter. Because shock compression is not an isothermal process, thermal parameters are necessary to convert the Hugoniot curve to isotherms or vice versa, to compare the isothermal pressure scale and shock Hugoniot. The MGD model is widely used for high-pressure and high-temperature EoS, and the Grüneisen parameter, $\gamma$, and molar heat capacity at constant volume, $c_{V,\mathrm{m}}$, are critical as they are directly related to thermal pressure. Within the MGD model, the Grüneisen parameter represents the effect of crystal lattice volume change on its vibrational properties (*1*). Therefore, the Grüneisen parameter can be derived from the $\rho$-dependence of $v_\mathrm{p}$ and $v_\mathrm{s}$. Especially for metals, $c_{V,\mathrm{m}}$ has contributions from both phonons and electrons. The detailed derivation of the $c_{V,\mathrm{m}}$ is given in Methods Sect. 8 (see also Fig. S17 and Table S5). We obtained MGD EoS parameters for rhenium with $\Theta$ = 369(±5) K, $\gamma_0$ = 1.94(±0.31), $\gamma_\infty = (3K'_\infty-1)/6$ (fixed), $q$ = 0.53(±0.30), parameters for hcp-iron with $\Theta$ = 515(±21) K, $\gamma_0$ = 1.97(±0.16), $\gamma_\infty = (3K'_\infty-1)/6$ (fixed), and $q$ = 0.37(±0.24), and parameters for MgO with $\Theta$ = 760(±135) K, $\gamma_0$ = 1.53(±0.26), $\gamma_\infty = (3K'_\infty-1)/6$ (fixed), and $q$ = 0.44(±0.68), where $\Theta$ is the Debye temperature, $\gamma_0$ and $\gamma_\infty$ are the Grüneisen parameters at ambient and infinite pressures, respectively, and $q$ gives its $\rho$-dependence (Methods Sects. 6 and 7, and Tables S2). Figure 3A shows simultaneous compression data of hcp-iron calibrated by our rhenium scale and calculated shock Hugoniot compared with experimental shock compression data in ref. (*42*). We reproduced the Hugoniot curve of iron based on our EoS of hcp-iron with the Grüneisen parameter determined by the isothermal bulk modulus from this work and Birch's law of hcp-iron in ref. (*8*) as shown in Fig. 3A and Fig. S18A. Detailed derivation of the calculated shock Hugoniot is given in Methods Sect. 8 and Tables S5. These figures show that our calculated shock Hugoniot is nicely consistent with experimental shock compression data in ref. (*42*). Figure S18B shows our Grüneisen parameter, derived from Birch's law of hcp-iron in ref. (*8*) with our EoS compared with the Grüneisen parameter used for the previous reference EoS (*5*). This figure indicates that our EoS is consistent with both shock compression data and the experimentally determined $v_p$ of hcp-iron (*8*), whereas the Grüneisen parameter in the previous reference EoS (*5*) is inconsistent with $v_p$ of hcp-iron. This provides additional strong evidence in favor of the present pressure scale.

Figure 3B shows the calculated shock temperature, $T_{Hug}$, from our EoS compared with the experimental $T_{Hug}$ of solid iron (*43–45*). Because of the difficulty in measuring $T_{Hug}$, and/or the solid–liquid transition effect, or superheating state of iron over the melting curve [e.g., ref. (*46*)], experimental estimates of $T_{Hug}$ show large variations (e.g., ~4000–6500 K around $\rho \sim 12$ g cm$^{-3}$ and ~6000–12000 K around $\rho \sim 12.5$ g cm$^{-3}$) in previous measurements and remain under debate. However, as shown in Fig. 3B, our calculated $T_{Hug}$ by a free electron model with eight valence electrons (FEM-8) is consistent with most of experimental $T_{Hug}$ of solid iron (*43–45*) within uncertainties (Note S12 and Fig. S17C). The shock Hugoniot of rhenium and MgO also can be reproduced from each isotherm based on our rhenium scale within the uncertainties (Figs. S19 and S20). Our calculated $T_{Hug}$ of MgO is also consistent with the experimental $T_{Hug}$ of MgO (*47–50*) within the uncertainties (Fig. S20B).

## Discussion

Our revised pressure scale has important implications in the context of the seismically determined compositional model of the Earth's interior, the preliminary reference Earth model (PREM) (*9*). Previously, a ~3–5% density deficit compared to hcp-iron was estimated for the Earth's inner core (*5, 6, 10*). Figure 4 shows the density deficits of the PREM inner core from hcp-iron at high pressure and high temperature. We used our thermal EoS of hcp-iron to model the iron density at Earth's inner core conditions in Fig. 4. We used the *K*-primed EoS, the MGD model, and the FEM-8 for the $c_{V,\mathrm{m}}$ of hcp-iron. The details of the procedure to derive the high-pressure and high-temperature EoS for hcp-iron are given in Methods Sects. 6 and 7, Figs. S17 and S18, and Tables S2 and S7 to S10.

In the range of 330–365 GPa and 6000 K, typical for estimations of the Earth's inner core conditions (*10–12*), the density deficit from hcp-iron via our rhenium scale is 8(±2)%, which is much larger than ~3–5% of the previously estimated density deficit. The detailed analyses of the density deficit of the Earth's inner core are also given in Methods Sects. 5 to 7, Fig. S17D, and Table S6. If the density deficit is constrained to ~3–5% as previously estimated, a much higher temperature around 9000 K is required [~3000 K higher than the previous estimate (*10–12*)]. In conclusion, to account for the density of the PREM inner core, our present rhenium scale requires approximately a factor of two more light elements in the Earth's inner core than previously estimated, or much higher core temperatures, or some combination thereof as shown in Fig. 4.

# Figures

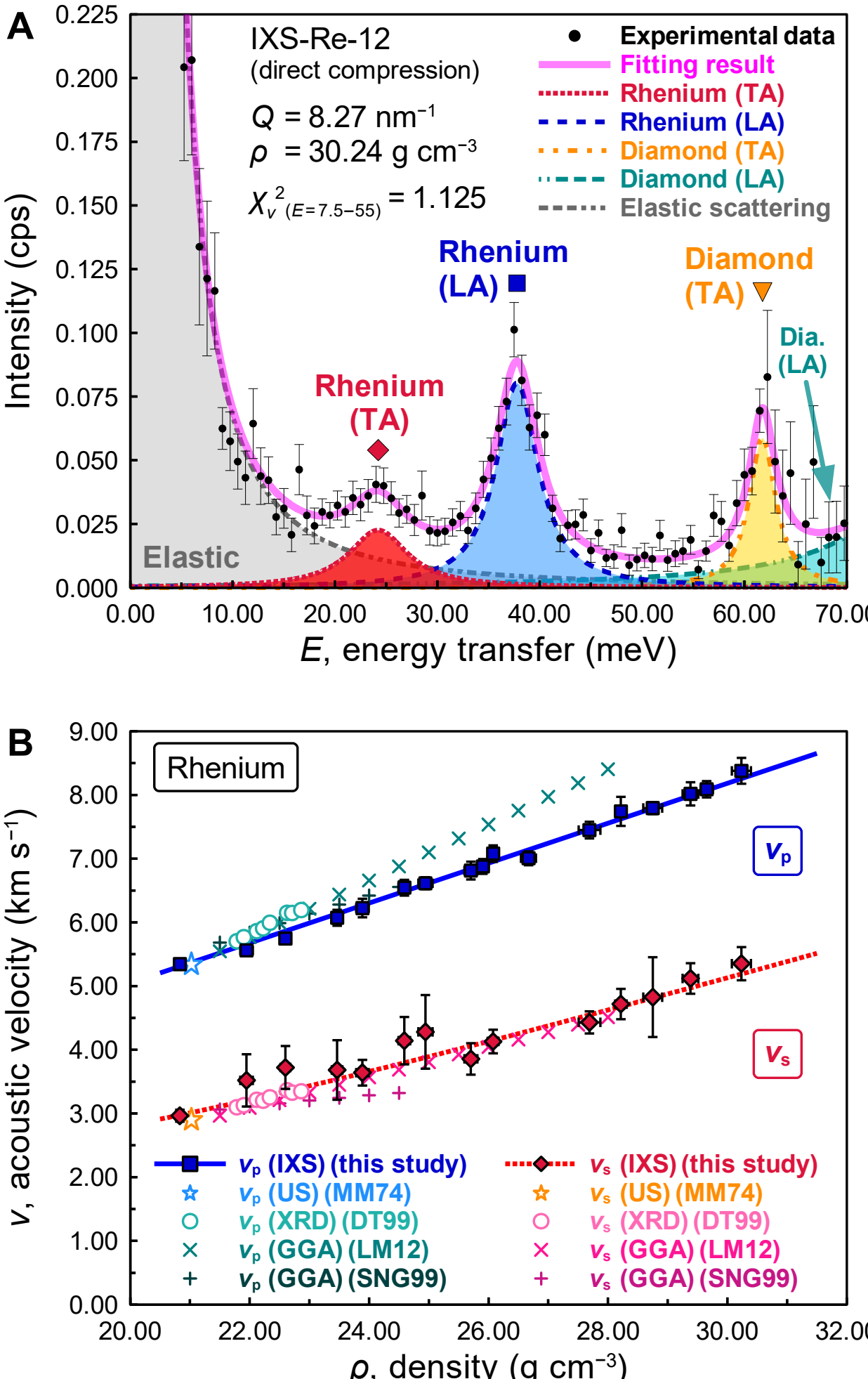

**Fig. 1. Results of acoustic velocity measurement for rhenium at high pressure.** (**A**) IXS spectrum and fitting results for rhenium at density, $\rho = 30.24$ g cm$^{-3}$ (230 GPa) and 300 K (IXS-Re-12). The black dots are the IXS data with one standard deviation (1σ) error bars. Other colored lines and areas are individual inelastic contributions of LA and TA modes as labeled, with colored symbols showing the fitted peak positions. (**B**) Acoustic velocities (compressional, $v_p$, and shear, $v_s$) for rhenium as a function of density (Table S1). The blue squares and red diamonds are $v_p$ and $v_s$ for rhenium determined from our IXS data with 1σ error bars. Other colored symbols are from previous studies [MM74 (*32*), DT99 (*35*), LM12 (*31*), and SNG99 (*36*)].

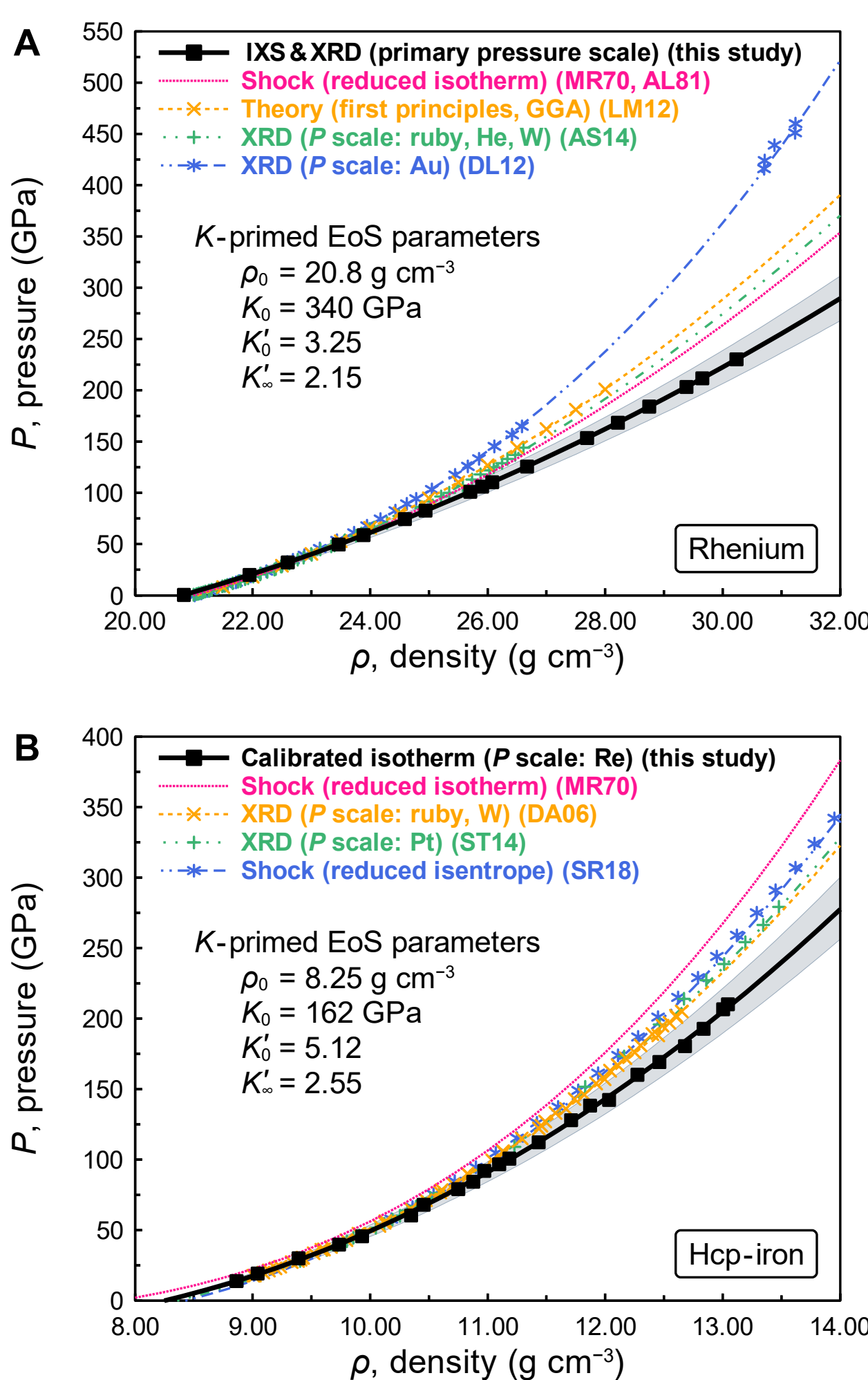

**Fig. 2. Primary pressure scale for rhenium and calibrated density–pressure relation for hcp-iron.** (**A**) Primary pressure scale for rhenium. The black curve is the compression curve of our rhenium scale with the density determined experimentally and the pressure evaluated by our rhenium scale determined from density and acoustic velocities measured in runs IXS-Re-01 to IXS-Re-16 and IXS-Re-foil (Fig. 1B). (**B**) Calibrated density–pressure relation for hcp-iron. The black curve with black squares is the compression curve of hcp-iron with the present simultaneous compression experiment (Table S4) calibrated by our rhenium scale (Table S2). The shaded areas around the black curves represent the 1σ uncertainty of each curve. Other colored curves and symbols are the compression curves of rhenium (A) and hcp-iron (B) with experimental data based on pressure scales from previous studies [MR70 (*29*), AL81 (*30*), LM12 (*31*), AS14 (*25*), DL12 (*24*), DA06 (*5*), ST14 (*6*), and SR18 (*23*)].

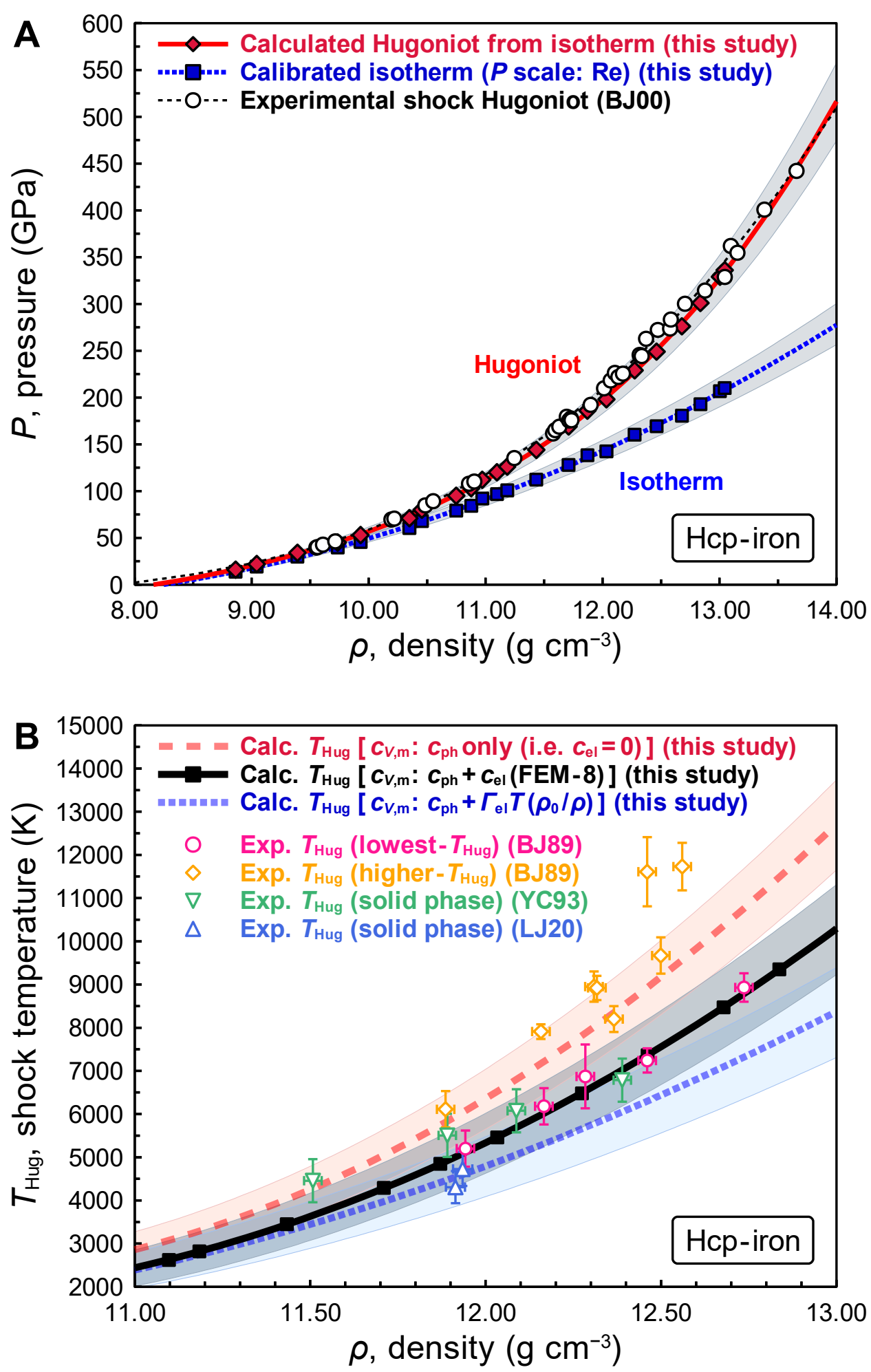

**Fig. 3. The isotherm of hcp-iron calibrated by the present rhenium scale, and calculated shock Hugoniot and its calculated shock temperature.** (**A**) Isothermal compression curve and calculated shock Hugoniot of hcp-iron. The blue dotted curve with squares represents the isothermal compression curve of hcp-iron based on the present simultaneous compression experiment and our rhenium scale. The black dashed curve with open circles represents the Hugoniot curve with experimental shock compression data (BJ00) (*42*). The red solid curve with diamonds represents the calculated shock Hugoniot from the isotherm of hcp-iron based on the present simultaneous compression experiment and our rhenium scale (Table S5). Our calculated shock Hugoniot can explain the experimental shock Hugoniot (*42*). (**B**) Comparison of calculated and experimental shock temperatures, $T_{\mathrm{Hug}}$, of hcp-iron based on three different molar heat capacity at constant volume, $c_{V,\mathrm{m}}$, models. The black squares are the calculated $T_{\mathrm{Hug}}$ with contributions of electrons to heat capacity, $c_{\mathrm{el}}$ by using the FEM-8 corresponding to the red diamonds in (A). The red dashed and blue dotted curves are the calculated $T_{\mathrm{Hug}}$ by the $c_{\mathrm{el}} = 0$ model and the linear temperature dependence model [expressed as $c_{\mathrm{el}} = \Gamma_{\mathrm{el}}T(\rho_0/\rho)$, where $\Gamma_{\mathrm{el}}$ is the electronic specific heat coefficient], respectively. The detailed description of the different approaches to determine $c_{V,\mathrm{m}}$ are presented in Methods Sects. 6 and 7, and Table S5. Other colored symbols are the experimentally measured $T_{\mathrm{Hug}}$ of solid iron from previous studies [BJ89 (*43*), YC93 (*44*), and LJ20 (*45*)]. The shaded areas around the curves in (A) and (B) represent the 1σ uncertainty of these curves.

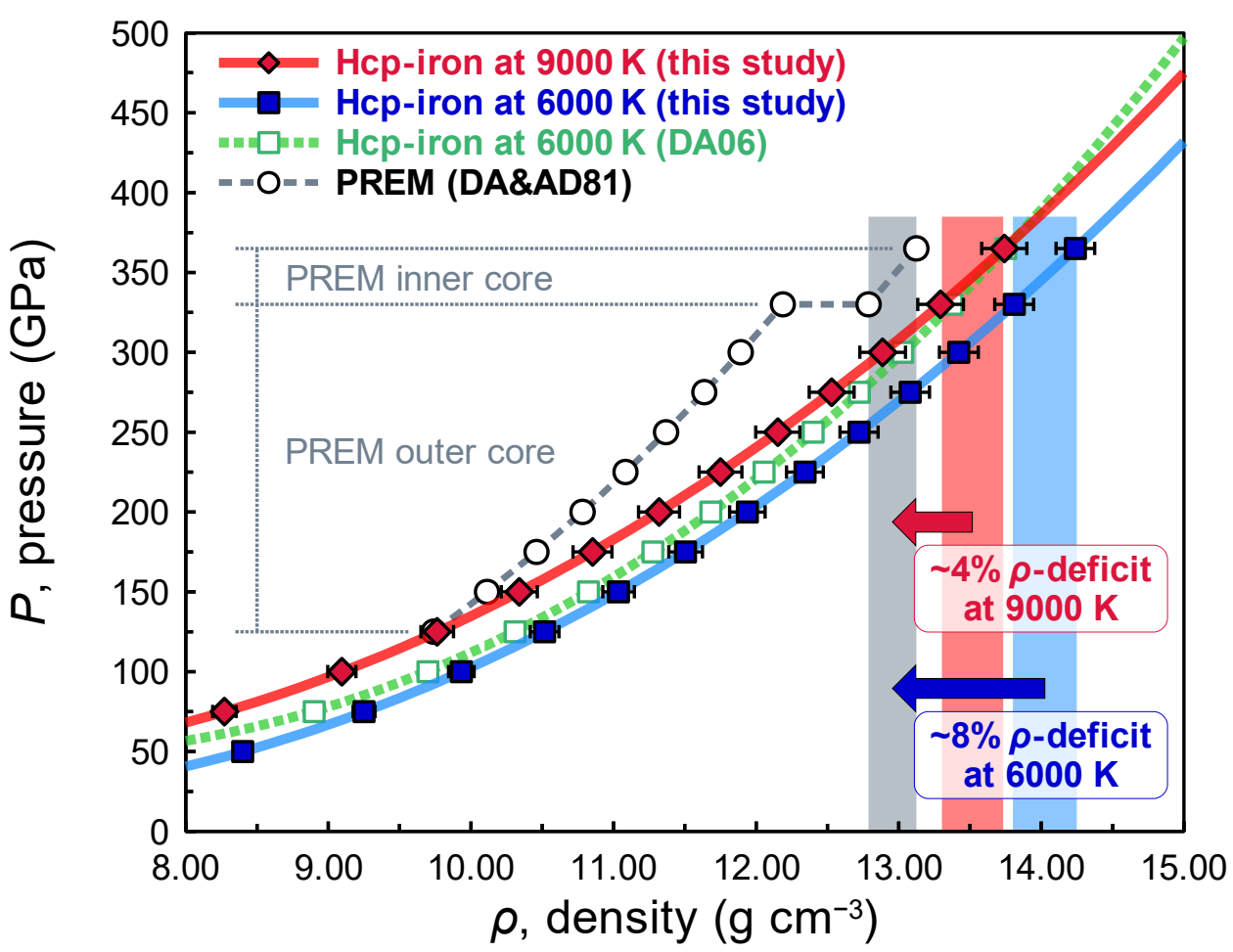

**Fig. 4. Density–pressure relations of hcp-iron at high temperature and PREM.** The red and blue curves with symbols are compression curves of hcp-iron at 9000 and 6000 K with 1σ error bars of density determined from the present work. The green dotted curve with open square symbols is the compression curve of hcp-iron at 6000 K determined by a previous pressure scale (DA06) (*5*). The gray dashed curve with open circles represents the density–pressure relation of the PREM (DA&AD81) (*9*). We used the heat capacity model by using the Debye model and the FEM-8 in this calculation (Methods Sect. 6 and 7). Details are given in Table S6. The red and blue arrows indicate the density deficits between hcp-iron and PREM for the compression curves of 9000 and 6000 K, respectively.

# Methods

## 1. Starting material and high-pressure generation

We performed the compression experiments both using direct compression without pressure medium and quasi-hydrostatic compression with a magnesium oxide (periclase, MgO) pressure medium for acoustic velocity measurements. For the experiments without pressure medium, we used a DAC with double-beveled diamond anvils with a culet of 30 μm in diameter. A rhenium metal foil (250 μm in thickness, 99.97% purity, Alfa Aesar) was pre-compressed to a thickness of ~20–30 μm and used as the starting material. We increased pressure in 12 compressional steps (IXS-Re-01 to IXS-Re-12). For the experiments with the MgO pressure medium, we used a DAC with single-beveled diamond anvils with a culet of 150 μm in diameter. A rhenium powder (−325 mesh, 99.99% purity, Alfa Aesar) was used as a starting material. The rhenium powder was pre-compressed to a thickness of ~15 μm and was surrounded by the MgO (>98% purity, Junsei Chemical) pellets of about 5-μm thickness, which served as the pressure medium. The sample and pressure medium were loaded into a sample hole with a diameter of ~50 μm of a pre-compressed rhenium gasket foil (250 μm in thickness, 99.97% purity, Alfa Aesar), which was ~25 μm in thickness. We increased pressure in four compressional steps (IXS-Re-13 to IXS-Re-16) and the sample was annealed at temperature over 1000 K after each compression (Methods Sect. 2). The acoustic velocity measurement of rhenium at ambient conditions (IXS-Re-foil, in air) were also performed using a rhenium metal foil (25 μm in thickness, Nilaco), which was pre-compressed to a thickness of ~10–15 μm.

We performed the simultaneous compression experiment of rhenium, iron, and MgO to establish the EoSs of iron and MgO based on our rhenium scale. We used a DAC with double-beveled diamond anvils with a culet of 50 μm and a rhenium gasket (250 μm in thickness, 99.97% purity, Alfa Aesar). A rhenium powder (−325 mesh, 99.99% purity, Alfa Aesar) and an iron powder (99.99% purity, Wako Chemicals) were mixed as a starting material and pre-compressed to a thickness of ~15 μm. The mixture was surrounded by the MgO (>98% purity, Junsei Chemical) pellets of about 5-μm thickness, which served as the pressure medium. The sample and pressure medium were loaded into a sample hole (with a diameter of ~15 μm) in a pre-compressed rhenium gasket foil (with a thickness of 25 μm).

## 2. IXS measurement

The acoustic velocity of rhenium was measured by IXS at BL43LXU (*27*) of the RIKEN SPring-8 Center. The Si (9 9 9) reflection at 17.79 keV provided a resolution of 2.8 meV (IXS-Re-01 to IXS-Re-16, high-pressure experiments) and the Si (11 11 11) reflection at 21.75 keV provided a resolution of 1.3 meV (IXS-Re-foil, in-air). The x-ray beam size for the high-pressure runs was focused to 5 μm × 5 μm at 17.79 keV by a multilayer Kirkpatrick-Baez mirror pair (*28*). The x-ray beam with 50 μm × 50 μm at 21.75 keV was used for the measurements under the ambient conditions (in air). To reduce the scattering background from the diamonds and improve the signal to noise ratio, a Soller screen (*28*) was installed downstream of the DAC at pressures over 150 GPa (IXS-Re-07 to IXS-Re-12) without pressure medium and also for all runs with the MgO pressure medium (IXS-Re-13 to IXS-Re-16). In runs IXS-Re-13 to IXS-Re-16, the rhenium sample with the MgO pressure medium was annealed at temperatures over 1000 K after each compression by a double-sided laser heating method using a fiber laser installed at BL43LXU (COMPAT system) (*51*) and quenched to ambient temperature before the IXS measurements, to minimize the deviatoric stress. A Soller screen was not used in the experiments in air without a DAC (IXS-Re-foil). The IXS spectra at several momentum transfers were collected simultaneously by 23 (12, with the Soller screen) analyzers, arranged in a two-dimensional 4 × 6 array in runs from IXS-Re-01 to IXS-Re-12, and 28 (16, with the Soller screen) analyzers, arranged in a two-dimensional 4 × 7 array in runs from IXS-Re-13 to IXS-Re-16 and IXS-Re-foil. The IXS spectra were collected for ~8–24 hours in each experimental run. The TA mode did not appear at some pressures (see Fig. S1) because the measurement time was limited for those cases and thus the data quality was insufficient to clearly isolate the TA mode in the spectra. The TA peaks (Figs. 1A and S2A) were larger than expected from calculations of perfect single crystals (Fig. S3). We discuss this and conclude that it is from a quite high (~0.1–1 $nm^{-2}$) defect density that appears after rhenium is pressurized (Fig. S4) in Note S2.

### 3. Phonon dispersion and fitting

The IXS spectra are characterized by elastic scattering near zero energy and inelastic contributions from the LA and TA modes of rhenium and diamond (Figs. 1A and S2A). The energy positions of the inelastic contribution were extracted by fitting with Lorentzian functions. The relation between the excitation energy, the velocity for each acoustic mode, and the momentum transfer of phonons is given by:

$$E = \frac{h v Q_{\max}}{\pi^2} \sin\left(\frac{\pi Q}{2 Q_{\max}}\right), \tag{2}$$

where $E$ is excitation energy, $h$ is Plank constant, $v$ is acoustic velocity for each LA and TA mode, $Q$ is momentum transfer, and $Q_{\max}$ is the averaged distance to the edge of the first Brillouin zone (BZ) including the effect of the preferred orientation. To obtain the acoustic velocity from the IXS results, the dispersion was fit with the sine function, Eq. 2, which was used to determine the long-wavelength (the limit as $Q$ approaches zero) acoustic velocity (*8, 52, 53*). A weighted least-squares method was used with $v$ and $Q_{\max}$ as free parameters (see also Note S4). Figs. S1 and S2B show the fitting results of the LA and TA phonon dispersions for rhenium. LA phonons were clear in all sixteen high-pressure experimental conditions and one ambient condition. In runs IXS-Re-05, 06, and 11, the TA phonons could not be clearly identified because of poor signal-to-noise ratio and/or insufficient exposure time (within the limitation of the experimental beamtime). However, TA peaks were observed in some of the spectra in these runs, and they were found to be consistent with the estimated TA phonon dispersion curves (gray dashed lines in Fig. S1) calculated by Eq. 2 with interpolated $v$ (red dotted line in Fig. S5B) and $Q_{\max}$ (averaged distance to the edge of the first BZ in Fig. S6A) values.

## 4. Two-dimensional XRD measurement

Two-dimensional XRD patterns were taken to measure the density and to characterize the texture of the sample and were done in the same optical setup at BL43LXU (*in situ*) using a flat-panel detector (FP, C9732DK, Hamamatsu Photonics). The XRD was measured *in situ*, with the same incident x-ray beam at the same position on the sample as for the IXS work. This allowed us to investigate the impact of hydrostatic/non-hydrostatic conditions, preferred orientation, and LSs on the same sample volume used to measure the acoustic velocities (Note S3 to S9 and Figs. S3 to S13). The distance between the sample and the flat-panel detector was calibrated by using a cerium dioxide standard (NIST, National Institute of Standards and Technology). The lattice parameters and densities, $\rho$, of rhenium in the hcp structure were obtained from XRD patterns as:

$$\frac{1}{d_{(hkl)}^2} = \frac{4}{3}\left(\frac{h^2+hk+k^2}{a^2}\right) + \frac{l^2}{c^2} , \tag{3}$$

$$\rho = \frac{Z\,M}{N_\mathrm{A}} \frac{2}{\sqrt{3}\,a^2 c} , \tag{4}$$

where $a$ and $c$ are lattice parameters, $h$, $k$, and $l$ are Miller indices, $d_{(hkl)}$ is the $d$-spacing value for a reflection indexed by $hkl$, $Z$ is the number of atoms in the rhenium lattice, $M$ is molar mass of rhenium, and $N_\mathrm{A}$ is Avogadro constant. The density determination was carried out by using six $d$-spacing values of different diffraction peaks ($hkl$: 100, 002, 101, 102, 110, and 103) as shown in Fig. S7. The XRD patterns were analyzed using the IPAnalyzer/PDindexer/ReciPro software package (*54, 55*).

For the simultaneous compression experiments of rhenium, iron, and MgO, we performed annealing the samples to measure reasonable density relations among rhenium, iron, and MgO for minimizing the deviatoric stress in the samples. The samples were annealed at temperatures over 1000 K after each compression by a double-sided laser heating method using a fiber laser installed at BL43LXU (COMPAT system) (*51*) and quenched to ambient temperature before taking XRD patterns. The lattice parameters and densities of rhenium, iron, and MgO were determined by XRD patterns (Table S4).

## 5. Primary pressure scale derivation

The primary pressure scale can be derived from compressional and shear wave velocities and density ($v_\mathrm{p}$, $v_\mathrm{s}$, and $\rho$) following the procedure in previous work of polycrystalline samples (*15, 16, 19, 20*). We employed the $K$-primed EoS to keep consistency of the pressure dependence of thermodynamic Grüneisen parameter, $\gamma_\mathrm{th}$, i.e., it must be greater than two-thirds because of the thermodynamic consistency, whereas other EoSs including Birch–Murnaghan EoS and Rydberg–Vinet EoS violate the consistency at extremely high pressures (*38, 39*). The isothermal bulk modulus, $K$, and the density, $\rho$, were fit with the $K$-primed EoS (*37–39*) as follows:

$$P = K_0 \left\{ \frac{K'_0}{K'^2_\infty} \left[ \left( \frac{\rho}{\rho_0} \right)^{K'_\infty} - 1 \right] - \left( \frac{K'_0}{K'_\infty} - 1 \right) \ln \left( \frac{\rho}{\rho_0} \right) \right\} , \tag{5}$$

$$K = K_0 \left\{ \frac{K'_0}{K'_\infty} \left[ \left( \frac{\rho}{\rho_0} \right)^{K'_\infty} - 1 \right] + 1 \right\} , \tag{6}$$

$$\gamma_\infty = \frac{3\, K'_\infty - 1}{6} , \tag{7}$$

where $K_0$ is the isothermal bulk modulus at ambient pressure, and $K'_0$ and $K'_\infty$ are its first pressure derivatives ($\partial K/\partial P$) at ambient and infinite pressures, respectively.

$v_p$ and $v_s$ are related to the adiabatic bulk modulus, $K_S$, and the shear modulus, $G$, as follows [the formulas used here can be found in ref. (*1*)]:

$$K_S = \rho \left( v_p^2 - \frac{4}{3} v_s^2 \right) , \tag{8}$$

$$G = \rho\, v_s^2 \, . \tag{9}$$

The isothermal bulk modulus, $K$, is related to the adiabatic bulk modulus, $K_S$, the thermodynamic Grüneisen parameter, $\gamma_{th}$, the molar heat capacity at constant volume, $c_{V,m}$, density, $\rho$, molar mass, $M$, and temperature, $T$,

$$K = K_S - \gamma_{th}^2 \frac{\rho\, c_{V,m}}{M} T \, . \tag{10}$$

In a Debye model (DM), the Debye velocity $v_D$ is defined:

$$\frac{3}{v_D^3} = \frac{1}{v_p^3} + \frac{2}{v_s^3} \, . \tag{11}$$

Debye temperature $\Theta$ is defined:

$$\Theta = \frac{h\, v_D}{2\, k_B} \left( \frac{6\, N_A\, \rho}{\pi\, M} \right)^{\frac{1}{3}} , \tag{12}$$

where $h$ is Plank constant, $k_B$ is Boltzman constant, $N_A$ is Avogadro's constant, and $M$ is the molar mass.

There are several definitions of the Grüneisen parameter. Under the quasi-harmonic approximation, thermodynamic (macroscopic) Grüneisen parameter, $\gamma_{th}$, is defined:

$$\gamma_{th} = \frac{M}{\rho\, c_{V,m}} \left( \frac{\partial P}{\partial T} \right)_V \, . \tag{13}$$

Debye–Grüneisen (microscopic) parameter, $\gamma_D$, can be expressed by the Debye temperature, which is related to the vibration energy by Eq. 12:

$$\gamma_D = \frac{\partial \ln(\Theta)}{\partial \ln(\rho)} \, . \tag{14}$$

In the Debye approximation, the macroscopic and microscopic thermodynamic properties are assumed to be the same, thus,

$$\gamma_{\mathrm{th}} = \gamma_{\mathrm{D}} \ . \tag{15}$$

On the other hand, the temperature dependence of the $\gamma_{\mathrm{th}}$ can be expressed as (*39*):

$$\frac{\partial \gamma_{\mathrm{th}}}{\partial T} = -\frac{1}{T}\left[\frac{\partial \ln\left(c_{V,\mathrm{m}}\right)}{\partial \ln\left(\rho\right)}\right]_S , \tag{16}$$

which indicates the temperature dependence of the $\gamma_{\mathrm{th}}$ is inversely proportional to the temperature. In addition, under the Debye approximation, $c_{V,\mathrm{m}}$ becomes almost equal to constant of $3nR$ above the Debye temperature (the Dulong–Petit law). Thus, the temperature dependence of the $\gamma_{\mathrm{th}}$ can be negligible.

The density dependence of the Grüneisen parameter $\gamma_{\mathrm{th}}$ is expressed as a function of density with negligible temperature dependence by Al'tshuler form (*56*) as:

$$\gamma = \gamma_\infty + \left(\gamma_0 - \gamma_\infty\right)\left(\frac{\rho_0}{\rho}\right)^q , \tag{17}$$

where $\gamma_\infty$ is the Grüneisen parameter at infinite pressure and $q$ gives the $\rho$-dependence. In Al'tshuler form, either $\gamma_\infty$ or $q$ was usually fixed. In this study, we fixed $\gamma_\infty$ as $(3K'_\infty-1)/6$, which is a theoretical constraint of Grüneisen parameter in the $K$-primed EoS (*37–39*). The $\rho$-dependence of the Debye temperature can be expressed from Eqs. 14 and 17 as:

$$\Theta = \Theta_0\left(\frac{\rho_0}{\rho}\right)^{-\gamma_\infty} \exp\left\{\frac{\gamma_0 - \gamma_\infty}{q}\left[1 - \left(\frac{\rho_0}{\rho}\right)^q\right]\right\} . \tag{18}$$

The parameters, $\gamma_0$, $\gamma_\infty$, and $q$ for thermodynamic Grüneisen parameter, and $\Theta_0$ for Debye temperature can be derived by fitting with the experimental dataset of $v_{\mathrm{p}}$, $v_{\mathrm{s}}$, and $\rho$ from Eqs. 11, 12, and 18.

$c_{V,\mathrm{m}}$ is assumed to be a sum of contributions from phonons ($c_{\mathrm{ph}}$) and electrons ($c_{\mathrm{el}}$) (*57*). However, $c_{\mathrm{el}}$ is assumed to be zero at ambient temperature, because the contribution by electrons is negligible compared to phonons at low temperature (e.g., $T < \Theta$). $c_{\mathrm{ph}}$ is derived using the DM:

$$c_{\mathrm{ph}} = 9nR\left(\frac{T}{\Theta}\right)^3 \int_0^{\Theta/T} \frac{x^4 \exp\left(x\right)}{\left[\exp\left(x\right)-1\right]^2} \mathrm{d}x , \tag{19}$$

where $n$ is the number of atoms per chemical formula unit, $R$ is the gas constant, $T$ is the temperature, and $\Theta$ is the Debye temperature expressed by Eq. 18. By using Eqs. 10, 16, and 19, the isothermal bulk modulus, $K$, can be derived from the dataset of $v_{\mathrm{p}}$, $v_{\mathrm{s}}$, and $\rho$ determined experimentally. Last, we can determine the parameters for the $K$-prime EoS, $K'_0$ and $K'_\infty$ with the fixed $K'_\infty$ from Eq. 7, by fitting $K$ and $\rho$ to Eq. 6 of the $K$-primed EoS. The parameters of $K$-primed EoS for rhenium are given in Table S2.

## 6. High-pressure and high-temperature EoSs for hcp-iron and MgO by the MGD model

We re-evaluated the EoSs of hcp-iron and MgO at high pressure and ambient temperature using the present rhenium pressure scale (Table S2) with the *K*-primed EoS based on our simultaneous compressional experiments of rhenium, iron, and MgO.

The parameters for the *K*-primed EoS at ambient temperature, $\rho_0$, $K_0$, $K'_0$, and $K'_\infty$, are derived from Eq. 5 using the measured densities and our rhenium scale (Table S4). To do this, the Grüneisen parameter, $\gamma_\infty$ was taken from the relation of the *K*-primed EoS, $\gamma_\infty = (3K'_\infty - 1)/6$. However, because $v_s$ for hcp-iron and $v_p$ for MgO at sufficiently high pressure conditions are not available, we performed following procedures to obtain the Grüneisen parameter using $v_p$ for hcp-iron and $v_s$ for MgO, together with our EoSs of hcp-iron and MgO determined in this study. Under ambient temperature, the differences between $K$ and $K_S$ are not large (e.g., the differences for rhenium in this study are less than 1%). Thus, if we assume that isothermal and adiabatic bulk moduli are equal ($K = K_S$) at ambient temperature, we could derive provisional $v_s$ of hcp-iron and $v_p$ of MgO from Eqs. 8 and 9 by using reference data of $v_p$ for hcp-iron (*8*) and $v_s$ for MgO (*58*) combined with our EoSs of hcp-iron and MgO, respectively. Here, we can derive the provisional values of Grüneisen parameter and Debye temperature from the Eqs. 17 and 18 as was derived for rhenium. By using those provisional parameters of Debye temperature and Grüneisen parameter with Eq. 10, we can derive the provisional isothermal bulk modulus $K$. Using this isothermal bulk modulus $K$, we derive updated values for $v_s$ of hcp-iron (and $v_p$ of MgO) and updated values for the Grüneisen parameter, $\gamma_D$, and the Debye temperature, $\Theta$. After several iterations, the isothermal bulk modulus, $K$, Grüneisen parameter, $\gamma_D$, and Debye temperature, $\Theta$, converge, giving a self-consistent set of values for hcp-iron and MgO shown in Table S2. The Grüneisen parameter of hcp-iron determined by the $K = K_S$ assumption and the converged result after iteration are shown in Fig. S18B. The difference between two values for the Grüneisen parameter is ~1%, consistent within the uncertainty of the parameter.

The thermal pressure of hcp-iron under high pressure and high temperature conditions are derived from the present *K*-primed EoS and the $\gamma_{th}$ of hcp-iron with the MGD model. The pressure at high temperature conditions is derived from the isothermal pressure at ambient conditions with the thermal pressure $P_{th}$ as:

$$P_{(\rho,T)} = P_{(\rho,300\,\mathrm{K})} + P_{\mathrm{th}(\rho,T)} - P_{\mathrm{th}(\rho,300\,\mathrm{K})}\,, \tag{20}$$

where the thermal pressure $P_{th}$ is derived from the quasi-harmonic Debye thermal pressure:

$$P_{\mathrm{th}(\rho,T)} = \gamma_{\mathrm{th}} \frac{\rho}{M} \int_0^T c_{V,\mathrm{m}(\rho,T)} \mathrm{d}T\,. \tag{21}$$

$c_{V,m}$ is assumed to be a sum of contributions from phonons and electrons (*57*) as:

$$c_{V,\mathrm{m}(\rho,T)} = c_{\mathrm{ph}(\rho,T)} + c_{\mathrm{el}(\rho,T)}\,. \tag{22}$$

As described in Methods Sect. 5, the $\gamma_{th}$ and the Debye temperature, $\Theta$, can be derived from $v_p$, $v_s$, and $\rho$ with Eqs. 11, 12, and 18, and $c_{ph}$ is derived by Eq. 19. We used the FEM-8 to the $c_{V,m}$ of iron for calculation of thermal pressure, $P_{th}$. The details of the electron contribution to the $c_{V,m}$ are given in Methods Sect. 7, and the details of the parameters of high-pressure and high-temperature EoS for hcp-iron are given in Tables S7 to S10.

## 7. Electronic contribution to heat capacity

The electronic contribution to the heat capacity is generally negligible compared to the phonon contribution at low temperatures, but it increases at higher temperature. It is important in the present context as we compare out results to shock Hugoniot done at high temperature. In particular, we consider a linear temperature dependence model (LTD) and the FEM.

The electronic contribution, $c_{el}$, may be expressed as a linear temperature relation by the electronic specific heat coefficient, $\Gamma_{el}$, combined with the density dependence [e.g., $\Gamma_{el}$ of rhenium is 2.29 mJ $K^{-2}$ $mol^{-1}$, and $\Gamma_{el}$ of iron is 4.90 mJ $K^{-2}$ $mol^{-1}$ (*59*), obtained from resistivity measurements at near absolute zero temperature]. Using the LTD, $c_{el}$ becomes comparable with $c_{ph}$ at density, $\rho \sim 12.8$ g $cm^{-3}$, and shock temperature, $T_{Hug} \sim 10000$ K on the Hugoniot curve, doubling the total heat capacity as shown in Fig. S17B. Therefore, the MGD-EoS (i.e., also shock temperature estimation), especially at high temperature, depends sensitively on how $c_{el}$ is estimated. On the other hand, recent experimental and theoretical studies of the resistivity for iron at high pressure and high temperature suggest that the resistivity of iron may be about one-half to one-third of previous estimates [e.g., refs. (*60–62*)]. In addition, it has been experimentally confirmed that there is a strong correlation between the temperature derivative of resistivity and heat capacity of iron [e.g., ref. (*63*)]. Thus, the recent low resistivity results may suggest that the actual electronic contributions to $c_{V,m}$ is lower than that estimated by the LTD. For example, Brown & McQueen (*64*) used a simplified model, FEM, to consider the electron contributions of iron theoretically and showed that the FEM proposed lower electrical heat capacity (although the FEM has been considered that the precise electron behavior cannot be estimated for transition metals, such as iron). In this work, we compare three different models of $c_{V,m}$, which is a sum of $c_{ph}$ derived by the DM and the $c_{el}$ models of $c_{el} = 0$, LTD, and FEM, as follows:

$$c_{V,\mathrm{m,DM\text{-}zero}} = c_{\mathrm{ph,DM}}\,, \tag{23}$$

$$c_{V,\mathrm{m,DM\text{-}LTD}} = c_{\mathrm{ph,DM}} + \Gamma_{\mathrm{el}} T \left( \frac{\rho_0}{\rho} \right), \tag{24}$$

$$c_{V,\mathrm{m,DM\text{-}FEM}} = c_{\mathrm{ph,DM}} + c_{\mathrm{el,\,FEM}}\,. \tag{25}$$

Deriving $c_{el,FEM}$ by using the FEM, the probability, $f_{el}$, that an energy level, $\varepsilon$, is occupied by electrons at a temperature, $T$, is expressed by the Fermi–Dirac distribution:

$$f_{\mathrm{el}(\varepsilon,\rho,T)} = \left\{ \exp\left[ \frac{\varepsilon - \mu_{(\rho,T)}}{k_{\mathrm{B}} T} \right] + 1 \right\}^{-1}, \tag{26}$$

where $k_B$ is the Boltzmann constant and $\mu_{(\rho,T)}$ is the chemical potential at density, $\rho$, and $T$. Under free electron approximation, which assumes that the valence electrons move freely among the atoms, the molar density of electron states, $D_{el}$, can be expressed as:

$$D_{\mathrm{el}(\varepsilon,\rho)} = \frac{8\pi M}{h^3 \rho} \sqrt{2 m_{\mathrm{e}}^3 \varepsilon}\,, \tag{27}$$

where $h$ is Plank constant, $M$ is the molar mass, and $m_e$ is the electron mass. The number of valence electrons per mole can be obtained by integrating the product of density of state, $D_{el}$, and effect of temperature, $f_{el}$, with respect to quasi-continuum of energies, $\varepsilon$, as:

$$N_A n_{el} = \int_{-\infty}^{+\infty} D_{el(\varepsilon,\rho)} f_{el(\varepsilon,\rho,T)} \mathrm{d}\varepsilon , \tag{28}$$

where $N_A$ is Avogadro constant and $n_{el}$ is the valence electrons in an atom [e.g., for iron, $n_{el} = 8$ ($4s^2$ $3d^6$)]. Because $n_{el}$ is independent of $\rho$ and $T$, the chemical potential $\mu_{(\rho,T)}$ can be obtained by numerically analyzing Eqs. 26 to 28 [e.g., ref. (*57*)]. Here, the total electronic contribution to the internal energy and the heat capacity by the FEM can be obtained as:

$$E_{el,\,FEM(\rho,T)} = \int_{-\infty}^{+\infty} \varepsilon\, D_{el,\,m(\varepsilon,\rho)} f_{el(\varepsilon,\rho,T)} \mathrm{d}\varepsilon , \tag{29}$$

$$c_{el,\,FEM(\rho,T)} = \left[\frac{\partial E_{el,\,FEM(\rho,T)}}{\partial T}\right]_\rho . \tag{30}$$

The differences among $c_{el}$ models have only small impact on the shock pressure, $P_{Hug}$, but have large impact on $c_{V,m}$ and the shock temperature, $T_{Hug}$ (see also Note S12, and Figs. S17 and S20). Thus, $T_{Hug}$ can be used to evaluate the validity of the Grüneisen parameter, its $\rho$-dependence, and $c_{el}$ model. We compare the calculated $T_{Hug}$ of iron (Fig. 3B) and MgO (Fig. S20B) to the experimentally available values for each material. The experimentally measured $T_{Hug}$ (*43–45*) of hcp-iron is better explained by using the FEM-8 model for $c_{el}$ (Fig. 3B). Because MgO is an insulator, the calculated $T_{Hug}$ of MgO (Fig. S20B) derived assuming $c_{el} = 0$ by Eq. 23 is also reasonable agreement with the experimentally measured $T_{Hug}$ (*47–50*). The details of the derivation of calculated $P_{Hug}$ and $T_{Hug}$ are given in Methods Sect. 8.

**8. Calculation of the shock Hugoniot from the isotherm**

Under shock compression, the conditions of the system can be derived by Rankine–Hugoniot equations as follows (*1*):

$$\rho_{Hug} = \rho_{init} \frac{U_s}{U_s - U_p} , \tag{31}$$

$$P_{Hug} = \rho_{init} U_s U_p , \tag{32}$$

$$\Delta E_{Hug} = -\frac{1}{2} P_{Hug} \left( \frac{M}{\rho_{Hug}} - \frac{M}{\rho_{init}} \right) = \frac{1}{2} M U_p^2 , \tag{33}$$

where $\rho_{init}$ is the density before shock compression; $\rho_{Hug}$, $P_{Hug}$, $\Delta E_{Hug}$ are the density, pressure, and the increase of internal energy after shock compression; $U_s$ and $U_p$ are the shock and particle velocities; and $M$ is the molar mass. A reversible path is necessary to estimate the shock energy deposited and therefore the shock temperature, $T_{Hug}$. The total increase of internal energy by shock compression is equal to the increase in the following adiabatic and isochoric processes:

$$\Delta E_{Hug} = \Delta E_S + \Delta E_V , \tag{34}$$

where $\Delta E_S$ is the increase of internal energy under the adiabatic compression from the molar volume at initial conditions, $V_{m,init}$, to the molar volume after shock compression, $V_{m,Hug}$, and $\Delta E_V$ is the increase of internal energy by the isochoric temperature increase from the temperature after the adiabatic compression, $T_S$, to the shock temperature, $T_{\mathrm{Hug}}$. In the adiabatic process, the $\Delta E_S$ can be derived as follows (*65*):

$$\Delta E_S = -\left( \int_{V_{\mathrm{m,init}}}^{V_{\mathrm{m,Hug}}} P_S \, \mathrm{d} V_{\mathrm{m}} \right)_S . \quad (35)$$

In an adiabatic process, the temperature changes while the entropy is constant giving:

$$T \, \mathrm{d} S = c_{V,\mathrm{m}} \mathrm{d} T + T \left( \frac{\partial P}{\partial T} \right)_V \mathrm{d} V_{\mathrm{m}} = 0 . \quad (36)$$

$T_S$ can be derived by integrating Eq. 36 using Eq. 16 as follows:

$$T_S = T_{\mathrm{init}} \exp\left( -\int_{V_{\mathrm{m,init}}}^{V_{\mathrm{m,Hug}}} \frac{\gamma_{\mathrm{th}}}{V_{\mathrm{m}}} \, \mathrm{d} V_{\mathrm{m}} \right) . \quad (37)$$

In the isochoric process, the $\Delta E_V$ can be derived as follows:

$$\Delta E_V = \int_{T_S}^{T_{\mathrm{Hug}}} c_{V,\mathrm{m}} \mathrm{d} T . \quad (38)$$

The shock temperature, $T_{\mathrm{Hug}}$, can be estimated using Eqs. 31 to 38. Thus, the calculated shock Hugoniot from the isotherm, or vice versa, the reduced isotherm from the shock Hugoniot, can be derived by the thermal pressure with the shock temperature and the MGD model.

## Additional information

**Acknowledgments**

The synchrotron x-ray experiments were performed at BL43LXU of the RIKEN SPring-8 Center (proposal No. 20170051, 20180055, 20190087, and 20200014). The authors thank Dr. T. Miyazaki (Tohoku Univ., Japan) for the TEM measurements.

**Funding**

JSPS KAKENHI Grant Numbers JP15H05748 and JP20H00187 (EO)
RIKEN SPring-8 Center (AB)

**Author contributions**

Conceptualization (high-pressure IXS): AB
Conceptualization (rhenium pressure scale): EO
Methodology: DIK, EO, HF, DIS, AB
Resource (sample): DIK, EO
Resource (x-ray measurement): HF, DIS, AB
Investigation: DIK (lead), EO, HF, TS, DIS, AB
Formal analysis: DIK
Validation: DIK, EO, RH, AB
Visualization: DIK
Writing—original draft: DIK, EO, AB
Writing—review & editing: DIK, EO, AB
Data curation: DIK
Project administration: DIK, EO, AB
Supervision: EO, AB
Funding acquisition: EO, AB

**Competing interests**

The authors declare that they have no competing interests.

**Data and materials availability**

All data needed to evaluate the conclusions of this research article are available in the main text and/or the supplementary materials. The supplementary data for IXS spectra and fittings can be accessed from the public repositories Zenodo (https://doi.org/10.5281/zenodo.11500515) and/or arXiv (https://doi.org/10.48550/arXiv.2104.02076).

Supplementary Materials

# Density deficit of the Earth's core revealed by a multi-megabar primary pressure scale

Daijo Ikuta*, Eiji Ohtani*, Hiroshi Fukui, Tatsuya Sakamaki,
Rolf Heid, Daisuke Ishikawa, and Alfred Q. R. Baron*

* Corresponding authors. Email: dikuta@okayama-u.ac.jp,
eohtani@tohoku.ac.jp, and baron@spring8.or.jp

***This PDF file includes:***

*Supplementary Note*
*Figures: S1–S21*
*Tables: S1–S10*
*References: 1–90*

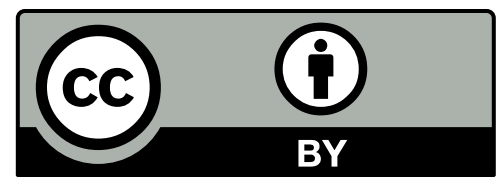

# *Supplementary Note*

## S1. Comparison of velocities between inelastic x-ray scattering and ultrasonic methods

To evaluate the validity of high-pressure experimental acoustic velocities ($v_p$ and $v_s$) obtained from phonon dispersion by inelastic x-ray scattering (IXS) method (Fig. S1), we measured $v_p$ and $v_s$ of a rhenium foil (25 μm in thickness, Nilaco) at ambient conditions (IXS-Re-foil, in air) which was pre-compressed to a thickness of ~10–15 μm. Figure S2A shows the typical IXS spectra at ambient conditions. The velocities, $v_p$ and $v_s$, of rhenium at ambient conditions measured using IXS (Fig. S2B) are consistent with ultrasonic (US) work (*32*) (Fig. 1B). Thus, we confirmed that the experimental $v_p$ and $v_s$ of rhenium with IXS method are reasonable.

## S2. Strong intensity of transverse acoustic mode

A crucial point in the present work is that the IXS spectra (Figs. 1A and S2A) showed clear peaks that we could consistently associate with both the longitudinal acoustic (LA) and transverse acoustic (TA) modes of rhenium, thereby allowing us to derive both acoustic velocities, $v_p$ and $v_s$. This is different than IXS measurements of iron-rich materials under pressure which generally only show the peak from the LA mode (e.g., *8, 52, 53*). As discussed below, experimental and calculational investigations suggest the appearance of the lower energy TA peak is the result of a quite high defect density in hexagonal close-packed (hcp) rhenium after it has been pressurized once. It is possible that these defects might shift the energies of the LA and TA peaks so that the peaks are not simply related to the relevant acoustic velocity. The extremely simplified models discussed below suggest such shifts are possible. However, ambient pressure measurements show that, in the real samples, it is reasonable to associate the position of these peaks with the TA and LA mode energies. In particular, we measured IXS spectra from a once-compressed rhenium foil and found that, after treating that data in the same way as we do the high-pressure data, we obtained the known ambient $v_p$ and $v_s$ (see Note S1). Thus, it is reasonable to associate the two peaks with the TA and LA modes at ambient pressure, and we continue to do so at high pressure. However, as this is important, we also discuss our work in more detail.

From the viewpoint of both single crystal calculations (Fig. S3) and previous IXS measurements of iron-rich materials by powder samples, the TA mode intensity at the momentum transfers used in the present work is expected to be quite weak, so the large intensity observed here requires explanation. While we initially speculated that complex (elliptical) eigen-polarizations for acoustic modes in rhenium might account for the relatively large TA intensity, in fact calculations using a first-principles model that was well validated by single crystal measurements showed this was not the case: the calculated intensity of the TA mode, in a Born approximation, was much too small to account for the observed spectra (see the black lines in Fig. S3). As the appearance of the intensity in this region and the association of that intensity with the TA mode is critical to our work, the calculation of negligible TA intensity spawned an investigation to determine the origin of the observed intensity. We should add that similar modelling was used to make estimates of the spectra in the presence of strong texture, or with a large (larger than observed) *c*-axis strain. In all cases the calculated spectra had much less intensity near the "TA" peak position than our measurements. This suggests then that either (see Note S2-A) we are not in the Born approximation limit, or (see Note S2-B) that the structure appropriate for describing the single crystal measurements was not appropriate for our powder samples. We discuss each of these in turn.

***S2-A.*** The possibility that the low energy, "TA", intensity originated from multiple scattering, or violation of the Born approximation, was ruled out by a mixture of experiment and calculation. We calculated the energy dependence of the largest multiple scattering correction, namely the double scattering process of powder diffraction (elastic Bragg scattering) followed by inelastic phonon scattering, or vice versa. Given, the x-ray diffraction (XRD) powder patterns from the sample indicate the presence of many grains within the beam spot, and the relatively small phonon cross section, we assumed no correlation between the orientation of the grain where the powder diffraction occurred and the grain where the phonon scattering takes place. That calculation, for an isotropic powder, using the well-validated phonon model mentioned in the previous paragraph, leads to a density-of-states-like contribution to the scattering that peaks strongly at ~14–15 meV and is nearly independent of momentum transfer. Both the shape and the momentum dependence do not agree with the observed intensity, suggesting any such multiple scattering contribution is small. Further, measurements at two different indentations on one foil, where the transmitted intensity varied by a factor of two, showed that the ratio of the intensity of the TA and LA peaks did not change so much when the sample thickness was changed, while one would expect that if the "TA" peak were due to multiple scattering, it would scale differently than the LA peak. These results lead us to rule out multiple scattering.

***S2-B.*** The real structure of once-compressed rhenium was determined to be different than that of a single crystal using electron microscopy. The microstructure of a once-compressed rhenium foil was observed by transmission electron microscopy (TEM) with electron diffraction patterns operated at 200 kV (JEM-2100F, JEOL) at Tohoku University. A TEM image is shown in Fig. S4 with defect locations as indicated. There is a high defect density, ~0.1–1 $nm^{-2}$. The full impact of such defects on phonon spectra is difficult to calculate, both because the exact defect structure is not easily determined and because the system it is necessary to calculate becomes very large. However, we made estimates using some simpler models with a twin boundary included and either 16 or 32 atoms per primitive cell. While these are dramatic simplifications, the calculations show (see Fig. S3) that the intensity in the TA region becomes larger, becoming similar in scale to the experimental IXS spectra. Further, cluster calculations using the force constants from the perfect crystal model were also tried and those also showed similar intensity in the TA region when the size of the clusters was reduced to a few nm scales. Thus, several calculations, albeit of simpler systems, suggest defects on a nm length scale can adequately account for the "TA" intensity. The measurements mentioned above at ambient pressure then suggest it is reasonable to use the observed peak position to derive the relevant acoustic velocities.

## S3. Comparison of compression and acoustic velocities with and without pressure medium

Figure S5A shows the experimental $c/a$ ratios of rhenium for direct compression without pressure medium and those for quasi-hydrostatic compression with periclase (MgO) pressure medium and laser annealing. The $c/a$ ratios with pressure medium are ~1.61, which are consistent with the calculated model $c/a$ ratio of rhenium (*31*). This is one of the proofs that the pressure medium and laser annealing were working well to keep quasi-hydrostatic compression. On the other hand, those without pressure medium show strong uniaxial compression in the $c$-axis direction. Figure S5B shows the relations between density, $\rho$, and $v_p$, and $\rho$ and $v_s$ of rhenium at high pressure and ambient temperature. As shown in Fig. S5B, both $v_p$ and $v_s$ measured with or without pressure medium and laser annealing are consistent within experimental errors, and the $v_p$ and $v_s$ measured at ambient conditions are also consistent with those of US measurement (*32*). The detailed explanations of small effects of uniaxial compression and uncertainty analysis are discussed in Notes S4 to S10.

### S4. Uncertainty of fitting for the phonon dispersion and acoustic velocity

Figure S6A shows the $Q_{max}$ value for the fitting of phonon dispersion of rhenium and the distance to the first Brillouin zone (BZ) of rhenium. To obtain $v_p$ or $v_s$ from the IXS results in this work, the dispersion was fitted by $v_p$ or $v_s$ and $Q_{max}$ as free parameters with Eq. 2. However, $Q_{max}$ is also related to the first BZ, as mentioned in Methods Sect. 3. To evaluate the fitting and $Q_{max}$ estimation, we compared the $v_p$ and $v_s$ derived by using $Q_{max}$ as a free parameter with those by using $Q_{max}$ fixed to the averaged value over the boundary of the first BZ in Fig. S6B. Both results are quite consistent within the errors for either $v_p$ or $v_s$ obtained by $Q_{max}$-free or $Q_{max}$-fixed fittings.

### S5. Preferred orientation analysis

In high-pressure experiments, hcp structure metals were known to have strong preferred orientations and lattice strains (*8, 53, 66*). Therefore, to estimate the accurate acoustic velocity, we should consider those effects. Figure S7 shows the experimental and calculated XRD patterns with the experimental preferred orientation or the calculated random orientation at 230 GPa (IXS-Re-12). As shown in Fig. S7A, observed intensities change depending on the azimuth angle in the experimental XRD pattern. Such intensity gradations indicate a strong preferred orientation (*8, 66*). To investigate the preferred orientation condition, we performed the XRD calculation based on the method of whole two-dimensional diffraction pattern fitting (*8, 55*). Considering the sample as an assembly of small crystal grains having individual orientations, the XRD pattern is expressed as the sum of the diffraction from all crystal grains. Therefore, the calculated XRD pattern is estimated as:

$$\mathrm{XRD}_{\mathrm{calc}} = \frac{I_0}{N}\left[\sum_{\varphi_1}\sum_{\varphi_2}\sum_{\varphi_3} w_{(\varphi_1,\varphi_2,\varphi_3)}\,\mathrm{XRD}_{(\varphi_1,\varphi_2,\varphi_3)}\right] + \mathrm{BKG}\ , \qquad \text{(S1)}$$

$$N = \sum_{\varphi_1}\sum_{\varphi_2}\sum_{\varphi_3} w_{(\varphi_1,\varphi_2,\varphi_3)}\ , \qquad \text{(S2)}$$

where $I_0$ is a constant which depends on the experimental conditions such as the sample thickness, exposure time, and sensitivity of the detector. $w_{(\varphi_1,\varphi_2,\varphi_3)}$ is the weight of a crystal grain having orientation given by the Euler angles $\varphi_1$, $\varphi_2$, and $\varphi_3$. $\mathrm{XRD}_{(\varphi_1,\varphi_2,\varphi_3)}$ is the diffraction pattern by a crystal grain having orientation given by $\varphi_1$, $\varphi_2$, and $\varphi_3$. BKG is the background of the XRD pattern. The XRD pattern is calculated by optimizing the weights, $w_{(\varphi_1,\varphi_2,\varphi_3)}$, to minimize the square of residuals between the experimental and calculated XRD patterns. Figure S7B shows the calculated XRD pattern in the preferred orientation conditions at 230 GPa (IXS-Re-12) with 5-degree increments for each of $\varphi_1$, $\varphi_2$, and $\varphi_3$ (i.e., 72×18×12 independent orientations by the hexagonal symmetry). Compared with the experimental and calculated XRD patterns, the goodness of fitting values, reduced chi-square ($\chi_v^2$) is 2.262 and weighted reliable factor ($R_w$) is 10.0%. Figure S7C shows the calculated XRD pattern by using lattice parameters at 230 GPa (IXS-Re-12) in a random orientation condition. Compared with the experimental and calculated XRD patterns, the goodness of fitting values, $\chi_v^2$ is 63.27 and $R_w$ is 69.0%. Figure S7, D and E shows the integrated XRD patterns of Fig. S7, A to C. The calculated XRD pattern with the preferred orientation is consistent with the experimental XRD pattern while the calculated XRD pattern with a random orientation is not.

Figure S8 shows the typical preferred orientation conditions at 32 GPa (IXS-Re-01) and 230 GPa (IXS-Re-12). The *c*-axis was generally close to the compression axis as shown in Fig. S8, A and B. However, this tendency was reduced with increasing pressure. The *a*-axis was also preferred in the plane that is perpendicular to the compression axis in lower pressure conditions, but again this tendency was reduced with increasing pressure as shown in Fig. S8, C and D. Figure S9 shows the concentration of *c*-axis in specific directions that is the direction inclined ~10 degrees in the vertical direction to compressional axis in all crystal grains as a function of density. For direct compression experiments, at 32 GPa ($\rho$ = 22.60 g cm$^{-3}$), *c*-axis in all crystals grains has been ten times concentrated in the specific direction around ±20 degrees regions compared with a random orientation, and even at 230 GPa, *c*-axis in all crystal grains still has been seven times concentrated in the specific direction. In contrast, for pressure medium experiments, there are still preferred orientation, but the concentrations are reduced to one-half or one-third of the value in the direct compression experiments.

**S6. Acoustic velocity anisotropy**

The hcp structure has five independent elastic moduli $C_{11}$, $C_{12}$, $C_{13}$, $C_{33}$, and $C_{44}$:

$$C_{ij} = \begin{pmatrix} C_{11} & C_{12} & C_{13} & 0 & 0 & 0 \\ C_{12} & C_{11} & C_{13} & 0 & 0 & 0 \\ C_{13} & C_{13} & C_{33} & 0 & 0 & 0 \\ 0 & 0 & 0 & C_{44} & 0 & 0 \\ 0 & 0 & 0 & 0 & C_{44} & 0 \\ 0 & 0 & 0 & 0 & 0 & C_{66} \end{pmatrix}, \quad \text{(S3)}$$

where

$$C_{66} = \frac{C_{11} - C_{12}}{2}. \quad \text{(S4)}$$

The anisotropy of $v_p$ and $v_s$ depends on the direction of the crystal lattice orientation and is derived from the $C_{ij}$ as follows (*67*):

$$v_{\mathrm{p}(\psi)} = \sqrt{\frac{C_{11}\sin^2\psi + C_{33}\cos^2\psi + C_{44} + C_{(\psi)}}{2\rho}}\ , \tag{S5}$$

$$v_{\mathrm{sv}(\psi)} = \sqrt{\frac{C_{11}\sin^2\psi + C_{33}\cos^2\psi + C_{44} - C_{(\psi)}}{2\rho}}\ , \tag{S6}$$

$$v_{\mathrm{sh}(\psi)} = \sqrt{\frac{C_{66}\sin^2\psi + C_{44}\cos^2\psi}{\rho}}\ , \tag{S7}$$

where

$$C_{(\psi)} = \sqrt{\left[\left(C_{11}-C_{44}\right)\sin^2\psi - \left(C_{33}-C_{44}\right)\cos^2\psi\right]^2 + \left(C_{13}-C_{44}\right)\sin^2 2\psi}\ , \tag{S8}$$

$\rho$ is density, $v_{\mathrm{p}(\psi)}$, $v_{\mathrm{sv}(\psi)}$, and $v_{\mathrm{sh}(\psi)}$ are compressional, vertically polarized shear, and horizontally polarized shear wave velocities in the direction of $\psi$, respectively, and the $\psi$ is the angle between the *c*-axis (in this study, approximately the compression axis as shown in Figs. S8 and S9) and the momentum transfer. As shown in Eqs. S3 to S8, the estimation for acoustic velocity anisotropy depends on the elastic moduli, $C_{ij}$. It is difficult to precisely estimate the $C_{ij}$ under non-hydrostatic conditions, however, as shown in Fig. 2, $v_p$ in this study is consistent with the first-principles generalized gradient approximation (GGA) calculation of ref. (*36*), and $v_s$ in this study is consistent with the GGA calculation of ref. (*31*), respectively. Therefore, we used two calculated $C_{ij}$ (*31, 36*) to evaluate the impact of the anisotropy. Figure S10 shows the anisotropy of the $v_p$ and $v_s$ for rhenium and the differences between the experimental $v_p$ and $v_s$ in the preferred orientation conditions and calculated $v_p$ and $v_s$ in a random orientation condition. Regardless of which of the two calculated $C_{ij}$ was used for the evaluation, both anisotropies of $v_p$ and $v_s$ for rhenium can be estimated as within ±10%, depend on the $\psi$ angle (Fig. S10, A and B). However, we should also consider the preferred orientation to estimate the experimental anisotropy of acoustic velocities, because as shown in Fig. S10, A and B, both acoustic velocities in hcp structure depending on the direction $\psi$, calculated from two calculated $C_{ij}$ (*31, 36*), are similar to the velocities in a random orientation at $\psi$ ~20–30 and ~70–80 degrees, and our experimental probability densities of *c*-axis were concentrated around $\psi$ = 80 degrees (Figs. S8 and S9). Therefore, the actual impacts of anisotropies were expected to be smaller than the maximum deviations of ±10%. The acoustic velocities in the preferred orientation, $v_{\mathrm{PO}}$, were estimated from the harmonic mean of calculated $v_{(\psi)}$ bases on the $C_{ij}$ (*31, 36*) weighted by the experimental probability densities (Figs. S8, S9, and S10, A and B) of *c*-axis as follows:

$$\frac{1}{v_{\mathrm{PO}}} = \frac{1}{\pi}\int_0^{\pi}\frac{x_{(\psi)}}{v_{(\psi)}}\,\mathrm{d}\psi\ , \tag{S9}$$

where $x_{(\psi)}$ is the normalized weight derived from the experimental preferred orientation as:

$$\frac{1}{\pi}\int_0^{\pi} x_{(\psi)}\,\mathrm{d}\psi = 1\ . \tag{S10}$$

Figure S10, C and D shows the estimated velocity differences between the experimentally observed acoustic velocities in the preferred orientation, $v_{\mathrm{PO}}$, and the velocities in a random orientation. As shown in Fig. S10, C and D, the effects of anisotropy on both $v_p$ and $v_s$ with pressure medium and laser annealing are less than ±0.5%. On the other hand, the effects of anisotropy on both $v_p$ and $v_s$ without pressure medium and laser annealing are slightly large, but still less than ±1.3%. Therefore, the effect of the preferred orientation, and also pressure medium and laser annealing are small, regardless of the differences for $C_{ij}$, and our experimental $v_p$ and $v_s$ in the preferred orientation can be estimated to be almost consistent with the $v_p$ and $v_s$ in a random orientation. As mentioned in the main text and Methods Sects. 5 to 7, our primary scale is based a Debye approximation under hydrostatic conditions. Though the part of our experimental $v_p$ and $v_s$ were under non-hydrostatic conditions and affected by the preferred orientation, we conclude that the hcp structure and the elastic properties of rhenium allow us to apply the procedure of the primary scale and the Debye approximation in the data set of $v_p$ and $v_s$ derived in the present IXS measurements.

### S7. Lattice strain analysis

Under hydrostatic pressure, bulk modulus, $K$, and shear modulus, $G$, of a strain constant average model (Voigt model) (*68*) and a stress constant average model (Reuss model) (*69*), $K_V$, $K_R$, $G_V$, and $G_R$ are:

$$K_{\mathrm{V}} = \frac{1}{9}\left[2\left(C_{11}+C_{12}\right)+4C_{13}+C_{33}\right], \tag{S11}$$

$$K_{\mathrm{R}} = \frac{C_{\mathrm{A}}}{C_{\mathrm{B}}}, \tag{S12}$$

$$G_{\mathrm{V}} = \frac{1}{30}\left[12\left(C_{44}+C_{66}\right)+C_{\mathrm{B}}\right], \tag{S13}$$

$$G_{\mathrm{R}} = \frac{5}{2}\left[\frac{C_{\mathrm{A}}C_{44}C_{66}}{C_{\mathrm{A}}\left(C_{44}+C_{66}\right)+3K_{\mathrm{V}}C_{44}C_{66}}\right], \tag{S14}$$

where

$$C_{\mathrm{A}} = C_{33}\left(C_{11}+C_{12}\right)-2C_{13}^{2}, \tag{S15}$$

$$C_{\mathrm{B}} = C_{11}+C_{12}-4C_{13}+2C_{33}. \tag{S16}$$

Because the Voigt and Reuss models are strain constant and stress constant average models, the actual bulk modulus and shear modulus have been assumed to lie between the two models as follows:

$$K_{\mathrm{VRH}} = \left(1-\alpha\right)K_{\mathrm{V}}+\alpha K_{\mathrm{R}}, \tag{S17}$$

$$G_{\mathrm{VRH}} = \left(1-\alpha\right)G_{\mathrm{V}}+\alpha G_{\mathrm{R}}, \tag{S18}$$

where $\alpha$ = 0.5, i.e., the Voigt–Reuss–Hill (VRH) average (*70*) ($\alpha$ = 0 and $\alpha$ = 1 correspond to the Voigt and Reuss models, respectively).

However, non-hydrostaticity should be considered in high-pressure experiments with a diamond anvil cell (DAC), especially without a pressure medium and laser annealing. Figure S11A shows the azimuth integrated XRD patterns at 230 GPa (IXS-Re-12) with azimuth angle, $\eta$. The position of each XRD peak, having *hkl* Miller index, is affected, depending on the azimuth angle, by lattice strains. The total deviatoric strain, $\varepsilon_{\rm exp}$, experimentally measured under non-hydrostatic pressure, is describes as follows (*71, 72*):

$$\varepsilon_{\rm exp} = \frac{d_{\rm exp} - d_{\rm hsp}}{d_{\rm hsp}} , \tag{S19}$$

where $d_{\rm exp}$ and $d_{\rm hsp}$ are experimentally measured *d*-spacing values and *d*-spacing values under hydrostatic pressure, respectively. In lattice strain theory, $d_{\rm hsp}$ can be estimated from the gradient of *d*-spacing values with azimuth angle. Here we account for following the treatment by an analytical method for lattice strains (*71, 72*). The angle between the compression axis and the normal to the diffracting crystallographic plane, $\psi$, is related to the azimuth angle, $\eta$, the angle between the compression axis and incident x-ray, $\zeta$, and the diffracting angle, $2\theta$, by the following relation (*66, 71, 72*):

$$\cos\psi = \cos\eta \sin\zeta \cos\theta - \cos\zeta \sin\theta \ . \tag{S20}$$

The deviatoric strain $\varepsilon_{\psi(hkl)}$ in the angle $\psi$ is describes as follows (*71, 72*):

$$\varepsilon_{\psi(hkl)} = \frac{d_{\psi(hkl)} - d_{{\rm hsp}(hkl)}}{d_{{\rm hsp}(hkl)}} = Q_{(hkl)}\left(1 - 3\cos^2\psi\right) , \tag{S21}$$

where $d_{\psi(hkl)}$ and $d_{{\rm hsp}(hkl)}$ are experimentally observed *d*-spacing value in the angle $\psi$ and *d*-spacing value under hydrostatic pressure, having *hkl* Miller index, respectively. Here, $d_{{\rm hsp}(hkl)}$ and $Q_{(hkl)}$ can be derived from a fit of $d_{\psi(hkl)}$ with Eq. S21, and the *c*- and *a*-axis lengths of hcp structure and the density under hydrostatic pressure can be estimated from azimuth integrated XRD patterns. Figure S11B shows the *c*- and *a*-axis lengths obtained from $d_{\psi(hkl)}$ as a function of $(1-3\cos^2\psi)$. Though the observed direction ranges are limited [$(1-3\cos^2\psi)$ ~0.4–1.0] and the direction of hydrostatic pressure [$(1-3\cos^2\psi) = 0$] could not be observed, due to the experimental optical setup, Fig. S11B clearly shows that the *c*-axis was strongly affected by the uniaxial stress and depends on the direction $\psi$, meanwhile, the *a*-axis was not affected compared with the *c*-axis. This is also consistent with the observed preferred orientation, which shows the *c*-axis being concentrated in the uniaxial compressional direction (Note S5 and Figs. S7 to S9). Figure S11C shows the density differences between experimentally observed densities and estimated densities under hydrostatic pressure calculated from $d_{{\rm hsp}(hkl)}$. A shown in Fig. S11C, the differences of the density between under hydrostatic pressure and present experimental conditions are within ±2%, and also, the density under hydrostatic pressure may be larger than experimentally observed densities in this study. This suggests that our rhenium scale may shift towards the high-density side and further away from the previous scales (see Fig. 3). Therefore, in the terms of density, we conclude that the impact of the lattice strains is small on our rhenium scale and does not affect the discussions for the Earth's inner core in the main text.

The effect of lattice strains on the shear modulus (directly related to $v_{\rm s}$ by Eq. 9, and also related to $v_{\rm p}$ by Eq. 8) should also be considered. In anisotropic linear elasticity theory, the averaged uniaxial stress component $t_{\rm avg}$ ($= \sigma_{33} - \sigma_{11}$) in the hcp structure, where $\sigma_{11}$ and $\sigma_{33}$ are radial and axial stress components, is given by the average of $t_{(hkl)}$ for all *hkl* reflections. However, because all *hkl* reflections cannot be observed, $t_{\mathrm{avg}}$ is estimated from the arithmetic mean of $t_{(hkl)}$ for all experimentally observed *hkl* reflections.

$$t_{\mathrm{avg}} = \overline{t_{(hkl)}} \, , \tag{S22}$$

where $t_{(hkl)}$ is given by $\alpha = 0.5$ (VRH average), Eqs. S11 to S21, and following equations (*71, 72*):

$$t_{(hkl)} = \frac{6}{5}\left\{\frac{G_{\mathrm{VRH}(hkl)} Q_{(hkl)}\left(3S_0^2 + 19S_0 + 3\right)}{3\alpha\left(S_0 - 1\right)\left[S_0 - \left(3S_0 + 2\right)H_{(hkl)}\right] + 5S_0}\right\} , \tag{S23}$$

$$G_{\mathrm{VRH}(hkl)} = \left(1-\alpha\right)G_{\mathrm{V}} + \alpha G_{\mathrm{R}(hkl)} \, , \tag{S24}$$

$$H_{(hkl)} = \frac{h^2k^2 + k^2l^2 + l^2h^2}{\left(h^2 + k^2 + l^2\right)^2} \, , \tag{S25}$$

$$D_{(hkl)} = \left[d_{\mathrm{hsp}(hkl)}\frac{l}{c}\right]^2 , \tag{S26}$$

$$\begin{aligned} G_{\mathrm{R}(hkl)} = \Big\{ & S_{11}\left[1 - D_{(hkl)}\right]\left[2 - 3D_{(hkl)}\right] - \left[S_{12} - 3S_{44}D_{(hkl)}\right]\left[1 - D_{(hkl)}\right] \\ & - \left[S_{13} - 2S_{13}D_{(hkl)} + S_{33}D_{(hkl)}\right]\left[1 - 3D_{(hkl)}\right]\Big\}^{-1} , \end{aligned} \tag{S27}$$

$$S_{11} = \frac{1}{2}\left(\frac{C_{33}}{C_{\mathrm{A}}} + \frac{1}{C_{11} - C_{12}}\right) , \tag{S28}$$

$$S_{12} = \frac{1}{2}\left(\frac{C_{33}}{C_{\mathrm{A}}} - \frac{1}{C_{11} - C_{12}}\right) , \tag{S29}$$

$$S_{13} = -\frac{C_{13}}{C_{\mathrm{A}}} \, , \tag{S30}$$

$$S_{33} = \frac{C_{11} + C_{12}}{C_{\mathrm{A}}} \, , \tag{S31}$$

$$S_{44} = \frac{1}{C_{44}} \, , \tag{S32}$$

$$S_0 = \frac{2\left(S_{11} - S_{12}\right)}{S_{44}} \, . \tag{S33}$$

Figure S11D shows the ratio of uniaxial stress component to shear modulus estimated from the azimuth XRD patterns (Fig. S11A) with the $C_{ij}$ from the GGA calculation of ref. (*32*). The ratio of uniaxial stress component to shear modulus in the direct compression experiments is ~2.7% at 32 GPa and 1.4% at 230 GPa, and those strains affect 1.8% and 1.0% uncertainties to $v_{\mathrm{p}}$ at 32 GPa and 230 GPa, respectively, and 1.4% and 0.8% uncertainties to $v_{\mathrm{s}}$ at 32 GPa and 230 GPa, respectively, from Eqs. 8 and 9, and decreasing with increasing pressure. On the other hand, the ratio of uniaxial stress component to shear modulus in the pressure medium and laser annealing experiments is ~0.5–1.1%, those strains affect only less than 1% uncertainties to $v_{\mathrm{p}}$ and $v_{\mathrm{s}}$. Although there are some differences in stress components in the experiments with and without pressure medium and laser annealing, the effects of lattice strains on density and acoustic velocity were expected to be canceled out each other. Therefore, we conclude that the effect of the uniaxial stress is small on our rhenium scale in this study.

### S8. Density gradient analysis

Due to non-hydrostatic high pressure conditions, the density gradient of the sample across the sample chamber should be considered. Figure S12 shows the density gradient at 230 GPa (IXS-Re-12) from the XRD patterns obtained by 2 μm steps in two direction scans perpendicular to the compression axis, i.e., vertical and horizontal scans of the DAC. Within the 5 μm region around the sample center position, which was irradiated by the full width half maximum (FWHM) size of x-ray beam, density gradients were only 0.05% in both vertical and horizontal direction of the sample chamber. Even in the 15 μm region around the sample center position which was irradiated by the x-ray beam with whole tails (*28*), density gradients were only 0.5%. Thus, density gradients in this study are negligibly small.

### S9. Relaxation of preferred orientation and lattice strain

The preferred orientation in the direct compression experiments was observed to be relaxed with increasing pressure (Figs. S8 and S9) and also the lattice strains decreased with increasing pressure (Fig. S11). One possibility to explain this relaxation of the preferred orientation and decreasing the lattice strains is that a uniaxial compression is relaxed with increasing pressure due to increased "cupping" of the diamonds. Under extreme pressures, the culet of diamonds could not keep the flatness and deformed to "cupping" shape (*73*). Figure S13A shows the transmitted x-ray intensity profiles at 32 GPa (IXS-Re-01) and 230 GPa (IXS-Re-12), respectively. The x-ray transmission increased by a factor of four to five from 32 GPa to 230 GPa. This indicates that the sample was compressed to a thickness of 10–20% with increasing pressure. The transmitted x-ray intensity was almost constant over the full sample area at 32 GPa. On the other hand, at 230 GPa, the transmitted x-ray intensity was reduced by ~25% at the center compared to the edge of the sample. Such large differences of x-ray transmission in high pressure conditions indicate large differences in the sample thickness between the sample center position and the sample edge position, i.e., the deformation of the diamond culet or "cupping". Figure S13B shows the estimation of difference in sample thickness between the sample center position and the sample edge position obtained from the transmitted x-ray intensity profiles. The differences were ~0.2 μm, and ~1% to whole sample thickness at 32 GPa ($\rho$ = 22.60 g cm$^{-3}$). However, the differences increased with increasing pressure and became more than 1 μm after 150 GPa ($\rho$ ~27.5 g cm$^{-3}$). At 230 GPa ($\rho$ = 30.24 g cm$^{-3}$), such large difference indicates the sample thickness in sample center position and sample edge position differs ~50% as shown in Fig. S13C. The difference of the sample thickness with compression has two discontinuities correlated with the experimental period of SPring-8 beamtime (2017A, 2017B, and 2018A) as shown in Fig. S13B. In this study, each pressurization was performed within 8 to 24 hours during each experimental period. On the other hand, experimental periods were several months apart. The sample shape might have been deformed from a cylinder shape to a prolate spheroid, and the state of the stress might change from uniaxial conditions toward semi-hydrostatic conditions with increasing pressure (Fig. S13C). Diamonds deformed slowly and reached equilibrium during a few months, though the pressure increase occurred instantly with compression. This may explain the relaxation of preferred orientation and lattice strain with increasing pressure.

### S10. Uncertainty analysis of the pressure scale

The impact of preferred orientation and lattice strain on our rhenium scale has been evaluated in Notes S5 to S9. Figure S14, A to C and Table S3 show contributions of uncertainty of $v_p$, $v_s$, and $\rho$ of rhenium by experimental error (Fig. S6), preferred orientation and anisotropy (Figs. S7 to S10), lattice strain (Fig. S11), density gradient (Fig. S12), and thermodynamic properties (e.g., Grüneisen parameter, see Methods Sects. 5 and 6) to calibrated pressure. If we assumed that each contribution is independent and takes its maximum value, the uncertainties of $v_p$, $v_s$, and $\rho$ become ~2–5%, ~5–16%, and ~1%, respectively. Figure S14D shows the uncertainty of calibrated pressure with these maximum deviations. Though it is a difficult task to compare our rhenium scale with previous scales (*24, 25*) because there are large discrepancies even among previous scales (*26*), our compression curve of rhenium is consistent with previous curves within those of respective experimental pressure ranges ($P$ <150 GPa). However, our compression curve differs from the curves using previous scales for rhenium (*24, 25, 29–31*) at higher pressure even considering the maximum uncertainty of our rhenium scale.

### S11. Simultaneous compression experiment

In simultaneous compression experiment of rhenium, iron and MgO, we annealed samples at temperatures over 1000 K by a double-sided laser heating method using a fiber laser (COMPAT system) (*51*) to minimize the deviatoric stress in the samples. Figure S15 shows the *c*/*a* ratios of rhenium and hcp-iron under the present simultaneous density measurement of rhenium and iron with the MgO pressure medium and laser annealing. Both experimental *c*/*a* ratios of rhenium and hcp-iron are consistent with the calculated model *c*/*a* ratios (*31, 36*) even over 200 GPa, which indicate that annealing of the sample worked well to release the deviatoric stress caused by uniaxial compression. Figure S16 show the present calibrated $K$-primed EoS of MgO. Our present compression curve of MgO is consistent with the curves based on previous scales (*14, 40, 41*) up to those respective experimental pressure range within the uncertainties of our scale (Fig. S16A). On the other hand, our compression curve of MgO cannot be accounted for by a single theoretical approximation. Figure S16B shows the comparison between our curve and several theoretical compression curves of MgO (*74–77*). Our compression curve is consistent with molecular dynamics (*74*) up to ~75 GPa, but it shows better agreement with local density approximation (LDA) with linearized augmented plane wave method (LAPW) (*77*) from ~75 GPa to ~200 GPa. At above 200 GPa, the difference from LDA+LAPW gradually increases, but on the other hand, it is closer to the curve by quantum Monte Carlo method (*76*) within uncertainty. The curve by LDA with pseudopotential (PP) (*75*) is inconsistent with our curve in most experimental range except low pressures ($P$ <20 GPa).

### S12. Consistency of our rhenium scale with shock Hugoniot

The reduced isotherms from Hugoniot curve were traditionally used as a primary pressure scale (*29, 30, 65, 78–80*). To consider the validity of our rhenium scale, we evaluated the consistency between our scale and shock Hugoniot. As shown in Figs. 3A and S17A, the calculated shock Hugoniot from the isotherm for hcp-iron calibrated our rhenium scale is consistent with shock Hugoniot of iron in ref. (*42*). The shock temperature, $T_{Hug}$, also can be used to evaluate the validity of our present pressure scale. Figure S17B shows the comparison of calculated $c_{V,\mathrm{m}}$ of hcp-iron on the Hugoniot curve by the three different $c_{\mathrm{el}}$ models shown in Eqs. 23 to 25. The $c_{V,\mathrm{m,DM\text{-}zero}}$ model is almost constant as $3R$ (where $R$ is the gas constant). The $c_{V,\mathrm{m,DM\text{-}LTD}}$ model shows about twice of the $c_{V,\mathrm{m,DM\text{-}zero}}$ at higher pressures. In the $c_{V,\mathrm{m,DM\text{-}FEM}}$ model, the contributions of electron is about one-half to one-third of contributions of phonon, that may be consistent with the low-resistivity of hcp-iron at high pressure and high temperature. As shown in Fig. S17A, the calculated Hugoniot curves of hcp-iron by three different $c_{\mathrm{el}}$ models show only small differences within the uncertainties. However, the shock temperature, $T_{\mathrm{Hug}}$, strongly depends on the $c_{\mathrm{el}}$ model as shown in Fig. S17C (calculation of $T_{\mathrm{Hug}}$ is described in Methods Sect. 8). The experimental $T_{\mathrm{Hug}}$ of hcp-iron has large variations (*43–45*), but most of the $T_{\mathrm{Hug}}$ can be accounted for by either $c_{V,\mathrm{m,DM\text{-}zero}}$ or $c_{V,\mathrm{m,DM\text{-}FEM}}$ models (Figs. 3B and S17C). Though the differences of $T_{\mathrm{Hug}}$ between three $c_{V,\mathrm{m}}$ models increases with pressures, the $\rho$-deficits at inner core boundary (ICB) conditions by $c_{V,\mathrm{m,DM\text{-}zero}}$ and $c_{V,\mathrm{m,DM\text{-}LTD}}$ models are consistent with 8(±2)% in the $c_{V,\mathrm{m,DM\text{-}FEM}}$ model within the uncertainties (Fig. S17D). Thus, to discuss the Earth's inner core, the differences among $c_{V,\mathrm{m}}$ models, especially between $c_{V,\mathrm{m,DM\text{-}zero}}$ and $c_{V,\mathrm{m,DM\text{-}FEM}}$ models, which can account for experimental $T_{\mathrm{Hug}}$, have only small impact on the present conclusions.

The MGD model is widely used for high-pressure and high-temperature EoS, and the Grüneisen parameter, $\gamma$, together with molar heat capacity at constant volume, $c_{V,\mathrm{m}}$, is critical as it is directly related to thermal pressure. Within the MGD model, the Grüneisen parameter represents the effect of crystal lattice volume change on its vibrational properties (*1, 39*). Therefore, the Grüneisen parameter can be derived from the $\rho$-dependence of $v_{\mathrm{p}}$ and $v_{\mathrm{s}}$ as shown in Eqs. 11, 12 and 18. The $\gamma_{\mathrm{th}}$ of hcp-iron was derived from the isothermal bulk modulus, $K$, by our EoS of hcp-iron with the experimental $v_{\mathrm{p}}$ of hcp-iron (*8*) by using $K = K_S$ assumption as the initial value, and iterating for optimization of the parameters of the Debye temperature and the Grüneisen parameter to be consistent with both isothermal and adiabatic bulk moduli, and $v_{\mathrm{p}}$ and $v_{\mathrm{s}}$. The details are given in Methods Sect. 6. Figure S18B shows the comparison between the initial values of Grüneisen parameter with $K = K_S$ assumption and the final values of Grüneisen parameter after the iterations of five times for optimizations. As shown in Fig. S18B, the difference between the Grüneisen parameters is quite small within the uncertainties and there is almost no effect on conversion from isotherm to the shock Hugoniot. On the other hand, large discrepancies in the Grüneisen parameter exist between the previously proposed value (*5*) and that derived from the experimental $v_{\mathrm{p}}$ (*8*) and the EoS of hcp-iron (*5*) as shown in Fig. S18B. This means that our EoS is consistent with both the shock Hugoniot and experimental $v_{\mathrm{p}}$ of hcp-iron.

Figure S19A shows our present isotherm, and calculated and experimental Hugoniot of rhenium. Because there is no experimental $T_{\mathrm{Hug}}$ data for rhenium, only the differences between the $c_{V,\mathrm{m,DM\text{-}zero}}$ and $c_{V,\mathrm{m,DM\text{-}LTD}}$ [the electronic specific heat coefficient, $\Gamma_{\mathrm{el}}$, of rhenium is fixed to 2.29 mJ $K^{-2}$ $mol^{-1}$ (*59*)] models are discussed here. Similar to hcp-iron, the two $c_{V,\mathrm{m}}$ models for rhenium reproduce almost the same Hugoniot curves, which are consistent with the experimental shock Hugoniot (*29, 30*) within the uncertainties, though there are large differences in calculated $c_{V,\mathrm{m}}$ and $T_{\mathrm{Hug}}$ on the Hugoniot curve (Fig. S19B). However, a larger difference between the calculated and experimental Hugoniots of rhenium is observed compared with that of hcp-iron, though it is still within the uncertainty. While hcp-iron shows a good agreement, one possible reason for deviation in rhenium is that the present estimate for $c_{\mathrm{el}}$ of rhenium, which is a high atomic number element, is not sufficient. The effect of electrons on the $c_{V,\mathrm{m}}$ and the effect of anharmonicity neglected in the approximation of quasi-harmonic motion in rhenium may be larger than hcp-iron. Because rhenium has many electrons and heavier atomic weight than iron, the temperature dependence of the Grüneisen parameter in rhenium may be too large to be ignored. Further experiments are necessary to discuss the high temperature state of rhenium.

Figure S20A shows the calculated shock Hugoniot of MgO (B1, rock salt type cubic structure), which is consistent with the experimental shock Hugoniot (*47, 79, 81*) within uncertainties. The $\gamma_{\mathrm{th}}$ of MgO was derived from the isothermal bulk modulus, $K$, by our EoS of MgO with the experimental $v_{\mathrm{s}}$ of MgO (*57*) by using $K = K_S$ assumption as the initial value, and iterating for optimization of the parameters of the Debye temperature and the Grüneisen parameter to be consistent with both isothermal and adiabatic bulk moduli, and $v_{\mathrm{p}}$ and $v_{\mathrm{s}}$. The details are given in Methods Sect. 6. Figure S20B shows the calculated $T_{\mathrm{Hug}}$ of MgO, which is consistent with the experimental $T_{\mathrm{Hug}}$ (*47–50*) within uncertainties.

## S13. Summary of our rhenium scale and previous scales

The differences in experimental conditions between our experiments and other experiments for rhenium (*24, 25, 29, 30, 35*) are summarized in Fig. S21. As shown in Fig. S21, previous experiments were calibrated by shock compression data (*29, 30, 78–80*), gas pressure gauge (*82*), and theoretical work (*83, 84*). In most high-pressure experiments, the pressure standard such as rhenium (*24, 25*), gold (*78, 82, 85*), helium (*86*), tungsten (*24, 85*), and ruby (*21, 85, 87*) have been used to calibrate pressure. Those pressure standards are mostly secondary pressure scales and are calibrated by other pressure scales derived by shock compression data (*29, 30, 78–80*) or theoretical work (*83, 84*). Pressures in several previous studies are calibrated by the pressure scales based on the density–velocity relations (*14, 20*) as this study (see also Methods Sect. 5). However, their applicable pressure ranges, up to 55 GPa (*14*) and 120 GPa (*20*), are below the core–mantle boundary pressures.

Shock compression data has been widely used as a primary pressure standard, but as mentioned in Methods Sect. 8 and Note S12, the shock compression data should be converted to isothermal conditions by Rankine–Hugoniot equations and density dependence of Grüneisen parameters. However, previous isotherms were derived from the estimated Grüneisen parameter, which is not based on acoustic velocity (see Methods Sects. 6 and 7, and Fig. S18B). The density, $\rho$, dependence of the Grüneisen parameter, $\gamma$, was assumed simply as $\gamma\rho$ = constant due to the lack of the acoustic velocity data at high pressure in previous studies (*29, 30, 78–80*) and the consistency with acoustic velocity was not considered to derive the pressure scale. This might explain the differences between our scale and previous scales.

## *Supplementary Figures*

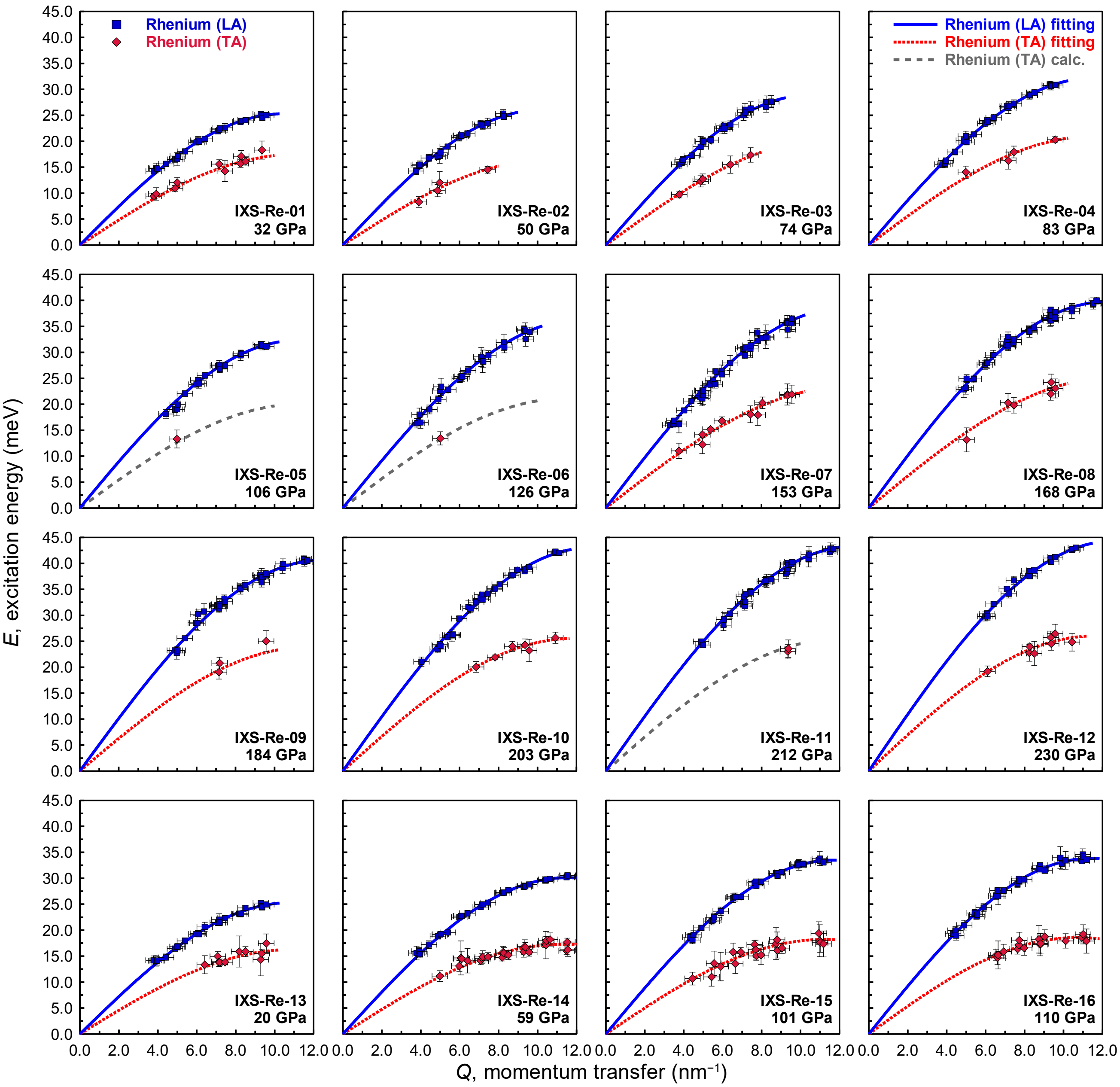

**Fig. S1.** The phonon dispersion and fits for rhenium in all high-pressure runs. Blue squares are LA modes and red diamonds are TA modes. The horizontal bars give the $Q$ resolution while the vertical bars indicate the one standard deviation ($1\sigma$) uncertainties from the fits. Colored lines are the fitting result of phonon dispersion by the sine function; blue solid: LA phonons, red dotted: TA phonons. Gray dashed lines show that partial data sets for TA dispersion are consistent with the interpolation in this work (see main text, Fig. 1B, Methods Sect. 3, and Note S4).

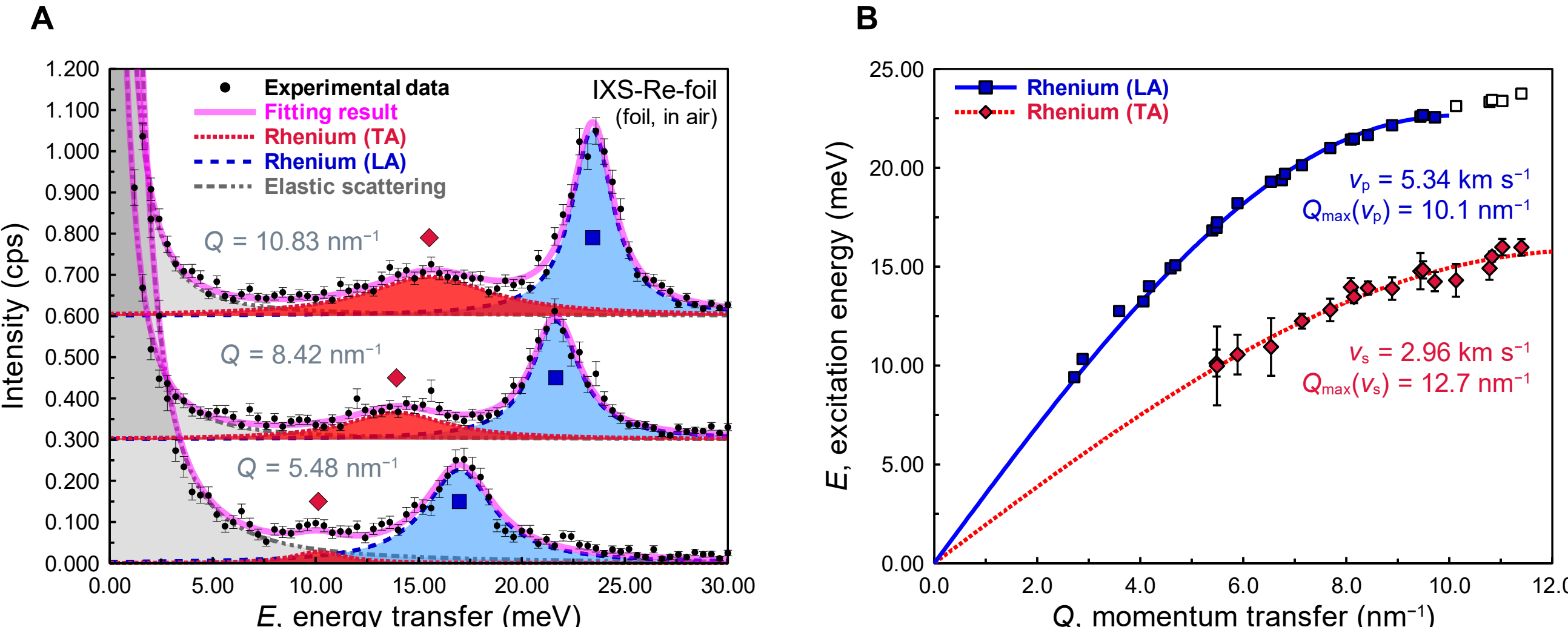

**Fig. S2.** (**A**) IXS spectra and fitting results for a rhenium foil in ambient conditions after indentation at three different momentum transfers ($Q$ = 5.48, 8.42, and 10.83 nm$^{-1}$). The black dots are the IXS data with 1σ error bars. Other colored lines and areas are individual inelastic contributions of LA and TA modes as labeled. (**B**) The phonon dispersion and fits for rhenium at ambient conditions. The colored symbols are individual LA and TA modes of rhenium at ambient conditions. The error bars represent the 1σ uncertainties. The colored lines are the fitting result of phonon dispersion by the sine function (Eq. 2); blue: LA phonons, red: TA phonons.

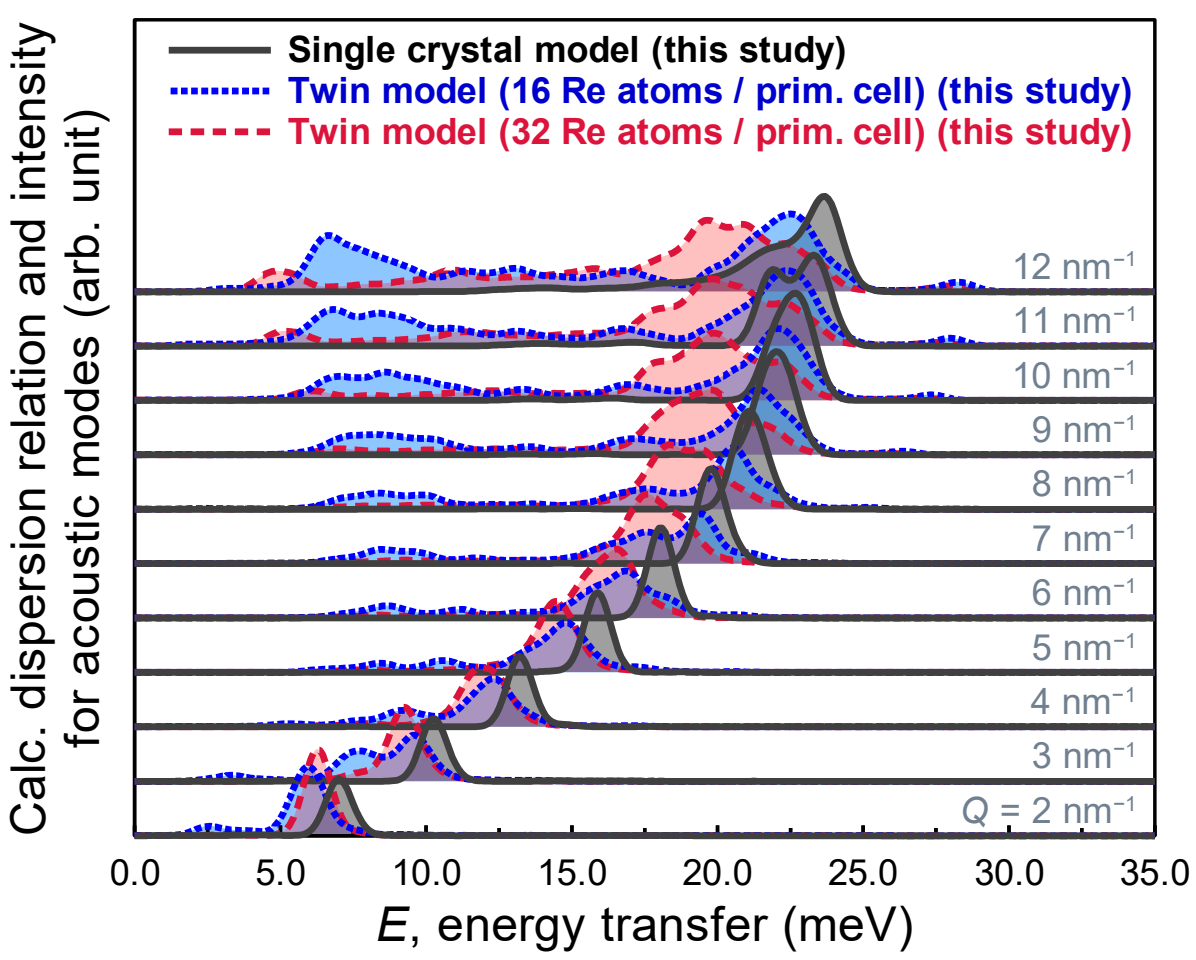

**Fig. S3.** Calculated IXS spectra for rhenium using three different models including a perfect single crystal model (black solid lines/areas) and two models with a twin boundary with 16 atoms per primitive cell (blue dotted lines/areas) and 32 atoms per primitive cell (red dashed lines/areas). The momentum transfer $Q$ of each line is indicated on the right side. In each case, the powder spectrum is calculated by keeping the scattering angle fixed and integrating over all possible crystal orientations.

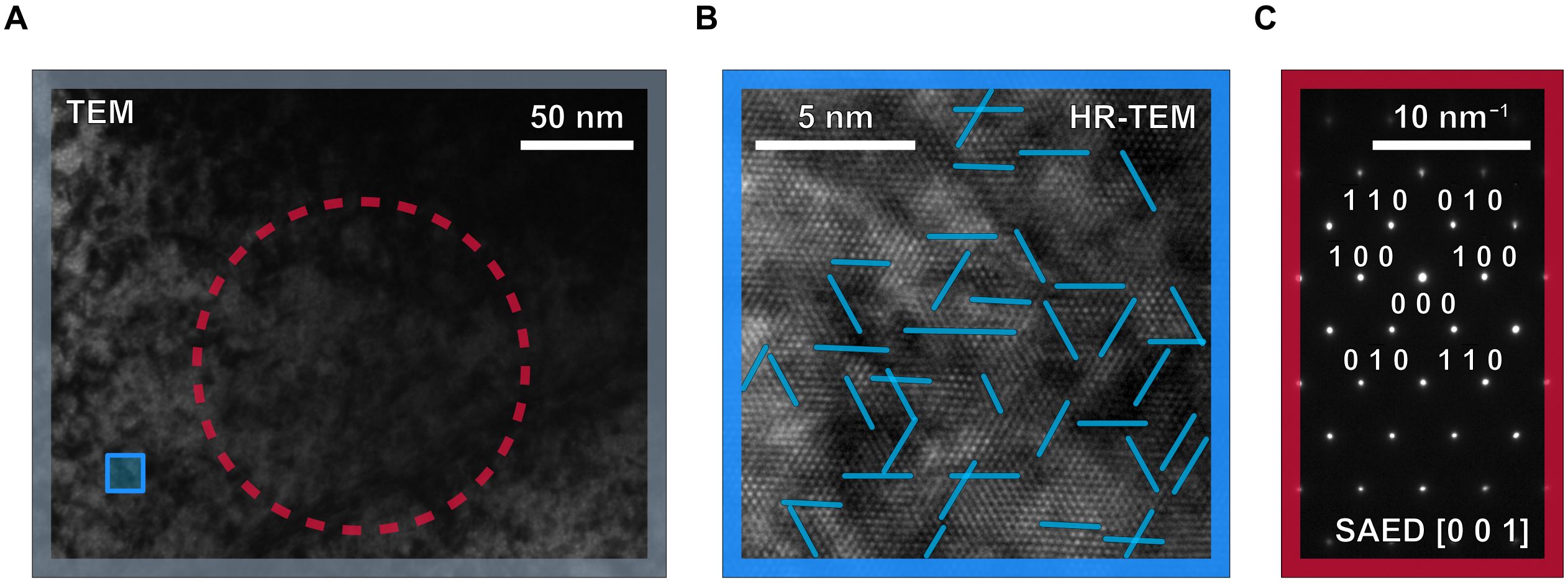

**Fig. S4.** (**A**) A transmission electron microscopic (TEM) image of ~15 μm indented rhenium foil. The blue square represents the areas of high-resolution (HR) TEM image in (B). The red dashed circle represents the area of the selected area electron diffraction (SAED) pattern measurement in (C). The white line is the scale bar of 50 nm. (**B**) HR-TEM images of the indented rhenium foil. The blue lines in the HR-TEM image show the dislocations observed in the crystallographic direction. The white line is the scale bar of 5 nm. (**C**) The SAED pattern along [0 0 1] zone axis of hcp rhenium. The white line is the scale bar of 10 nm$^{-1}$.

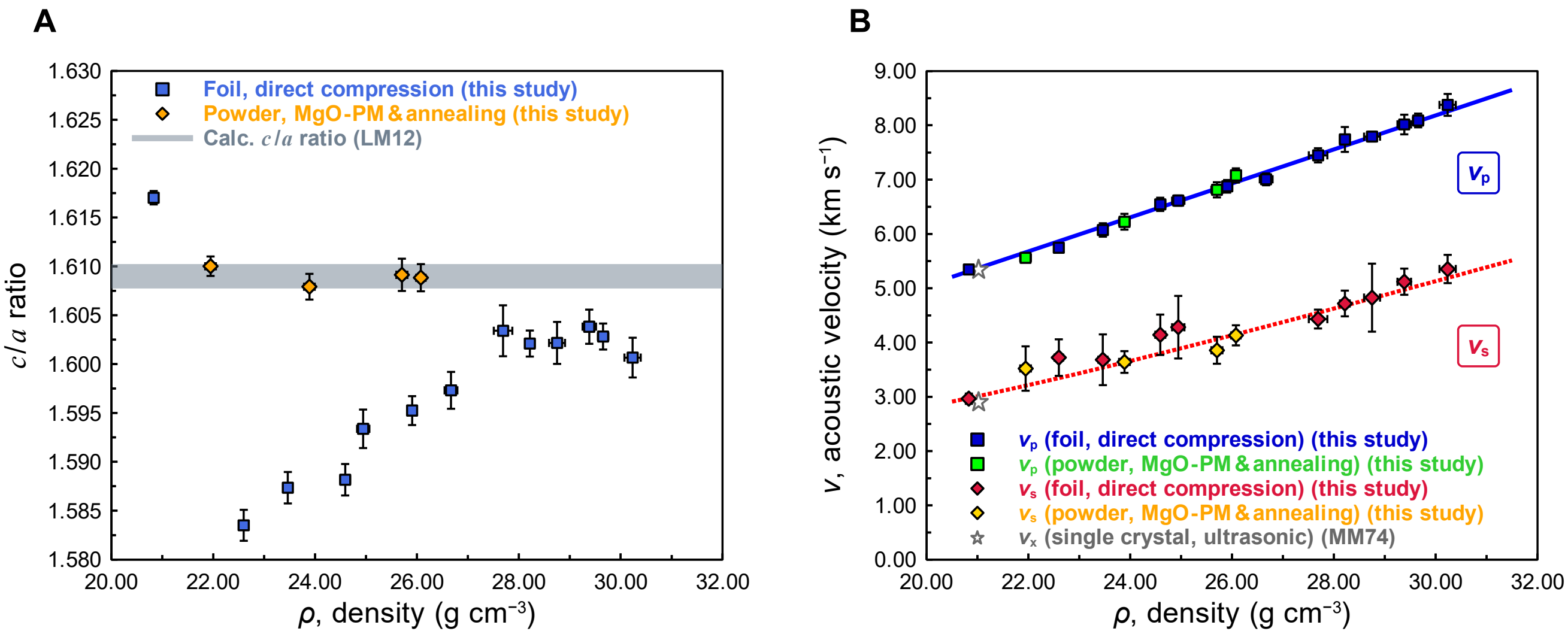

**Fig. S5.** (**A**) The *c*/*a* ratio of rhenium as a function of density. The color symbols represent the experimental *c*/*a* ratio, blue squares: direct compression experiments (IXS-Re-01 to IXS-Re-12, and IXS-Re-foil) and orange diamonds: experiments with MgO pressure medium and laser annealing (IXS-Re-13 to IXS-Re-16) with 1σ error bars. The gray broad line represents the calculated model *c*/*a* ratio of rhenium (*31*). (**B**) Compressional ($v_p$) and shear ($v_s$) wave velocities for rhenium as a function of density. The blue squares and red diamonds are $v_p$ and $v_s$ for rhenium, respectively, by direct compression experiments (IXS-Re-01 to IXS-Re-12, and IXS-Re-foil) with 1σ error bars. The green squares and yellow diamonds are $v_p$ and $v_s$ for rhenium, respectively, by experiments with MgO pressure medium and laser annealing (IXS-Re-13 to IXS-Re-16) with 1σ error bars. The results of two-type experiments are consistent within the errors. Gray star symbols are from the ultrasonic data at ambient conditions (MM74) (*32*).

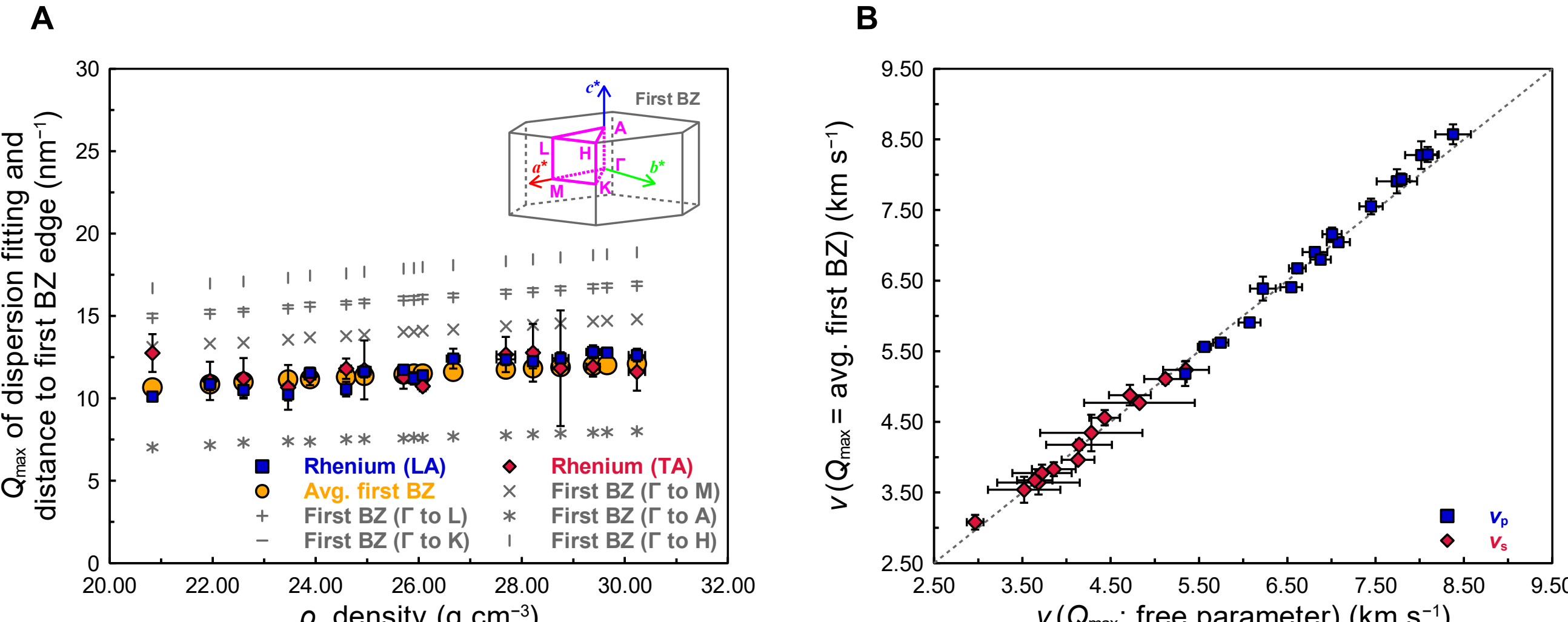

**Fig. S6.** (**A**) The $Q_{max}$ value used in the fitting. The $Q_{max}$ value from the fitting as shown in Figs. S1 and S2A, and the first BZ calculated from the lattice parameters. The blue square and red diamond symbols represent the $Q_{max}$ value from the fitting for longitudinal acoustic (LA) and transverse acoustic (TA) phonons of rhenium, respectively. The yellow circles represent the averaged first BZ. Each gray symbol represents the first BZ at the critical points (A, M, K, L, and H) of the first BZ that has high symmetry in the hcp structure. The inset represents the first BZ of the hcp structure. Colored arrows represent the reciprocal axis direction ($a^*$, $b^*$, and $c^*$) of the hcp structure and magenta line represents the asymmetric unit for the first BZ of the hcp structure. Typical high symmetry points of the first BZ are Γ(0, 0, 0), A(0, 0, 1/2), M(1/2, 0, 0), K(1/3, 1/3, 0), L(1/2, 0, 1/2), and H(1/3, 1/3, 1/2) in the reciprocal lattice (*88*). (**B**) Comparison of $v_p$ and $v_s$ determined by fitting with $Q_{max}$ as a free parameter or $Q_{max}$ fixed to the averaged value over the boundary of the first BZ. The error bars represent 1 σ uncertainties.

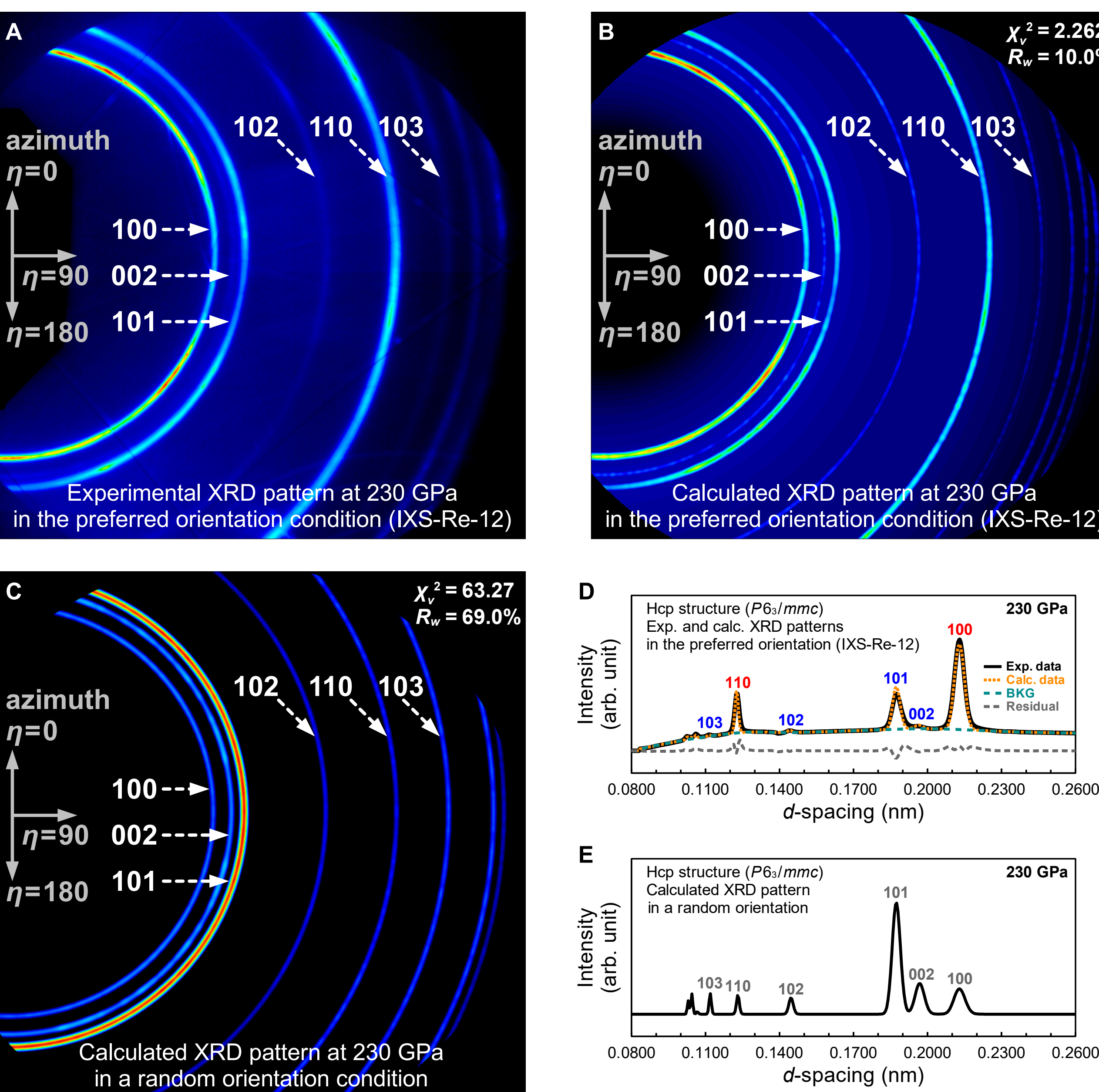

**Fig. S7.** XRD patterns of rhenium at 230 GPa (IXS-Re-12). (**A**) The experimental XRD pattern in the preferred orientation condition at 230 GPa (IXS-Re-12). The gray arrows represent the azimuth angle of the diffraction. The white numbers give the *hkl* Miller indices of ring. (**B**) The calculated XRD pattern in the preferred orientation condition at 230 GPa (IXS-Re-12). The goodness of fitting parameters compared with the experimental XRD pattern are shown in figures. (**C**) The calculated XRD pattern by using the lattice parameters obtained from experimental XRD pattern and assumed as a random orientation condition at 230 GPa (IXS-Re-12). (**D**) The integrated experimental and calculated XRD profiles in the preferred orientation condition. Black line represents the integrated experimental XRD profile. Each colored dashed and/or dotted line represents the calculated XRD profiles as mentioned in the figure. The colored numbers represent the *hkl* Miller indices of each diffraction. The red and blue numbers in the experimental XRD pattern indicate that the relative intensities are increasing and decreasing compared with the calculated XRD pattern of a random orientation, respectively. (**E**) The integrated calculated XRD profile in a random orientation condition. The gray numbers represent the *hkl* Miller indices of each diffraction.

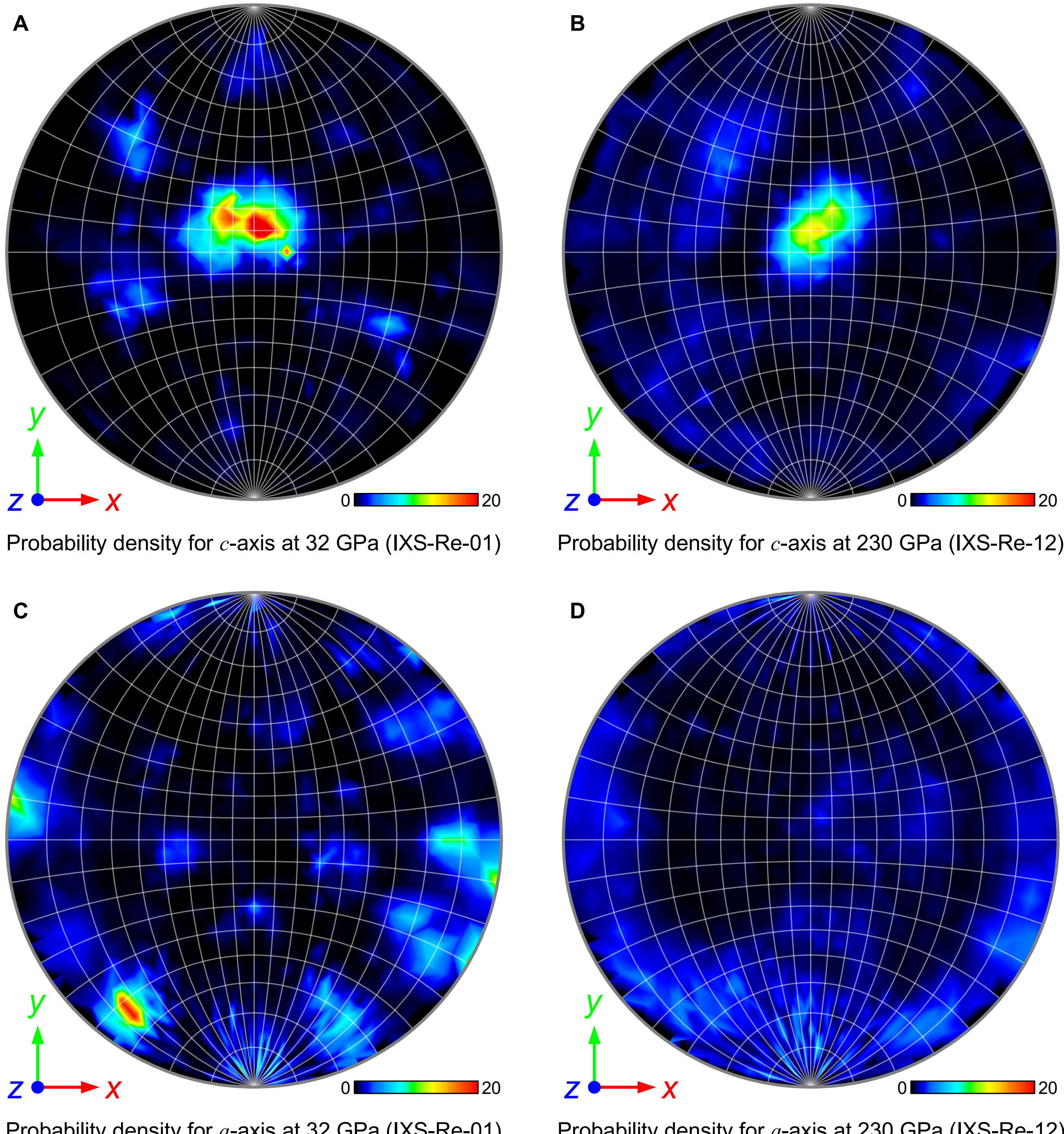

**Fig. S8.** Experimentally determined stereographic projection of the probability density for *c*-axis and *a*-axis of rhenium showing the preferred orientation. The probability density for *c*-axis at (**A**) 32 GPa (IXS-Re-01) and (**B**) 230 GPa (IXS-Re-12), and *a*-axis at (**C**) 32 GPa and (**D**) 230 GPa. The *x*-*y*-*z* arrows represent each direction of the experimental apparatus; *x*: horizontal, *y*: vertical, and *z*: compressional directions of a DAC, respectively. The color contour bar represents the probability density value for each axis. Under a random orientation condition, the probability density in all directions is 1. The probability density less/greater than 1 means lower/higher probability for crystal grains having the specific directions than at random condition, and 0 means that there is no crystal grain of the direction.

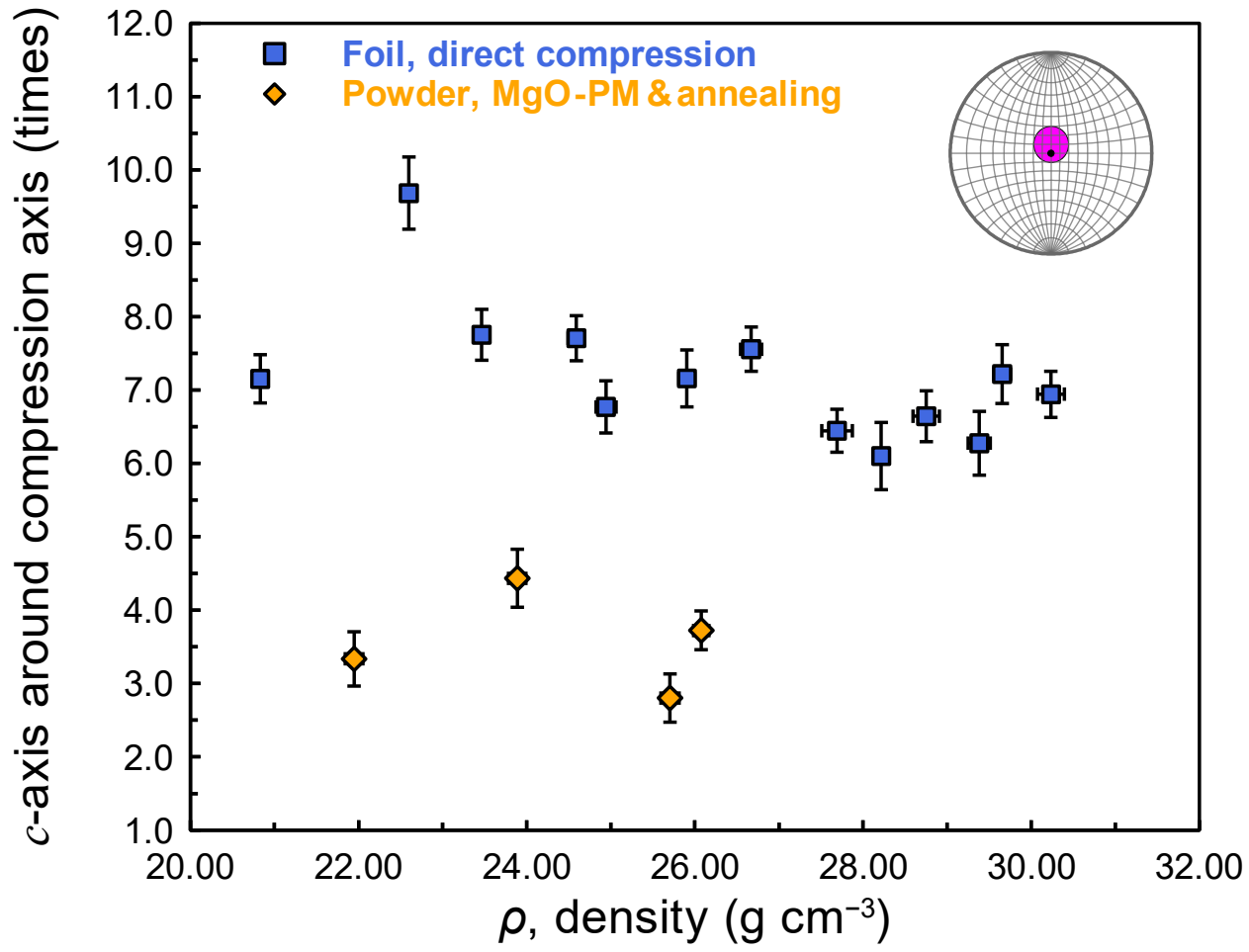

**Fig. S9.** The concentration of *c*-axis in specific directions for all crystal grains as a function of density. The *c*-axis is concentrated in the direction inclined ~10 degrees in the vertical direction (see Fig. S8). The color symbols represent the concentration of *c*-axis around the specific directions that are shown in the inset which represents the ±20 degrees integration area of the concentration of *c*-axis. The error bars represent the 1σ uncertainties.

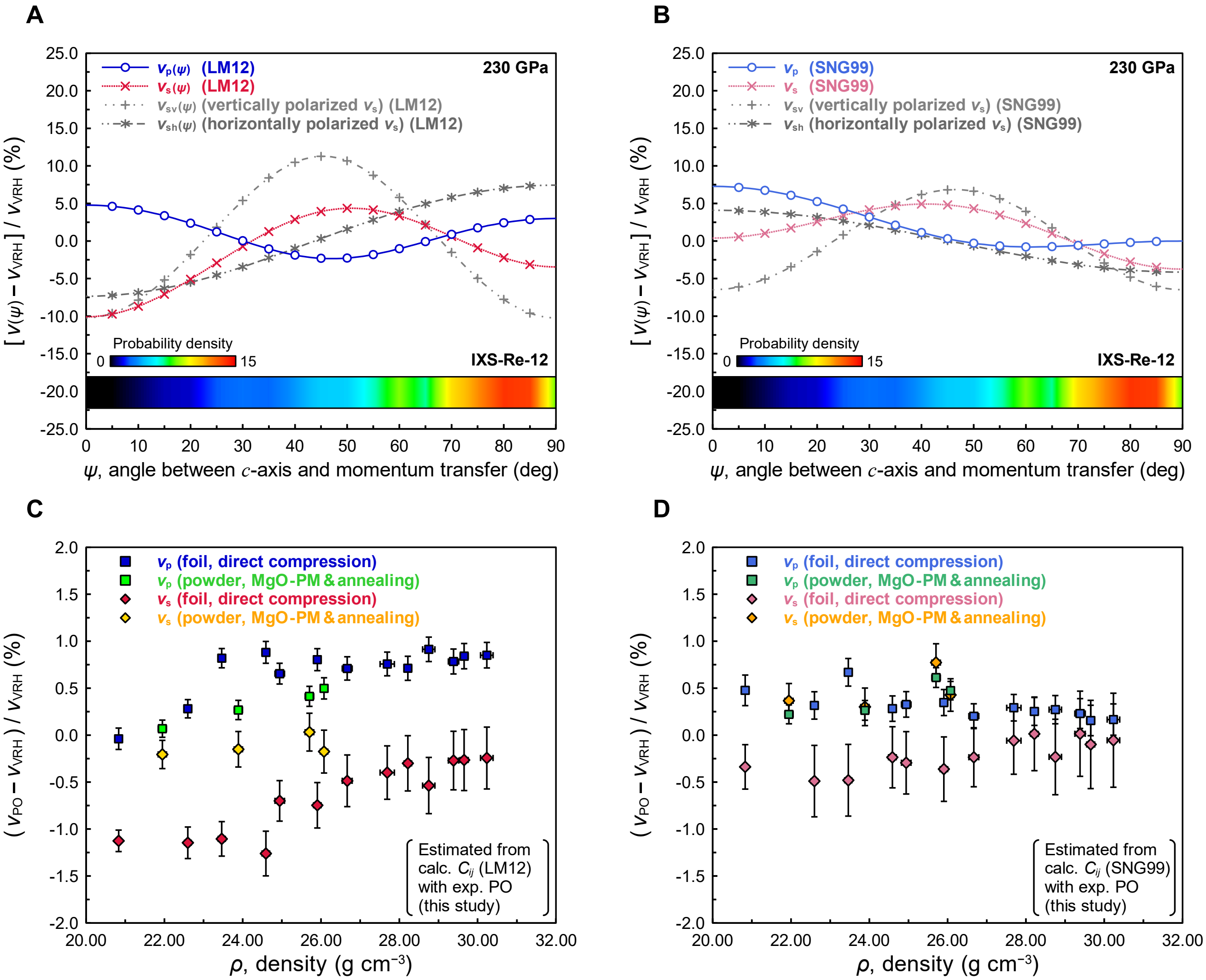

**Fig. S10.** Estimation of the anisotropy of acoustic velocity in the experimental preferred orientation (PO) conditions. Calculated anisotropies of $v_p$ and $v_s$ based on the $C_{ij}$ by using GGA calculations (LM12) of ref. (*31*) (**A**) and GGA calculations (SNG99) of ref. (*36*) (**B**), respectively, compared with the average $v_p$ and $v_s$ of Voigt–Reuss–Hill average (VRH) as a function of angle, $\psi$, that is the angle between the $c$-axis (approximately to the compression axis as shown in Figs. S8 and S9) and the lattice vibration direction due to inelastic x-ray scattering. The color bars indicate our experimentally determined distribution of PO. The acoustic velocity difference, $(v_{PO}-v_{VRH})/v_{VRH}$ for $v_p$ and $v_s$ along $c$-axis as a function of density. The $v_{(\psi)}$ and $v_{VRH}$ was calculated from the $C_{ij}$ by using GGA calculations (LM12) of ref. (*31*) (**C**) and GGA calculations (SNG99) of ref. (*36*) (**D**), respectively, with VRH for random orientation conditions. The $v_{PO}$ was calculated with the integral of $v_{(\psi)}$ weighted by the probability density in the direction of $\psi$ (see Eqs. S9 and S10) based on the PO derived from the present XRD patterns (Figs. S7 to S9). Each colored symbol and line are mentioned in the figure. The error bars represent the 1σ uncertainties.

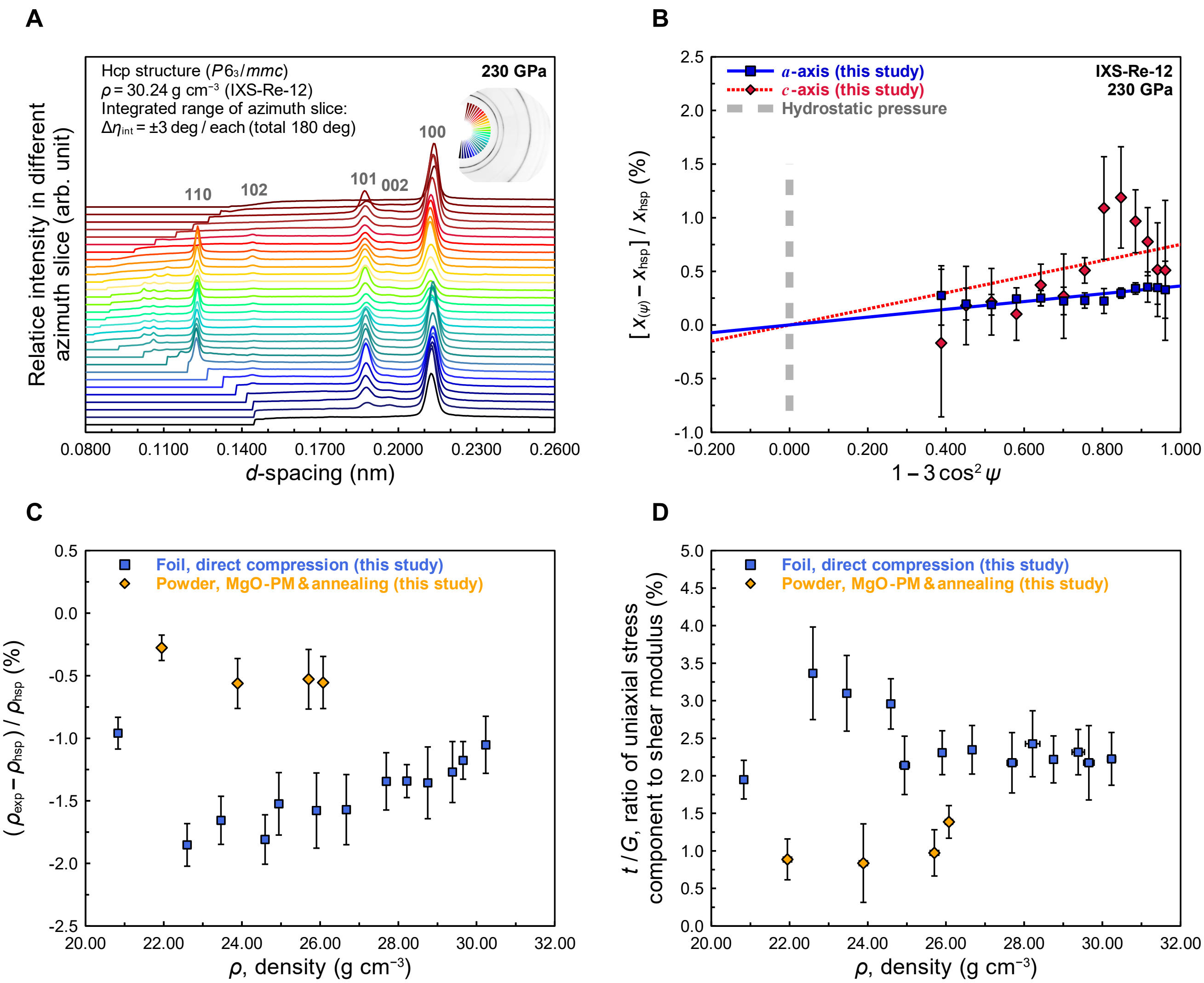

**Fig. S11.** (**A**) Series of thirty integrated XRD profiles with 6-degree intervals for different integrated azimuth angle range ($\Delta\eta_{\mathrm{int}}$ = ±3 degrees) from $\eta$ = +8 to +189 degrees at 230 GPa (IXS-Re-12). The analyzed XRD pattern is given in Fig. S7A. The colors in the upper right inset represents the integrated directions in different azimuth slice, corresponding the colored integrated XRD profiles. The numbers in the figure represent the *hkl* Miller indices. The abrupt intensity changes at small *d*-spacing are due to the opening angle of the DAC and the geometry of the flat panel detector (see Fig. S7A). (**B**) *a*- and *c*- axis length differences, $[x_{(\psi)}{-}x_{\mathrm{hsp}}]/x_{\mathrm{hsp}}$, between the experimentally observed *a*- and *c*-axis lengths, $a_{\psi(hkl)}$, and $c_{\psi(hkl)}$ in the direction $\psi$, and the *a*- and *c*-axis lengths under hydrostatic pressure, $a_{\mathrm{hsp}}$ and $c_{\mathrm{hsp}}$, estimated from analyzed XRD pattern of rhenium (Figs. S7A and S11A), as a function of $(1{-}3\cos^2\psi)$, where $\psi$ is the angle between the compression axis and the normal to the diffracting crystallographic plane, blue squares and red diamonds: the *a*- and *c*-axis of rhenium, respectively. The direction of $(1{-}3\cos^2\psi) = 0$ means the hydrostatic pressure in anisotropic linear elasticity theory (*71, 72*). (**C**) The density difference, $(\rho_{\mathrm{exp}}{-}\rho_{\mathrm{hsp}})/\rho_{\mathrm{hsp}}$, between the experimentally observed density, $\rho_{\mathrm{exp}}$, and the density under hydrostatic pressure, $\rho_{\mathrm{hsp}}$, estimated from analyzed XRD pattern (Figs. S7A and S11A) as a function of density, blue squares: direct compression without pressure medium experiments (IXS-Re-01 to IXS-Re-12, and IXS-Re-foil) and orange diamonds: experiments with MgO pressure medium and laser annealing (IXS-Re-13 to IXS-Re-16). The error bars represent the 1σ uncertainties. As shown in figure, $(\rho_{\mathrm{exp}}{-}\rho_{\mathrm{hsp}})/\rho_{\mathrm{hsp}}$ in all experiments are negative, which means that the experimentally observed density may be smaller than the density under hydrostatic pressure due to the lattice strains. (**D**) Ratio of uniaxial stress component, *t*, to shear modulus, *G*, estimated from the azimuth slices of XRD pattern (Fig. S10A) as a function of density, blue squares: direct compression without pressure medium experiments (IXS-Re-01 to IXS-Re-12, and IXS-Re-foil) and orange diamonds: experiments with MgO pressure medium and laser annealing (IXS-Re-13 to IXS-Re-16). The error bars represent the 1σ uncertainties.

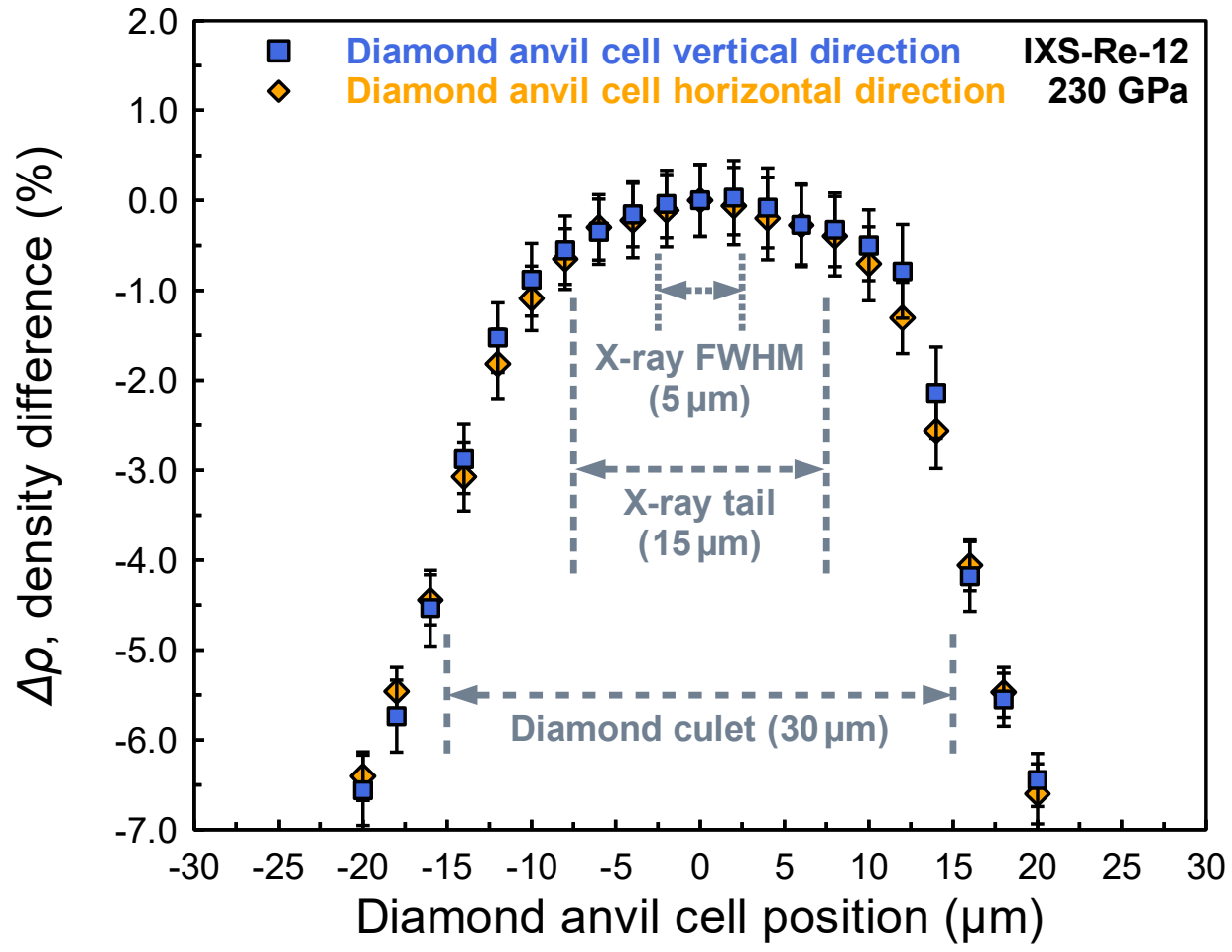

**Fig. S12.** Density gradient across the sample center at 230 GPa (IXS-Re-12), blue squares and orange diamonds: two-directional (i.e., vertical and horizontal) scans perpendicular to the compression axis of the DAC, respectively. The error bars represent the 1σ uncertainties.

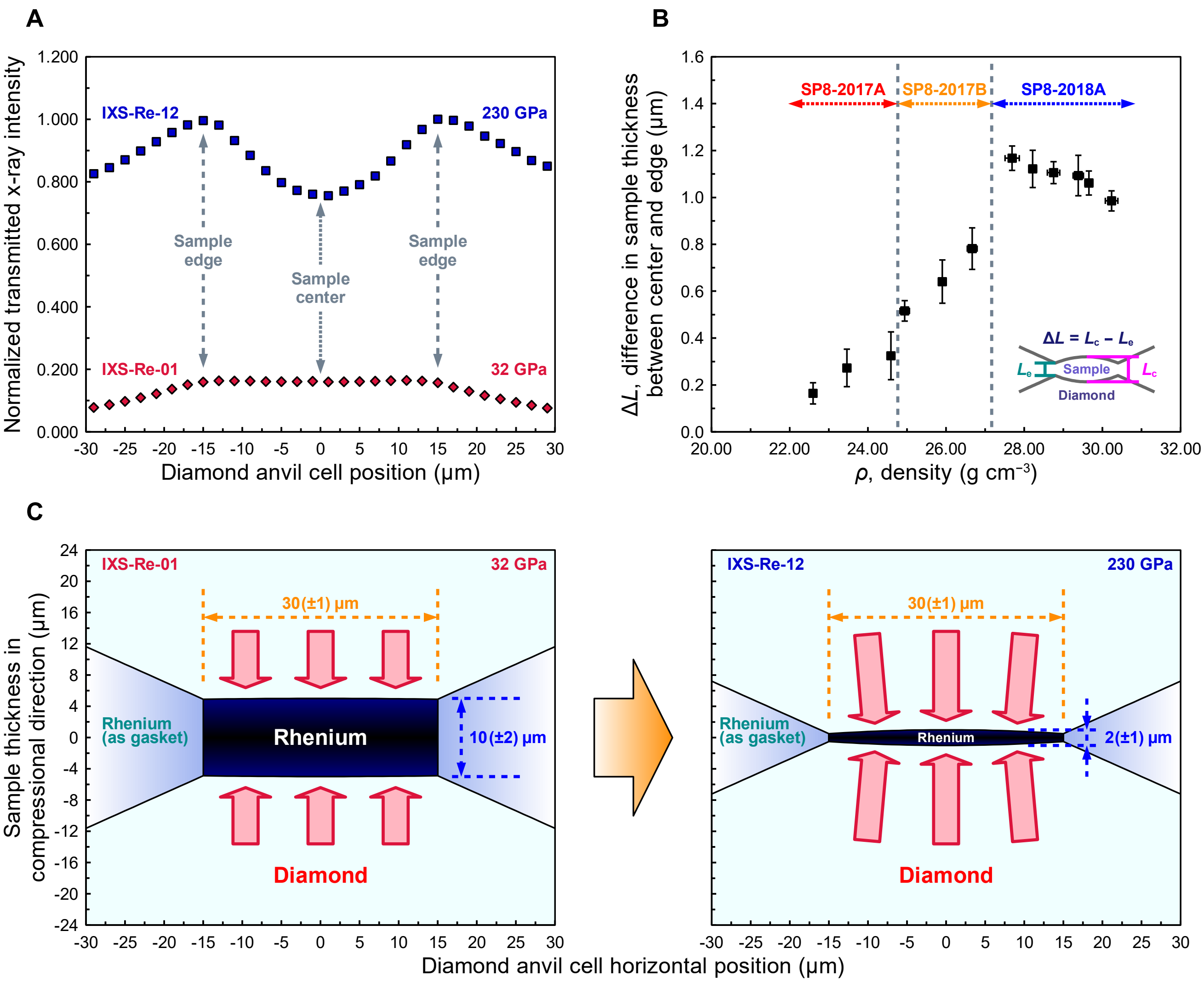

**Fig. S13.** Deformation of diamond culet. (**A**) X-ray transmission profiles, red diamond: at 32 GPa (IXS-Re-01) and blue square: at 230 GPa (IXS-Re-12). (**B**) Differences in sample thickness between sample center position and the sample edge position as a function of density. The error bars represent the 1σ uncertainties. The colored arrows indicate the SPring-8 beamtime periods as mentioned in the figure. The inset represents the schematic of the difference in sample thickness between sample center position and the sample edge position. (**C**) Schematic images of the diamond deformation at 32 GPa (IXS-Re-01) and 230 GPa (IXS-Re-12), respectively.

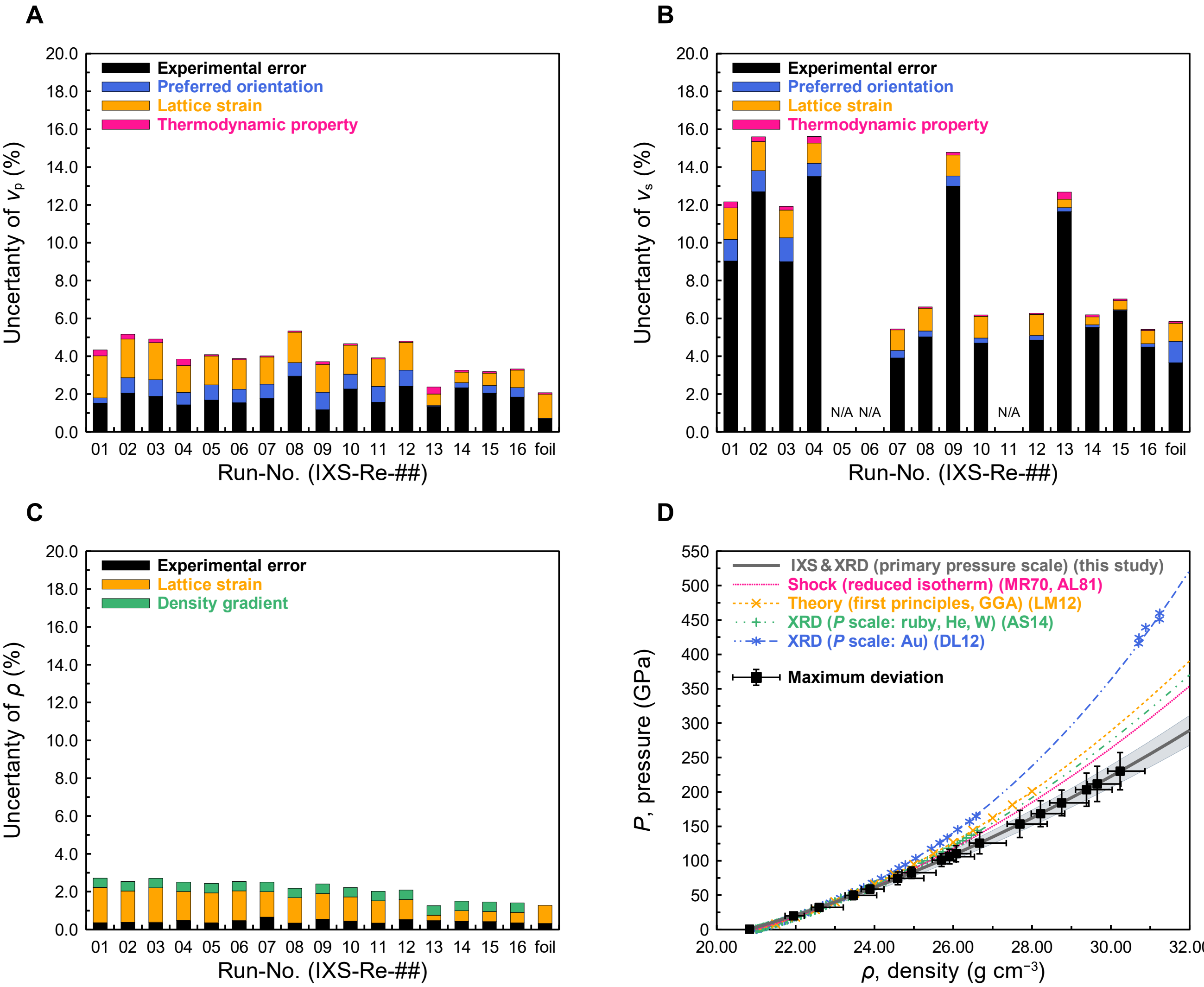

**Fig. S14.** Error budget contribution of the uncertainties to the rhenium pressure scale. (**A**) Upper bound on the uncertainty of $v_p$ and (**B**) that of $v_s$ from indicated sources. (**C**) Upper bound on the uncertainty of $\rho$ from indicated sources. (**D**) The resulting uncertainty of calibrated pressure (same as Fig. 3A in the main text).

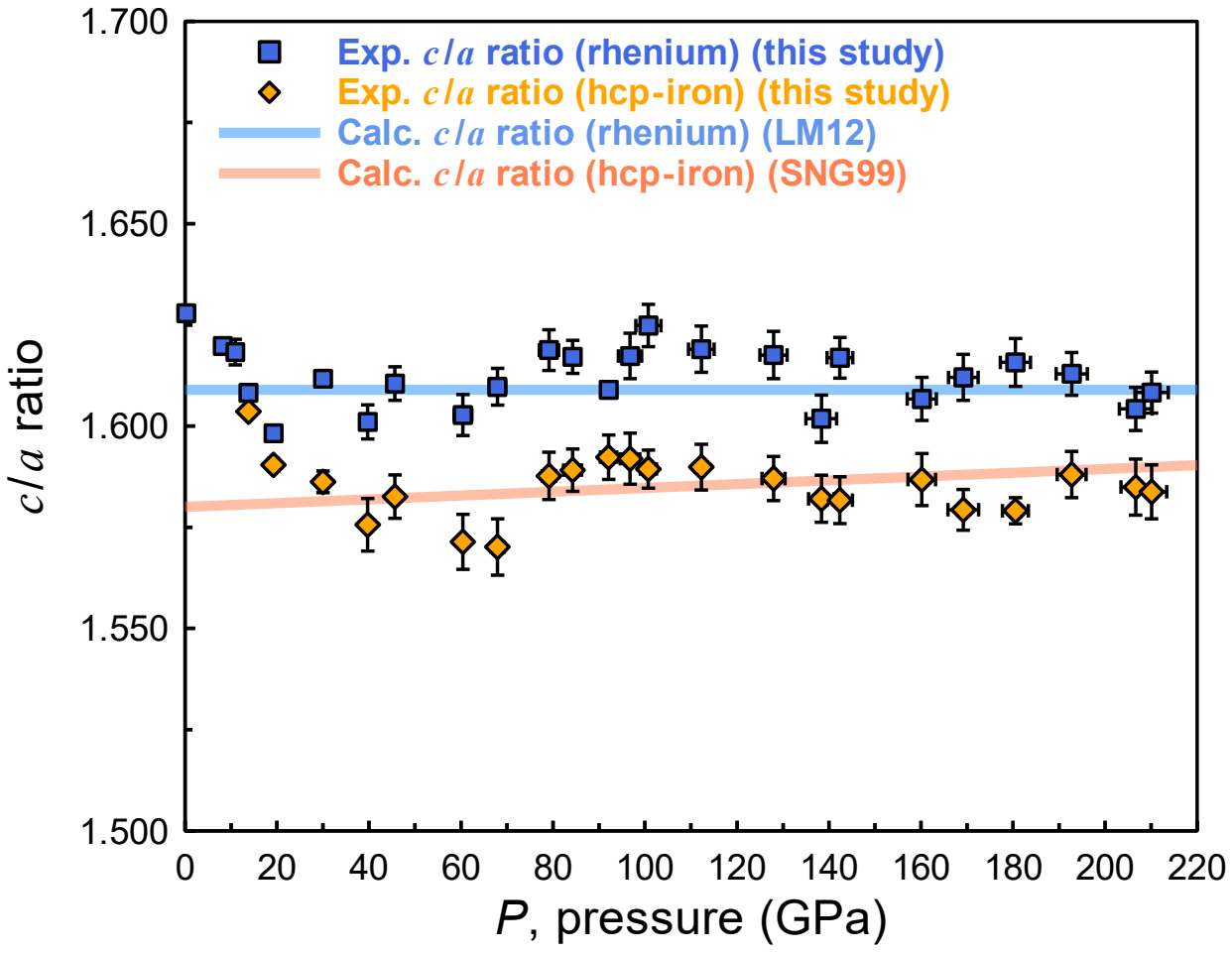

**Fig. S15.** *c*/*a* ratios of rhenium and hcp-iron under the present simultaneous density measurement as a function of pressure. The color symbols represent the experimental *c*/*a* ratio, blue squares: rhenium and orange diamonds: hcp-iron with 1σ error bars. The blue and orange lines represent the calculated model *c*/*a* ratios of rhenium (*31*) and iron (*36*), respectively.

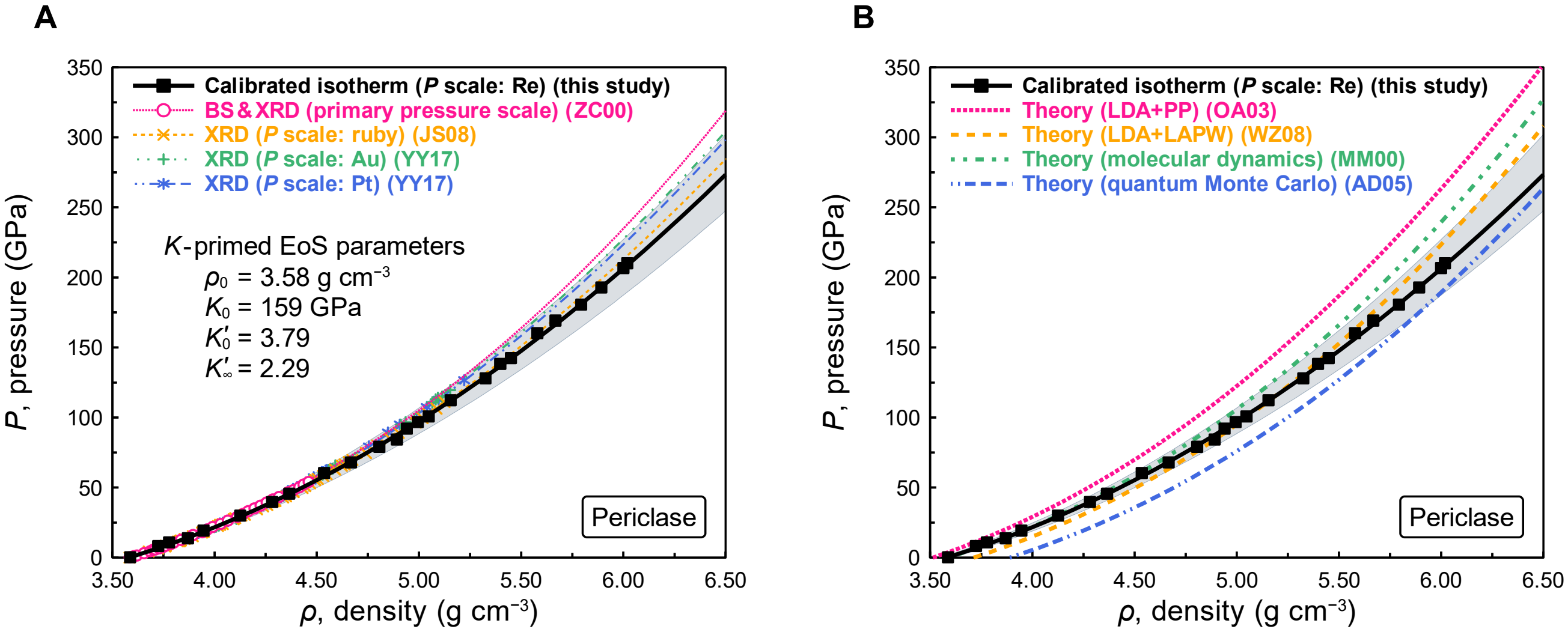

**Fig. S16.** Re-evaluated density–pressure relation for MgO. The black curve with black squares is the compression curve of MgO re-evaluated by the present simultaneous compression experiment based on our rhenium scale (Tables S2 and S4). The shaded areas around the black curve represent the 1σ uncertainty of our compression curve. (**A**) Comparison with previous experimental compression curves of MgO. Colored curves and symbols are the compression curves of MgO based on pressure scales with experimental data from previous studies [ZC00 (*14*), JS08 (*40*), and YY17 (*41*)]. (**B**) Comparison with previous theoretical compression curves of MgO. Colored curves are the compression curves of MgO based on theoretical studies (LDA: local density approximation, PP: pseudopotential, LAPW: linearized augmented plane wave method) [OA03 (*75*), WZ08 (*77*), MM00 (*74*), and AD05 (*76*)].

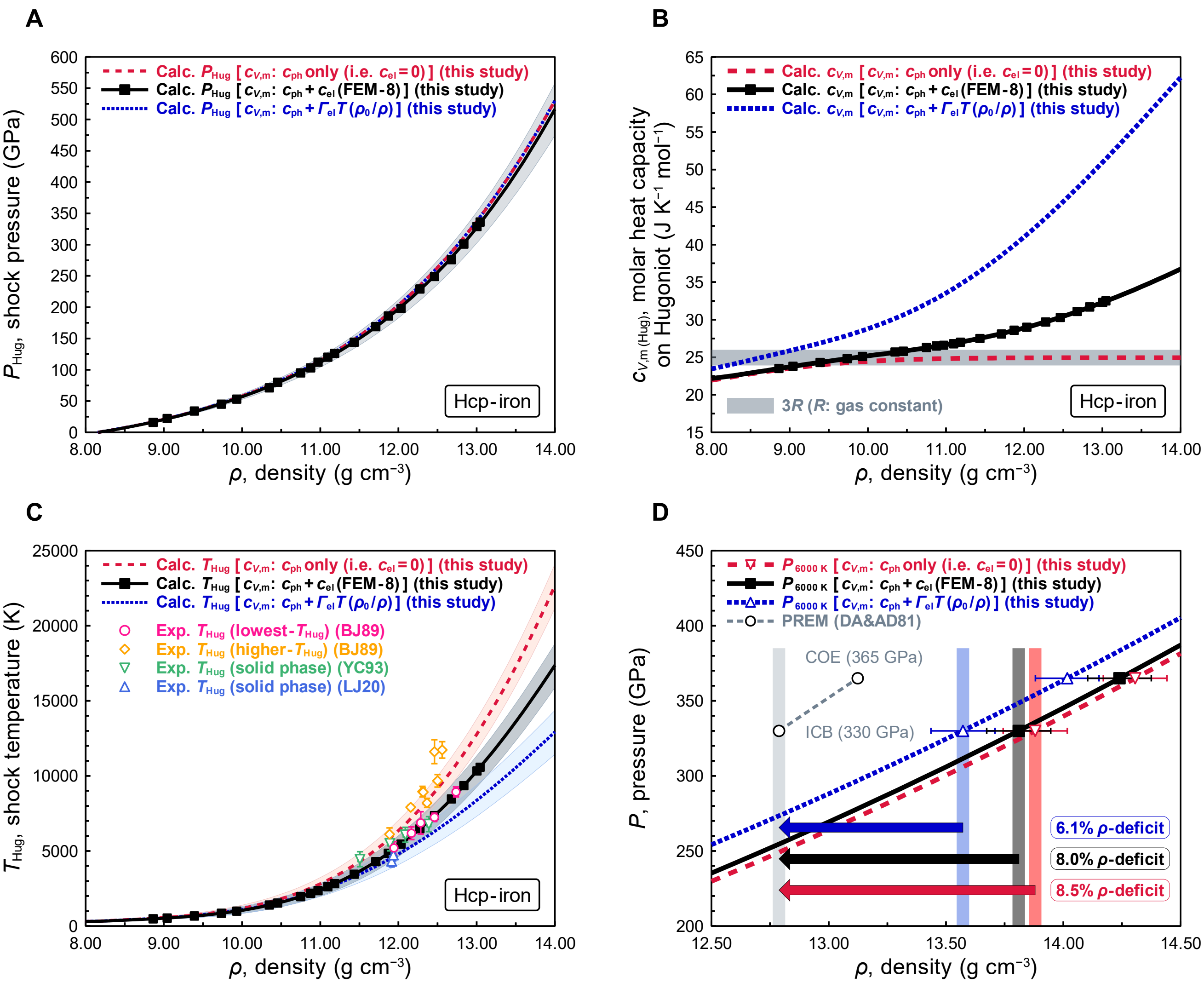

**Fig. S17.** Summary of thermodynamic properties of hcp-iron on shock Hugoniots and the density deficit of the inner core at 330 GPa and 6000 K with three different $c_{el}$ models. Details of the $c_{el}$ models are given in Methods Sect. 7. The parameters used for calculation are given in Table S2. (**A**) Comparison of calculated shock Hugoniots of hcp-iron based on three different $c_{V,m}$ models. The shaded areas around the black curve represent the 1σ uncertainty of calculated shock pressure, $P_{Hug}$ based on $c_{V,m,DM\text{-}FEM8}$ model. There are no large differences on $P_{Hug}$ between the three models. (**B**) Comparison of calculated $c_{V,m}$ of iron on the Hugoniot curves based on $c_{V,m,DM\text{-}zero}$ (red dashed), $c_{V,m,DM\text{-}LTD}$ (LTD: linear temperature dependence model) (blue dotted), and $c_{V,m,DM\text{-}FEM8}$ (FEM-8: free electron model with eight valence electrons) (black solid) models corresponding to Eqs. 23 to 25 in Methods Sect. 7, respectively. The gray bold line represents 3$R$ (where $R$ is the gas constant), the converged value of the contribution of phonons to heat capacity derived from the Debye model. (**C**) Comparison of calculated shock temperature, $T_{Hug}$ of hcp-iron based on three different $c_{V,m}$ models, compared with the experimental $T_{Hug}$ from previous studies [BJ89 (*43*), YC93 (*44*), LJ20 (*45*)]. The symbols and lines are the same as those in Fig. 3B. (**D**) Comparison of density–pressure relations of hcp-iron based on three different $c_{V,m}$ models at 6000 K and the preliminary reference Earth model (PREM). The gray dashed curve with open circles represents the density–pressure relation of the PREM (DA&AD81) (*9*). Other colored lines represent the same as those in (A). The red ($c_{V,m,DM\text{-}zero}$ model), blue ($c_{V,m,DM\text{-}LTD}$ model), and black ($c_{V,m,DM\text{-}FEM8}$ model) arrows indicate the density deficits between hcp-iron and PREM inner core for the compression curves of 6000 K. This analysis of $c_{el}$ models gives the density deficit from hcp-iron via our pressure scale is 8(±2)% in the range 330–365 GPa and 6000 K (ICB: inner core boundary, COE: center of the Earth/core) of the typical estimated Earth's inner core conditions as discussed in the main text.

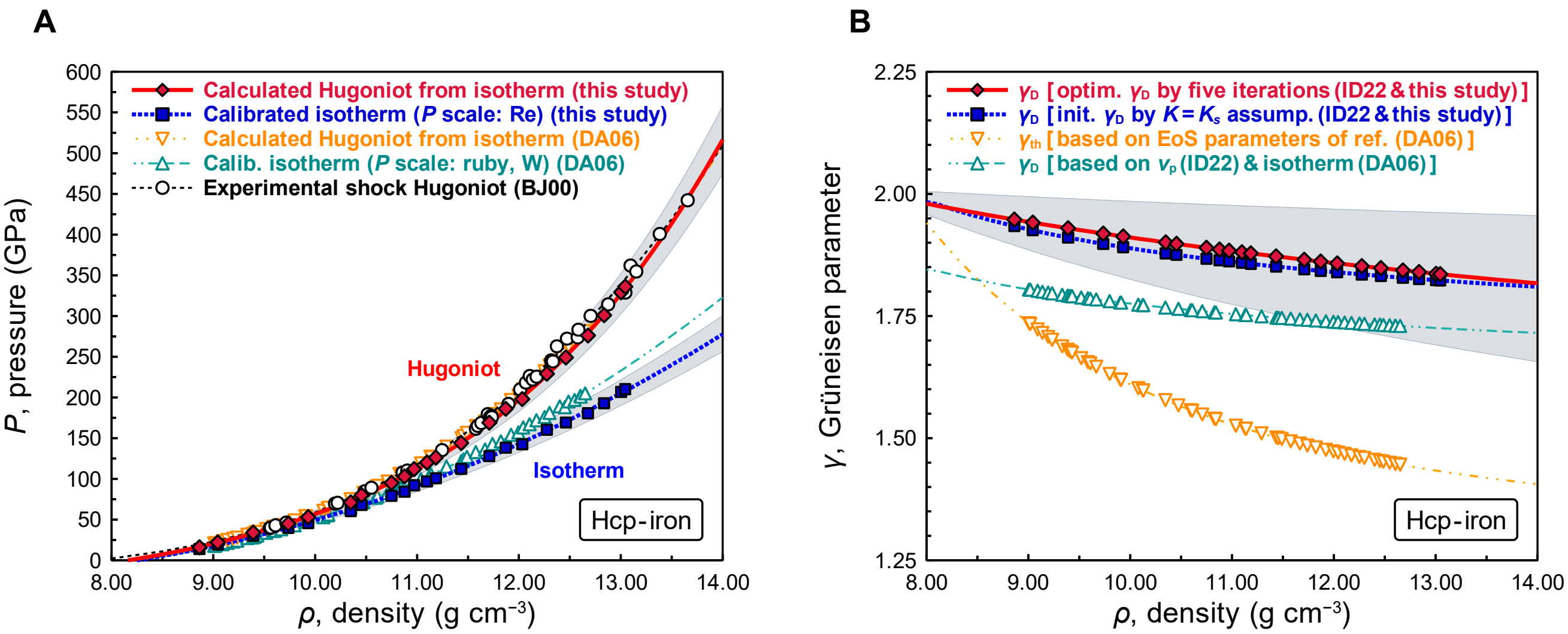

**Fig. S18.** The isotherm of hcp-iron based on our rhenium scale and calculated shock Hugoniot (**A**) and the Grüneisen parameter used for conversion (**B**). The blue dotted curve with squares in (A) represents the isothermal compression curve of hcp-iron based on our rhenium scale, whereas the green dashed-dotted curve with up-pointing triangles represents the isothermal compression curve based on the previous scale (DA06) (*5*). The black dashed curve with open circles represents the shock Hugoniot with experimental data (BJ00) (*42*). The red solid curve with diamonds represents the calculated shock Hugoniot from the isothermal compression curve of hcp-iron based on our rhenium scale (Tables S5 and S6), whereas the orange dashed-dotted curve with down-pointing triangles represents the calculated shock Hugoniot from the isotherm of the previous scale (DA06) (*5*). Each colored symbol represents density and pressure calculated on each scale, corresponding to the experimental shock compression data (BJ00) (*42*). Our calculated shock Hugoniot based on the Grüneisen parameter derived from the experimental $v_p$ of hcp-iron (ID22) (*8*) can explain the experimental shock Hugoniot (BJ00) (*42*), as well as calculated shock Hugoniot by the previous scale (DA06) (*5*) which is parameterized to account for the shock Hugoniot. Grüneisen parameter of hcp-iron determined from our EoS and experimental $v_p$ of hcp-iron (ID22) (*8*) is shown in (B). We performed iteration for optimization to derive the Grüneisen parameter (details are given in Methods Sect. 6). Blue squares show the initial Debye–Grüneisen parameter, $\gamma_D$, with $K = K_S$ assumption and red diamonds show the final Debye–Grüneisen parameter (which is equal to the thermodynamic Grüneisen parameter, $\gamma_{th}$, in the Debye approximation) after five times iterations for optimization. The difference between both Grüneisen parameters is small within uncertainty and has small effect on the conversion of the isotherm to the shock Hugoniot. On the other hand, a large discrepancy exists between the $\gamma_{th}$ (yellow down-pointing triangles) proposed by previous EoS (DA06) (*5*) and $\gamma_D$ (green up-pointing triangles) derived from the EoS (DA06) (*5*) and $v_p$ (ID22) (*8*) of hcp-iron. This means that our EoS of hcp-iron is consistent with both experimental shock Hugoniot and $v_p$, whereas the previous EoS is inconsistent with $v_p$ determined experimentally.

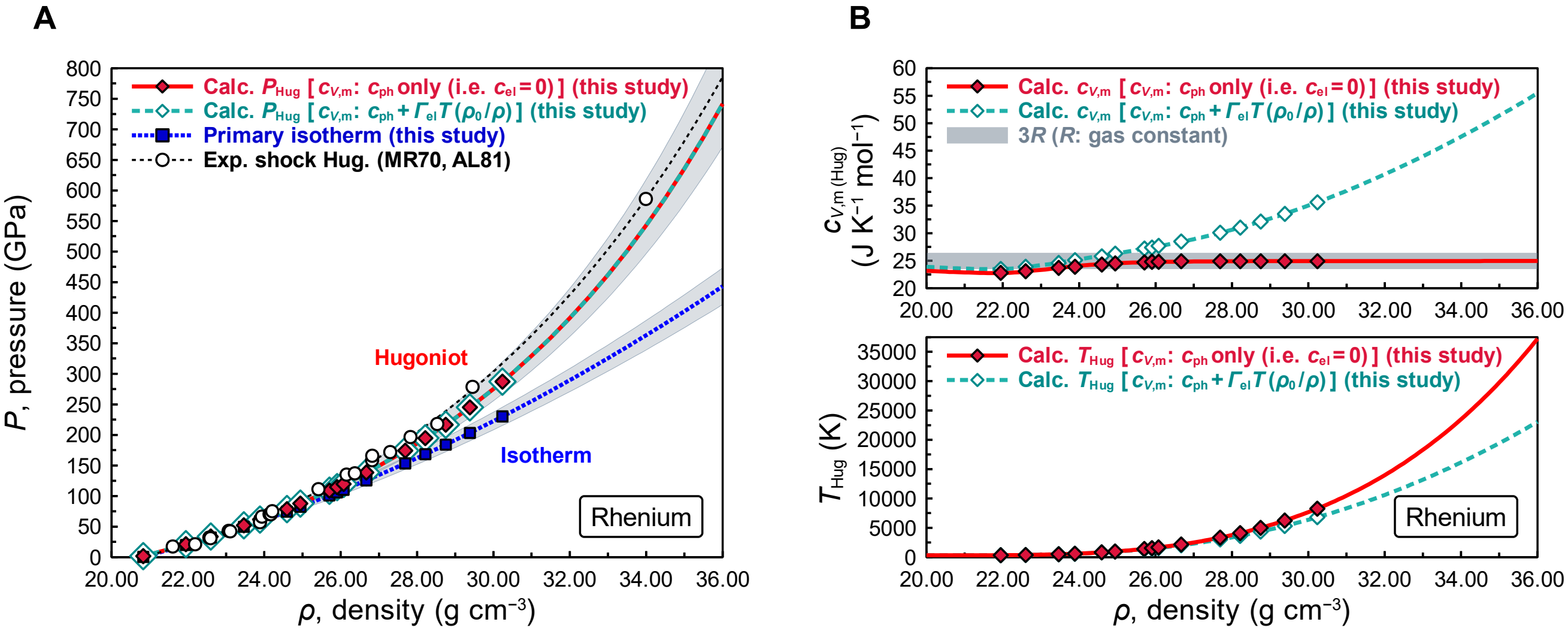

**Fig. S19.** (**A**) The calibrated isotherm (blue dotted curve with squares) and calculated shock Hugoniot (red solid curve with diamonds and green dashed curve) for rhenium. The shaded area represents the 1σ uncertainty of each curve. The red ($c_{V,\mathrm{m,DM\text{-}zero}}$ model) and green ($c_{V,\mathrm{m,DM\text{-}LTD}}$ model) curves are derived from two different $c_{\mathrm{el}}$ models. The open circle symbols and black dotted line represent the experimental shock compressional data and shock Hugoniot (*29, 30*). (**B**) Comparison of calculated $c_{V,\mathrm{m}}$ and $T_{\mathrm{Hug}}$ of rhenium on the Hugoniot curves based on $c_{V,\mathrm{m,DM\text{-}zero}}$ (red diamond) and $c_{V,\mathrm{m,DM\text{-}LTD}}$ (green diamond) models. The gray bold line represents $3R$ (where $R$ is the gas constant), converged value of the contribution of phonons to heat capacity derived from the Debye model. The parameters used for calculation are given in Table S2.

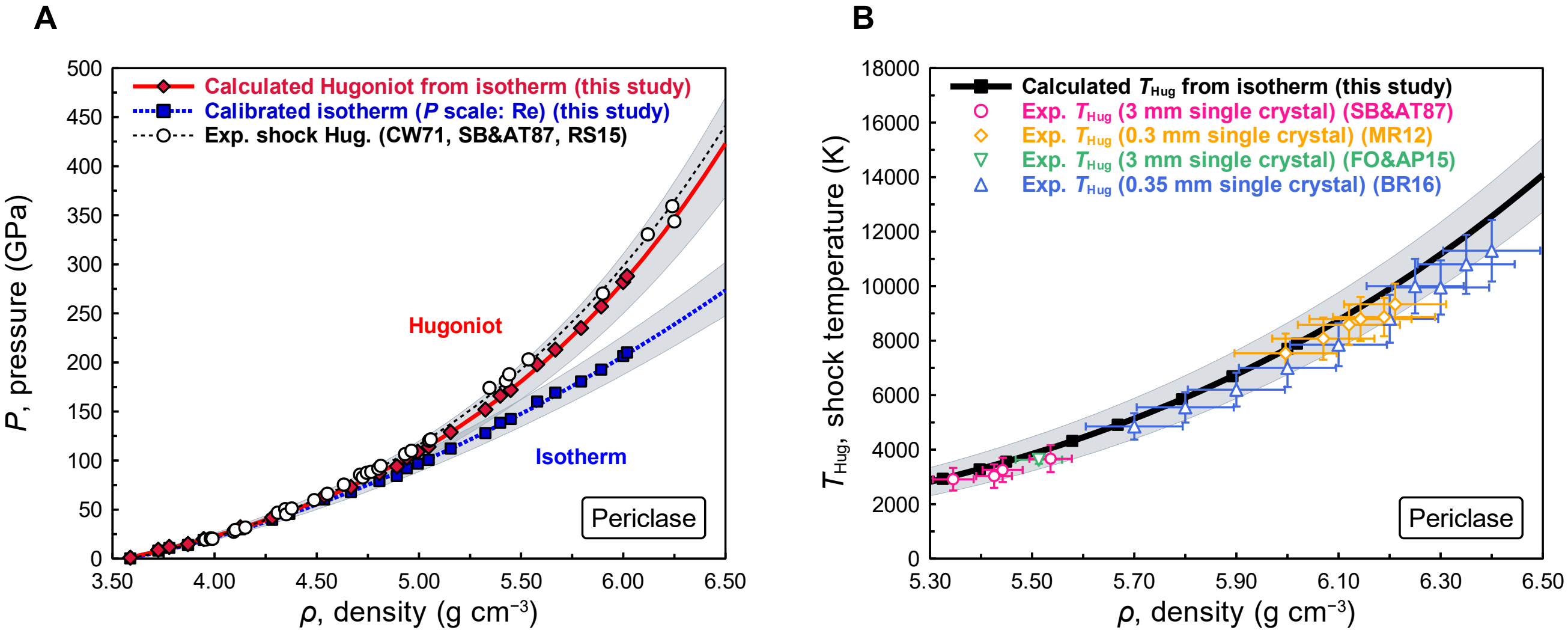

**Fig. S20.** (**A**) The calibrated isotherm (blue dotted curve with squares) of MgO based on our rhenium scale and calculated shock Hugoniot (red solid curve with diamonds). The shaded area represents the 1σ uncertainty of each curve. The black dotted curve and open circle symbols represent the shock Hugoniot and experimental shock compression data (*47, 79, 80*). (**B**) Calculated $T_{\mathrm{Hug}}$ of MgO. The $T_{\mathrm{Hug}}$ was derived assuming $c_{\mathrm{el}} = 0$ for the molar heat capacity, $c_{V,\mathrm{m}}$. The detailed parameters used for calculation are given in Table S2. Black squares are the calculated $T_{\mathrm{Hug}}$ corresponding to the red diamonds in (A). Other colored symbols are the experimentally measured $T_{\mathrm{Hug}}$ of MgO (B1 structure) from previous studies [SB&AT89 (*47*), MR12 (*48*), FO&AP15 (*49*), BR16 (*50*)]. The shaded area around the curve represents the 1σ uncertainty.

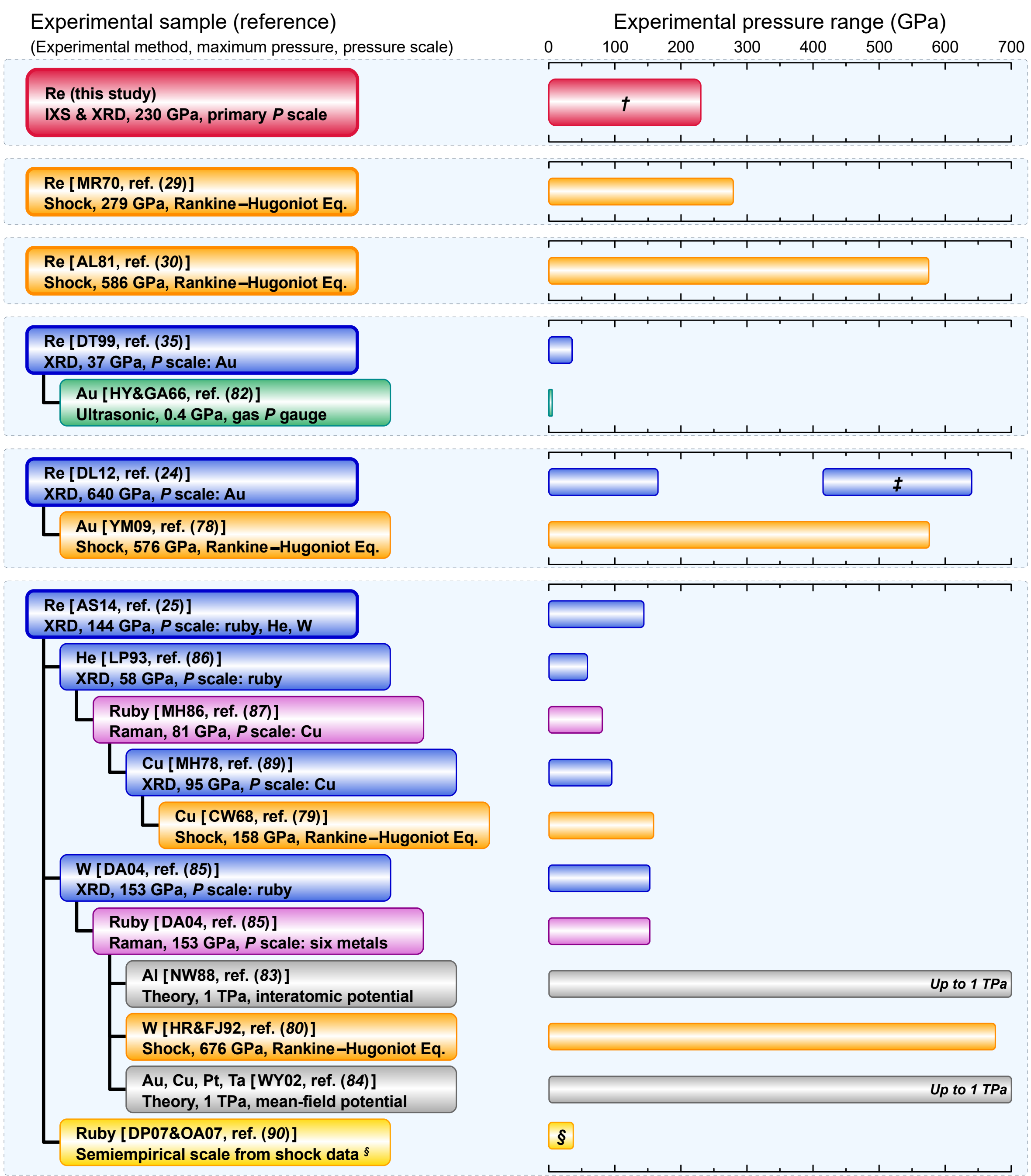

**Fig. S21.** Experimental methods for the compression of rhenium in this and previous studies (*24, 25, 29, 30, 35*) with pressure scales used in those studies (*78–80, 82–87, 89, 90*). Colors indicate the experimental method; red: IXS and XRD with DAC, yellow: shock compression measurement, light yellow: semiempirical scale from shock compression data (e.g., *29, 30, 80*), blue: XRD with DAC, green: US with gas compression, magenta: Raman spectroscopy with DAC, gray: theoretical work. Experimental conditions are also given within the boxes. Each system diagram represents the flow of the secondary and/or primary pressure scales based on the experiments. *Notes*; *†*; The highest experimental pressures by using previous pressure scales for rhenium are 274 GPa (*29, 30*), 285 GPa (*25*), 300 GPa (*31*), and 380 GPa (*24*). *‡*; the experimental pressure achieved by the work was discussed in refs. (*26*). *§*: this ruby scale re-evaluated from numerous shock compression data (e.g., *29, 30, 80*) has been used to calibrate only for low pressure data up to 37 GPa in ref. (*25*).

# *Supplementary Tables*

**Table S1.** Results of acoustic velocity measurement and calibrated pressures for rhenium.

| Run-No. | $\rho$, density (g cm$^{-3}$) | $v_p$, compressional wave velocity (km s$^{-1}$) | $v_s$, shear wave velocity (km s$^{-1}$) | $Q_{max}$ ($v_p$) (nm$^{-1}$) | $Q_{max}$ ($v_s$) (nm$^{-1}$) | $P$, calibrated pressure (GPa) § |
|---|---|---|---|---|---|---|
| IXS-Re-01 * | 22.60 (±0.08) | 5.75 (±0.09) | 3.72 (±0.34) | 10.5 (±0.4) | 11.2 (±1.2) | 32 (±3) |
| IXS-Re-02 * | 23.46 (±0.09) | 6.07 (±0.12) | 3.68 (±0.47) | 10.2 (±0.4) | 10.7 (±1.4) | 50 (±4) |
| IXS-Re-03 * | 24.59 (±0.10) | 6.54 (±0.12) | 4.14 (±0.37) | 10.6 (±0.4) | 11.8 (±0.6) | 74 (±6) |
| IXS-Re-04 * | 24.94 (±0.12) | 6.61 (±0.09) | 4.28 (±0.58) | 11.6 (±0.4) | 11.7 (±1.8) | 83 (±6) |
| IXS-Re-05 * | 25.91 (±0.09) | 6.88 (±0.12) | - | 11.2 (±0.4) | - | 106 (±8) |
| IXS-Re-06 * | 26.67 (±0.13) | 7.01 (±0.11) | - | 12.4 (±0.6) | - | 126 (±9) |
| IXS-Re-07 * | 27.69 (±0.18) | 7.45 (±0.13) | 4.43 (±0.17) | 12.4 (±0.4) | 12.7 (±1.1) | 153 (±11) |
| IXS-Re-08 * | 28.22 (±0.10) | 7.74 (±0.23) | 4.72 (±0.24) | 12.3 (±0.4) | 12.8 (±1.8) | 168 (±12) |
| IXS-Re-09 * | 28.75 (±0.16) | 7.79 (±0.09) | 4.83 (±0.63) | 12.4 (±0.4) | 11.8 (±3.5) | 184 (±13) |
| IXS-Re-10 * | 29.38 (±0.13) | 8.02 (±0.18) | 5.12 (±0.24) | 12.8 (±0.4) | 11.9 (±0.6) | 203 (±15) |
| IXS-Re-11 * | 29.65 (±0.10) | 8.09 (±0.13) | - | 12.8 (±0.2) | - | 212 (±15) |
| IXS-Re-12 * | 30.24 (±0.16) | 8.38 (±0.20) | 5.35 (±0.26) | 12.6 (±0.4) | 11.6 (±1.1) | 230 (±17) |
| IXS-Re-13 † | 21.95 (±0.10) | 5.56 (±0.07) | 3.52 (±0.41) | 10.9 (±0.4) | 11.1 (±1.2) | 20 (±3) |
| IXS-Re-14 † | 23.89 (±0.10) | 6.22 (±0.15) | 3.64 (±0.20) | 11.5 (±0.2) | 11.3 (±0.5) | 59 (±5) |
| IXS-Re-15 † | 25.70 (±0.11) | 6.81 (±0.14) | 3.86 (±0.25) | 11.7 (±0.3) | 11.3 (±0.7) | 101 (±7) |
| IXS-Re-16 † | 26.08 (±0.09) | 7.08 (±0.13) | 4.13 (±0.19) | 11.4 (±0.2) | 10.7 (±0.3) | 110 (±8) |
| IXS-Re-foil ‡ | 20.83 (±0.07) | 5.34 (±0.04) | 2.96 (±0.09) | 10.1 (±0.3) | 12.7 (±1.1) | 0.5 (±0.5) |

*Note*: * Direct compression without pressure medium. † Quasi-hydrostatic compression with periclase (MgO) pressure medium. ‡ Pre-compressed foil in air. § Pressures were derived from $v_p$, $v_s$, and $\rho$ with the $K$-primed Mie–Grüneisen–Debye model (details are in the Methods section, and the parameters of the equation of state for rhenium are in Table S2).

**Table S2.** Parameters of $K$-primed MGD EoS of rhenium, iron, and MgO.

| | Rhenium (Re) (hcp-phase) | Iron (Fe) (hcp-phase) | Periclase (MgO) (B1-phase) |
|---|---|---|---|
| | Equation of state (EoS), $K$-primed Mie–Grüneisen–Debye (MGD) model (*refs. 1, 37–39*) | | |
| $\rho_0$ (g cm$^{-3}$) | 20.8 (±0.1) | 8.25 (±0.05) | 3.58 (±0.03) |
| $K_0$ (GPa) | 340 (±9) | 162 (±5) | 159 (±6) |
| $K'_0$ ($= \partial K_0/\partial P_0$) | 3.25 (±0.12) | 5.12 (±0.08) | 3.79 (±0.08) |
| $K'_\infty$ ($= \partial K_\infty/\partial P_\infty$) | 2.15 (±0.11) | 2.55 (±0.09) | 2.29 (±0.12) |
| | Grüneisen parameter, Al'tshuler fixed $\gamma_\infty$ model (*ref. 56*) : $\gamma=\gamma_\infty+(\gamma_0-\gamma_\infty)(\rho_0/\rho)^q$ | | |
| $\Theta_0$ (K) | 369 (±5) | 515 (±21) | 760 (±135) |
| $\gamma_0$ | 1.94 (±0.31) | 1.97 (±0.16) | 1.53 (±0.26) |
| $\gamma_\infty$ | $(3K'_\infty-1)/6$ (fixed) * | $(3K'_\infty-1)/6$ (fixed) * | $(3K'_\infty-1)/6$ (fixed) * |
| $q$ | 0.53 (±0.30) | 0.37 (±0.24) | 0.44 (±0.68) |
| | Molar heat capacity, sum of phonon and electron contributions model (*ref. 57*) : $c_{V,m}=c_{ph}+c_{el}$ | | |
| $c_{V,m}$ (J K$^{-1}$ mol$^{-1}$) | $c_{ph}$ (DM) only (i.e. $c_{el}=0$) | $c_{ph}$ (DM) + $c_{el}$ (FEM-8) | $c_{ph}$ (DM) only (i.e. $c_{el}=0$) |
| | Isothermal compressional wave velocity, Birch's law (*ref. 34*) : $v_p=v_{p,0}+(\partial v_p/\partial\rho)(\rho-\rho_0)$ | | |
| $v_{p,0}$ (km s$^{-1}$) | 5.30 (±0.04) | 6.14 (fixed) † | 9.85 (fixed) § |
| $\partial v_p/\partial\rho$ (m$^4$ kg$^{-1}$ s$^{-1}$) | 0.313 (±0.002) | 1.16 (fixed) † | 2.27 (fixed) § |

*Note*: * Theoretical constraint of the Grüneisen parameter in the $K$-primed MGD model (*ref. 39*). † Fixed to the reference data of $\rho$–$v_p$ (*ref. 8*). § Calculated from the reference data of $\rho$–$v_s$ (*ref. 58*) with the EoS parameters (this study). *Abbreviations*: hcp: hexagonal close-packed, B1: rock salt type cubic structure, DM: Debye model, FEM-8: free electron model with eight valence electrons.

**Table S3.** Results of uncertainty analysis.

| Run-No. | $\sigma(\rho)$, density uncertainty | | | $\sigma(v_p)$, compressional wave velocity uncertainty | | | | $\sigma(v_s)$, shear wave velocity uncertainty | | | | Pressure medium |
|---|---|---|---|---|---|---|---|---|---|---|---|---|
| | EX | LS | DG | EX | PO | LS | TP | EX | PO | LS | TP | |
| IXS-Re-01 | 0.4% | 1.9% | 0.5% | 1.5% | 0.3% | 2.2% | 0.3% | 9.0% | 1.1% | 1.7% | 0.3% | - |
| IXS-Re-02 | 0.4% | 1.7% | 0.5% | 2.0% | 0.8% | 2.0% | 0.3% | 12.7% | 1.1% | 1.5% | 0.3% | - |
| IXS-Re-03 | 0.4% | 1.8% | 0.5% | 1.9% | 0.9% | 2.0% | 0.2% | 9.0% | 1.3% | 1.5% | 0.2% | - |
| IXS-Re-04 | 0.5% | 1.5% | 0.5% | 1.4% | 0.7% | 1.4% | 0.3% | 13.5% | 0.7% | 1.1% | 0.3% | - |
| IXS-Re-05 | 0.4% | 1.6% | 0.5% | 1.7% | 0.8% | 1.5% | 0.1% | | - | | | - |
| IXS-Re-06 | 0.5% | 1.6% | 0.5% | 1.5% | 0.7% | 1.6% | 0.1% | | - | | | - |
| IXS-Re-07 | 0.7% | 1.3% | 0.5% | 1.8% | 0.8% | 1.4% | 0.1% | 3.9% | 0.4% | 1.1% | 0.1% | - |
| IXS-Re-08 | 0.3% | 1.3% | 0.5% | 2.9% | 0.7% | 1.6% | 0.1% | 5.0% | 0.3% | 1.2% | 0.1% | - |
| IXS-Re-09 | 0.5% | 1.4% | 0.5% | 1.2% | 0.9% | 1.5% | 0.1% | 13.0% | 0.5% | 1.1% | 0.1% | - |
| IXS-Re-10 | 0.5% | 1.3% | 0.5% | 2.3% | 0.8% | 1.5% | 0.1% | 4.7% | 0.3% | 1.2% | 0.1% | - |
| IXS-Re-11 | 0.3% | 1.2% | 0.5% | 1.6% | 0.8% | 1.4% | 0.1% | | - | | | - |
| IXS-Re-12 | 0.5% | 1.1% | 0.5% | 2.4% | 0.9% | 1.5% | 0.1% | 4.9% | 0.2% | 1.1% | 0.1% | - |
| IXS-Re-13 | 0.5% | 0.3% | 0.5% | 1.3% | 0.1% | 0.6% | 0.4% | 11.7% | 0.2% | 0.4% | 0.4% | MgO |
| IXS-Re-14 | 0.4% | 0.6% | 0.5% | 2.3% | 0.3% | 0.6% | 0.1% | 5.5% | 0.2% | 0.4% | 0.1% | MgO |
| IXS-Re-15 | 0.4% | 0.5% | 0.5% | 2.0% | 0.4% | 0.6% | 0.1% | 6.4% | 0.1% | 0.5% | 0.1% | MgO |
| IXS-Re-16 | 0.4% | 0.6% | 0.5% | 1.8% | 0.5% | 0.9% | 0.1% | 4.5% | 0.2% | 0.7% | 0.1% | MgO |
| IXS-Re-foil | 0.3% | 1.0% | 0.0% | 0.7% | 0.0% | 1.3% | 0.1% | 3.7% | 1.1% | 1.0% | 0.1% | in air |

*Abbreviations*: EX: experimental error, LS: lattice strain, DG: density gradient in the sample region irradiated by x-ray beam with tails, PO: preferred orientation, TP: thermodynamic property, MgO: periclase (magnesium oxide).

**Table S4.** Results of simultaneous compression experiment for rhenium, iron, and MgO.

| Run-No. | $P$, calibrated pressure (GPa) * | Rhenium (Re) (hcp-phase) | | | Iron (Fe) (hcp-phase) | | | Periclase (MgO) (B1-ph.) | |
|---|---|---|---|---|---|---|---|---|---|
| | | Lattice param. | | $\rho$, density (g cm$^{-3}$) | Lattice param. | | $\rho$, density (g cm$^{-3}$) | Latt. param. | $\rho$, density (g cm$^{-3}$) |
| | | $a$ (nm) | $c$ (nm) | | $a$ (nm) | $c$ (nm) | | $a$ (nm) | |
| XRD-Re-01 | 0.3 (±0.4) | 0.2762 | 0.4496 | 20.82 (±0.03) | 0.2862 † | | 7.91 (±0.01) † | 0.4210 | 3.59 (±0.02) |
| XRD-Re-02 | 8 (±2) | 0.2746 | 0.4448 | 21.29 (±0.04) | 0.2822 † | | 8.25 (±0.01) † | 0.4158 | 3.72 (±0.02) |
| XRD-Re-03 | 11 (±2) | 0.2740 | 0.4434 | 21.45 (±0.06) | 0.2809 † | | 8.37 (±0.01) † | 0.4138 | 3.78 (±0.02) |
| XRD-Re-04 | 14 (±2) | 0.2739 | 0.4405 | 21.61 (±0.03) | 0.2470 | 0.3961 | 8.86 (±0.01) | 0.4105 | 3.87 (±0.02) |
| XRD-Re-05 | 19 (±3) | 0.2732 | 0.4366 | 21.91 (±0.04) | 0.2460 | 0.3913 | 9.04 (±0.02) | 0.4078 | 3.95 (±0.02) |
| XRD-Re-06 | 30 (±3) | 0.2701 | 0.4353 | 22.48 (±0.04) | 0.2432 | 0.3857 | 9.39 (±0.03) | 0.4018 | 4.13 (±0.02) |
| XRD-Re-07 | 40 (±4) | 0.2688 | 0.4303 | 22.98 (±0.07) | 0.2408 | 0.3794 | 9.74 (±0.05) | 0.3969 | 4.28 (±0.03) |
| XRD-Re-08 | 46 (±4) | 0.2671 | 0.4302 | 23.27 (±0.07) | 0.2389 | 0.3780 | 9.93 (±0.04) | 0.3943 | 4.37 (±0.03) |
| XRD-Re-09 | 60 (±5) | 0.2649 | 0.4246 | 23.96 (±0.09) | 0.2362 | 0.3711 | 10.35 (±0.05) | 0.3894 | 4.54 (±0.02) |
| XRD-Re-10 | 68 (±5) | 0.2633 | 0.4238 | 24.30 (±0.08) | 0.2354 | 0.3696 | 10.45 (±0.05) | 0.3857 | 4.67 (±0.02) |
| XRD-Re-11 | 79 (±6) | 0.2611 | 0.4226 | 24.79 (±0.09) | 0.2324 | 0.3689 | 10.75 (±0.05) | 0.3819 | 4.81 (±0.02) |
| XRD-Re-12 | 84 (±6) | 0.2604 | 0.4211 | 25.01 (±0.07) | 0.2314 | 0.3677 | 10.88 (±0.04) | 0.3797 | 4.89 (±0.03) |
| XRD-Re-13 | 92 (±7) | 0.2597 | 0.4178 | 25.34 (±0.05) | 0.2306 | 0.3671 | 10.97 (±0.04) | 0.3784 | 4.94 (±0.02) |
| XRD-Re-14 | 97 (±7) | 0.2586 | 0.4182 | 25.53 (±0.10) | 0.2297 | 0.3657 | 11.10 (±0.05) | 0.3770 | 4.99 (±0.02) |
| XRD-Re-15 | 101 (±7) | 0.2577 | 0.4186 | 25.69 (±0.11) | 0.2292 | 0.3644 | 11.18 (±0.04) | 0.3757 | 5.05 (±0.02) |
| XRD-Re-16 | 112 (±8) | 0.2564 | 0.4152 | 26.15 (±0.11) | 0.2275 | 0.3618 | 11.43 (±0.05) | 0.3731 | 5.16 (±0.02) |
| XRD-Re-17 | 128 (±9) | 0.2546 | 0.4118 | 26.76 (±0.11) | 0.2259 | 0.3585 | 11.71 (±0.05) | 0.3691 | 5.33 (±0.02) |
| XRD-Re-18 | 138 (±10) | 0.2542 | 0.4072 | 27.14 (±0.12) | 0.2251 | 0.3561 | 11.87 (±0.05) | 0.3674 | 5.40 (±0.02) |
| XRD-Re-19 | 142 (±10) | 0.2529 | 0.4090 | 27.29 (±0.10) | 0.2241 | 0.3544 | 12.03 (±0.05) | 0.3662 | 5.45 (±0.03) |
| XRD-Re-20 | 160 (±11) | 0.2515 | 0.4041 | 27.93 (±0.11) | 0.2224 | 0.3528 | 12.27 (±0.06) | 0.3634 | 5.58 (±0.02) |
| XRD-Re-21 | 169 (±12) | 0.2503 | 0.4035 | 28.24 (±0.11) | 0.2216 | 0.3500 | 12.46 (±0.05) | 0.3615 | 5.67 (±0.02) |
| XRD-Re-22 | 181 (±13) | 0.2490 | 0.4023 | 28.63 (±0.11) | 0.2203 | 0.3479 | 12.68 (±0.04) | 0.3588 | 5.79 (±0.02) |
| XRD-Re-23 | 193 (±14) | 0.2480 | 0.3999 | 29.04 (±0.11) | 0.2190 | 0.3478 | 12.84 (±0.05) | 0.3568 | 5.89 (±0.04) |
| XRD-Re-24 | 207 (±15) | 0.2471 | 0.3964 | 29.49 (±0.12) | 0.2182 | 0.3459 | 13.00 (±0.07) | 0.3547 | 6.00 (±0.02) |
| XRD-Re-25 | 210 (±15) | 0.2466 | 0.3966 | 29.60 (±0.12) | 0.2180 | 0.3453 | 13.04 (±0.06) | 0.3543 | 6.02 (±0.02) |

*Note*: * Pressure was calibrated by the *K*-primed Mie–Grüneisen–Debye equation of state for rhenium (Table S2). † Body-centered cubic (bcc) phase. *Abbreviations*: hcp: hexagonal close-packed, B1: rock salt type cubic structure.

**Table S5.** Calculated shock Hugoniot for hcp-iron.

| Iron (Fe) | | | $P_{Hug}$, calculated shock pressure, and $T_{Hug}$, calculated shock temperature * | | | | | |
|---|---|---|---|---|---|---|---|---|
| | | | $c_{V,m}$: $c_{ph}$ (DM) + $c_{el}$ (FEM-8) | | $c_{V,m}$: $c_{ph}$ (DM) only (i.e. $c_{el}=0$) | | $c_{V,m}$: $c_{ph}$ (DM) + $\Gamma_{el}T(\rho_0/\rho)$ † | |
| Run-No. | $\rho$ (g cm$^{-3}$) | $P_{300\,K}$ (GPa) | $P_{Hug}$ (GPa) | $T_{Hug}$ (K) | $P_{Hug}$ (GPa) | $T_{Hug}$ (K) | $P_{Hug}$ (GPa) | $T_{Hug}$ (K) |
| XRD-Re-01 | 7.91 | 0.3 | 2 (±1) § | 280 (±1) § | 2 (±1) § | 280 (±1) § | 2 (±1) § | 290 (±1) § |
| XRD-Re-02 | 8.25 | 8 | 10 (±2) § | 340 (±3) § | 10 (±2) § | 350 (±3) § | 10 (±2) § | 350 (±3) § |
| XRD-Re-03 | 8.37 | 11 | 13 (±2) § | 370 (±10) § | 13 (±2) § | 380 (±10) § | 13 (±2) § | 370 (±10) § |
| XRD-Re-04 | 8.86 | 14 | 16 (±3) | 480 (±50) | 17 (±3) | 510 (±50) | 17 (±3) | 500 (±50) |
| XRD-Re-05 | 9.04 | 19 | 22 (±3) | 530 (±70) | 22 (±3) | 580 (±70) | 23 (±3) | 560 (±70) |
| XRD-Re-06 | 9.39 | 30 | 34 (±4) | 670 (±110) | 35 (±4) | 730 (±110) | 35 (±4) | 700 (±110) |
| XRD-Re-07 | 9.74 | 40 | 45 (±5) | 850 (±170) | 46 (±5) | 950 (±170) | 46 (±5) | 890 (±170) |
| XRD-Re-08 | 9.93 | 46 | 53 (±5) | 990 (±200) | 53 (±5) | 1120 (±200) | 54 (±5) | 1040 (±200) |
| XRD-Re-09 | 10.35 | 60 | 71 (±7) | 1400 (±270) | 72 (±7) | 1590 (±270) | 73 (±7) | 1440 (±270) |
| XRD-Re-10 | 10.45 | 68 | 80 (±7) | 1530 (±300) | 81 (±7) | 1750 (±300) | 82 (±7) | 1560 (±300) |
| XRD-Re-11 | 10.75 | 79 | 95 (±8) | 1960 (±360) | 97 (±8) | 2260 (±360) | 98 (±8) | 1970 (±360) |
| XRD-Re-12 | 10.88 | 84 | 103 (±8) | 2180 (±390) | 105 (±8) | 2530 (±390) | 105 (±8) | 2180 (±390) |
| XRD-Re-13 | 10.97 | 92 | 112 (±9) | 2360 (±410) | 115 (±9) | 2740 (±410) | 115 (±9) | 2340 (±410) |
| XRD-Re-14 | 11.10 | 97 | 120 (±9) | 2620 (±440) | 123 (±9) | 3050 (±440) | 123 (±9) | 2570 (±440) |
| XRD-Re-15 | 11.18 | 101 | 126 (±9) | 2820 (±460) | 129 (±9) | 3290 (±460) | 129 (±9) | 2740 (±460) |
| XRD-Re-16 | 11.43 | 112 | 144 (±10) | 3450 (±530) | 148 (±10) | 4050 (±530) | 148 (±10) | 3290 (±530) |
| XRD-Re-17 | 11.71 | 128 | 169 (±12) | 4290 (±610) | 174 (±12) | 5070 (±610) | 174 (±12) | 3970 (±610) |
| XRD-Re-18 | 11.87 | 138 | 186 (±13) | 4850 (±650) | 192 (±13) | 5750 (±650) | 192 (±13) | 4420 (±650) |
| XRD-Re-19 | 12.03 | 142 | 198 (±15) | 5460 (±700) | 204 (±15) | 6520 (±700) | 204 (±15) | 4890 (±700) |
| XRD-Re-20 | 12.27 | 160 | 229 (±17) | 6480 (±780) | 236 (±17) | 7790 (±780) | 236 (±17) | 5660 (±780) |
| XRD-Re-21 | 12.46 | 169 | 249 (±19) | 7360 (±850) | 257 (±19) | 8910 (±850) | 257 (±19) | 6300 (±850) |
| XRD-Re-22 | 12.68 | 181 | 276 (±21) | 8470 (±920) | 284 (±21) | 10340 (±920) | 285 (±21) | 7100 (±920) |
| XRD-Re-23 | 12.84 | 193 | 301 (±23) | 9350 (±980) | 310 (±23) | 11490 (±980) | 310 (±23) | 7710 (±980) |
| XRD-Re-24 | 13.00 | 207 | 329 (±25) | 10330 (±1040) | 339 (±25) | 12800 (±1040) | 339 (±25) | 8380 (±1040) |
| XRD-Re-25 | 13.04 | 210 | 336 (±25) | 10580 (±1060) | 346 (±25) | 13130 (±1060) | 346 (±25) | 8550 (±1060) |

*Note*: * $P_{Hug}$ and $T_{Hug}$ were calculated based on the parameters of *K*-primed Mie–Grüneisen–Debye (MGD) equation of state (EoS) for hexagonal close-packed (hcp) iron (Table S2) and the experimental conditions of the simultaneous compression (Table S4) with the experimental shock compression data of iron (*ref. 42*). † $\Gamma_{el} = 4.90$ (mJ K$^{-2}$ mol$^{-1}$), fixed to the reference data (*ref. 59*).
§ $P_{Hug}$ and $T_{Hug}$ for hcp-iron were estimated from the parameters of *K*-primed MGD-EoS for hcp-iron (Table S2), while the experimentally observed phase at ambient temperature was body-centered cubic iron (Table S4).
*Abbreviations*: DM: Debye model, FEM-8: free electron model with eight valence electrons.

**Table S6.** Pressure–density relations for hcp-iron and PREM.

| Earth's model (PREM) | $P$, pressure (GPa) | $\rho$-PREM (g cm$^{-3}$) | Hexagonal close-packed (hcp) iron (this study) $\rho$, density (g cm$^{-3}$) | | | hcp-iron (DA06) $\rho$, density (g cm$^{-3}$) | |
|---|---|---|---|---|---|---|---|
| | | | 300 K | 6000 K | 9000 K | 300 K | 6000 K |
| Lower mantle | 25 | 4.41 | 9.26 (±0.05) | - | - | 9.28 | - |
| Lower mantle | 50 | 4.76 | 10.02 (±0.05) | 8.40 (±0.06) | - | 9.98 | - |
| Lower mantle | 75 | 5.04 | 10.64 (±0.06) | 9.25 (±0.08) | 8.27 (±0.08) | 10.55 | 8.90 |
| Lower mantle | 100 | 5.29 | 11.19 (±0.07) | 9.93 (±0.09) | 9.09 (±0.10) | 11.04 | 9.70 |
| Mantle (CMB) | 125 | 5.50 | 11.68 (±0.07) | 10.52 (±0.11) | 9.76 (±0.12) | 11.48 | 10.31 |
| Outer (CMB) | 125 | 9.74 | 11.68 (±0.07) | 10.52 (±0.11) | 9.76 (±0.12) | 11.48 | 10.31 |
| Outer core | 150 | 10.11 | 12.13 (±0.08) | 11.04 (±0.12) | 10.34 (±0.13) | 11.87 | 10.82 |
| Outer core | 175 | 10.46 | 12.54 (±0.08) | 11.51 (±0.12) | 10.85 (±0.14) | 12.24 | 11.28 |
| Outer core | 200 | 10.78 | 12.93 (±0.08) | 11.94 (±0.13) | 11.32 (±0.15) | 12.58 | 11.68 |
| Outer core | 225 | 11.08 | 13.29 (±0.08) | 12.34 (±0.14) | 11.75 (±0.16) | 12.90 | 12.06 |
| Outer core | 250 | 11.37 | 13.64 (±0.09) | 12.72 (±0.14) | 12.15 (±0.16) | 13.20 | 12.40 |
| Outer core | 275 | 11.64 | 13.97 (±0.09) | 13.08 (±0.14) | 12.53 (±0.17) | 13.49 | 12.72 |
| Outer core | 300 | 11.89 | 14.28 (±0.09) | 13.42 (±0.14) | 12.89 (±0.17) | 13.76 | 13.03 |
| Outer (ICB) | 330 | 12.19 | 14.65 (±0.10) | 13.81 (±0.14) | 13.29 (±0.17) | 14.07 | 13.37 |
| Inner (ICB) | 330 | 12.79 | 14.65 (±0.10) | 13.81 (±0.14) | 13.29 (±0.17) | 14.07 | 13.37 |
| Inner core | 365 | 13.12 | 15.05 (±0.11) | 14.24 (±0.14) | 13.74 (±0.17) | 14.42 | 13.75 |

*Note*: * Densities were derived from the *K*-primed Mie–Grüneisen–Debye model using the parameters for hcp-iron and the molar electronic heat capacity by the free electron model with eight valence electrons (FEM-8) (details are in the Methods section, and the parameters of the equation of state for hcp-iron are in Table S2).
*Abbreviations*: PREM: preliminary reference Earth model (*ref. 9*), CMB: core–mantle boundary, ICB: inner core boundary, DA06: (*ref. 5*).

**Table S7.** Isotherms by the *K*-primed MGD EoS for hcp-iron.

| $\rho$, density (g cm$^{-3}$) | *P*, pressure (GPa) * | | | | | | | | | | |
|---|---|---|---|---|---|---|---|---|---|---|---|
| | 300 K | 1000 K | 2000 K | 3000 K | 6000 K | 9000 K | 12000 K | 15000 K | 18000 K | 21000 K | 24000 K |
| 8.25 | 0 | 5 | 13 | 21 | 46 | 74 | 105 | 138 | 174 | 212 | 253 |
| 8.50 | 5 | 10 | 18 | 26 | 52 | 81 | 112 | 146 | 183 | 221 | 263 |
| 8.75 | 11 | 16 | 24 | 33 | 59 | 89 | 120 | 155 | 192 | 231 | 273 |
| 9.00 | 17 | 23 | 31 | 39 | 67 | 97 | 129 | 164 | 202 | 242 | 284 |
| 9.25 | 25 | 30 | 38 | 47 | 75 | 105 | 138 | 174 | 212 | 253 | 296 |
| 9.50 | 32 | 38 | 46 | 55 | 84 | 115 | 148 | 184 | 223 | 265 | 308 |
| 9.75 | 40 | 46 | 55 | 64 | 93 | 124 | 159 | 195 | 235 | 277 | 321 |
| 10.00 | 49 | 55 | 64 | 73 | 103 | 135 | 170 | 207 | 247 | 290 | 335 |
| 10.25 | 59 | 65 | 74 | 83 | 113 | 146 | 181 | 219 | 260 | 303 | 349 |
| 10.50 | 69 | 75 | 84 | 94 | 124 | 158 | 194 | 232 | 273 | 317 | 364 |
| 10.75 | 80 | 86 | 95 | 105 | 136 | 170 | 206 | 246 | 287 | 332 | 379 |
| 11.00 | 91 | 97 | 107 | 117 | 148 | 183 | 220 | 260 | 302 | 347 | 395 |
| 11.25 | 103 | 109 | 119 | 129 | 161 | 196 | 234 | 274 | 317 | 363 | 411 |
| 11.50 | 116 | 122 | 132 | 142 | 175 | 210 | 249 | 290 | 333 | 379 | 428 |
| 11.75 | 129 | 135 | 145 | 156 | 189 | 225 | 264 | 305 | 350 | 396 | 446 |
| 12.00 | 143 | 149 | 159 | 170 | 204 | 240 | 280 | 322 | 367 | 414 | 464 |
| 12.25 | 157 | 164 | 174 | 185 | 219 | 256 | 296 | 339 | 384 | 433 | 483 |
| 12.50 | 173 | 179 | 189 | 200 | 235 | 273 | 314 | 357 | 403 | 451 | 503 |
| 12.75 | 188 | 195 | 205 | 217 | 252 | 290 | 331 | 375 | 422 | 471 | 523 |
| 13.00 | 205 | 211 | 222 | 233 | 269 | 308 | 350 | 394 | 441 | 491 | 544 |
| 13.25 | 222 | 228 | 239 | 251 | 287 | 327 | 369 | 414 | 462 | 512 | 565 |
| 13.50 | 240 | 246 | 257 | 269 | 306 | 346 | 389 | 434 | 483 | 534 | 587 |
| 13.75 | 258 | 265 | 276 | 288 | 325 | 366 | 409 | 455 | 504 | 556 | 610 |
| 14.00 | 277 | 284 | 295 | 307 | 345 | 386 | 430 | 477 | 526 | 578 | 633 |

*Note*: * Pressures were derived from the *K*-primed Mie–Grüneisen–Debye model using the parameters for hexagonal close-packed (hcp) iron (details are in the Methods section, and the parameters of the equation of state for hcp-iron are in Tables S2, S8, and S9).

**Table S8.** Molar phonon heat capacity for hcp-iron by the Debye model.

| $\rho$, density (g cm$^{-3}$) | $c_{ph}$ (DM), molar phonon heat capacity by the Debye model (J K$^{-1}$ mol$^{-1}$) * | | | | | | | | | | |
|---|---|---|---|---|---|---|---|---|---|---|---|
| | 300 K | 1000 K | 2000 K | 3000 K | 6000 K | 9000 K | 12000 K | 15000 K | 18000 K | 21000 K | 24000 K |
| 8.25 | 21.622 | 24.616 | 24.861 | 24.907 | 24.934 | 24.939 | 24.941 | 24.942 | 24.942 | 24.943 | 24.943 |
| 8.50 | 21.254 | 24.575 | 24.851 | 24.902 | 24.933 | 24.939 | 24.941 | 24.942 | 24.942 | 24.943 | 24.943 |
| 8.75 | 20.864 | 24.532 | 24.840 | 24.897 | 24.932 | 24.938 | 24.940 | 24.942 | 24.942 | 24.942 | 24.943 |
| 9.00 | 20.454 | 24.485 | 24.828 | 24.892 | 24.930 | 24.938 | 24.940 | 24.941 | 24.942 | 24.942 | 24.943 |
| 9.25 | 20.025 | 24.434 | 24.815 | 24.886 | 24.929 | 24.937 | 24.940 | 24.941 | 24.942 | 24.942 | 24.942 |
| 9.50 | 19.577 | 24.380 | 24.801 | 24.880 | 24.927 | 24.936 | 24.939 | 24.941 | 24.942 | 24.942 | 24.942 |
| 9.75 | 19.112 | 24.321 | 24.786 | 24.873 | 24.926 | 24.936 | 24.939 | 24.941 | 24.941 | 24.942 | 24.942 |
| 10.00 | 18.632 | 24.259 | 24.770 | 24.866 | 24.924 | 24.935 | 24.939 | 24.940 | 24.941 | 24.942 | 24.942 |
| 10.25 | 18.137 | 24.193 | 24.753 | 24.858 | 24.922 | 24.934 | 24.938 | 24.940 | 24.941 | 24.942 | 24.942 |
| 10.50 | 17.629 | 24.123 | 24.735 | 24.850 | 24.920 | 24.933 | 24.938 | 24.940 | 24.941 | 24.941 | 24.942 |
| 10.75 | 17.111 | 24.049 | 24.715 | 24.842 | 24.918 | 24.932 | 24.937 | 24.939 | 24.941 | 24.941 | 24.942 |
| 11.00 | 16.584 | 23.970 | 24.695 | 24.832 | 24.916 | 24.931 | 24.936 | 24.939 | 24.940 | 24.941 | 24.942 |
| 11.25 | 16.050 | 23.887 | 24.673 | 24.823 | 24.913 | 24.930 | 24.936 | 24.939 | 24.940 | 24.941 | 24.941 |
| 11.50 | 15.510 | 23.799 | 24.650 | 24.812 | 24.911 | 24.929 | 24.935 | 24.938 | 24.940 | 24.941 | 24.941 |
| 11.75 | 14.967 | 23.706 | 24.626 | 24.801 | 24.908 | 24.928 | 24.934 | 24.938 | 24.939 | 24.940 | 24.941 |
| 12.00 | 14.423 | 23.610 | 24.600 | 24.790 | 24.905 | 24.926 | 24.934 | 24.937 | 24.939 | 24.940 | 24.941 |
| 12.25 | 13.879 | 23.508 | 24.573 | 24.778 | 24.902 | 24.925 | 24.933 | 24.937 | 24.939 | 24.940 | 24.941 |
| 12.50 | 13.338 | 23.401 | 24.545 | 24.765 | 24.899 | 24.923 | 24.932 | 24.936 | 24.938 | 24.940 | 24.941 |
| 12.75 | 12.800 | 23.290 | 24.515 | 24.752 | 24.895 | 24.922 | 24.931 | 24.936 | 24.938 | 24.939 | 24.940 |
| 13.00 | 12.268 | 23.174 | 24.483 | 24.737 | 24.892 | 24.920 | 24.930 | 24.935 | 24.938 | 24.939 | 24.940 |
| 13.25 | 11.744 | 23.053 | 24.451 | 24.723 | 24.888 | 24.919 | 24.930 | 24.935 | 24.937 | 24.939 | 24.940 |
| 13.50 | 11.228 | 22.927 | 24.416 | 24.707 | 24.884 | 24.917 | 24.929 | 24.934 | 24.937 | 24.939 | 24.940 |
| 13.75 | 10.722 | 22.796 | 24.380 | 24.691 | 24.880 | 24.915 | 24.927 | 24.933 | 24.936 | 24.938 | 24.939 |
| 14.00 | 10.227 | 22.660 | 24.343 | 24.674 | 24.876 | 24.913 | 24.926 | 24.933 | 24.936 | 24.938 | 24.939 |

*Note*: * $c_{ph}$ (DM) were derived from the Debye model (DM) using the parameters for hexagonal close-packed (hcp) iron (details are in the Methods section, and the parameters of the equation of state for hcp-iron are in Table S2).

**Table S9.** Molar electronic heat capacity for hcp-iron by the FEM-8.

| $\rho$, density (g cm$^{-3}$) | $c_{el}$ (FEM-8), molar electronic heat capacity by the free electron model with eight valence electrons (J K$^{-1}$ mol$^{-1}$) * | | | | | | | | | | |
|---|---|---|---|---|---|---|---|---|---|---|---|
| | 300 K | 1000 K | 2000 K | 3000 K | 6000 K | 9000 K | 12000 K | 15000 K | 18000 K | 21000 K | 24000 K |
| 8.25 | 0.243 | 0.924 | 1.897 | 2.870 | 5.785 | 8.691 | 11.585 | 14.462 | 17.317 | 20.145 | 22.941 |
| 8.50 | 0.238 | 0.906 | 1.860 | 2.813 | 5.671 | 8.520 | 11.358 | 14.180 | 16.982 | 19.758 | 22.504 |
| 8.75 | 0.234 | 0.889 | 1.824 | 2.760 | 5.562 | 8.358 | 11.142 | 13.912 | 16.662 | 19.389 | 22.086 |
| 9.00 | 0.230 | 0.872 | 1.790 | 2.708 | 5.459 | 8.203 | 10.937 | 13.656 | 16.357 | 19.036 | 21.688 |
| 9.25 | 0.225 | 0.857 | 1.758 | 2.659 | 5.360 | 8.055 | 10.740 | 13.411 | 16.065 | 18.699 | 21.306 |
| 9.50 | 0.221 | 0.841 | 1.727 | 2.612 | 5.266 | 7.914 | 10.552 | 13.177 | 15.786 | 18.376 | 20.941 |
| 9.75 | 0.218 | 0.827 | 1.697 | 2.568 | 5.176 | 7.778 | 10.372 | 12.953 | 15.519 | 18.067 | 20.591 |
| 10.00 | 0.214 | 0.813 | 1.669 | 2.525 | 5.089 | 7.649 | 10.199 | 12.738 | 15.263 | 17.770 | 20.255 |
| 10.25 | 0.210 | 0.800 | 1.642 | 2.483 | 5.006 | 7.524 | 10.034 | 12.532 | 15.017 | 17.485 | 19.933 |
| 10.50 | 0.207 | 0.787 | 1.616 | 2.444 | 4.927 | 7.405 | 9.875 | 12.334 | 14.781 | 17.211 | 19.623 |
| 10.75 | 0.204 | 0.775 | 1.590 | 2.406 | 4.850 | 7.290 | 9.722 | 12.144 | 14.554 | 16.948 | 19.324 |
| 11.00 | 0.201 | 0.763 | 1.566 | 2.369 | 4.776 | 7.179 | 9.575 | 11.961 | 14.335 | 16.695 | 19.037 |
| 11.25 | 0.198 | 0.752 | 1.543 | 2.334 | 4.706 | 7.073 | 9.433 | 11.784 | 14.124 | 16.450 | 18.760 |
| 11.50 | 0.195 | 0.741 | 1.521 | 2.300 | 4.637 | 6.970 | 9.296 | 11.614 | 13.921 | 16.215 | 18.492 |
| 11.75 | 0.192 | 0.730 | 1.499 | 2.267 | 4.571 | 6.871 | 9.165 | 11.450 | 13.725 | 15.987 | 18.235 |
| 12.00 | 0.190 | 0.720 | 1.478 | 2.236 | 4.508 | 6.776 | 9.038 | 11.292 | 13.536 | 15.768 | 17.985 |
| 12.25 | 0.187 | 0.710 | 1.458 | 2.205 | 4.446 | 6.683 | 8.915 | 11.139 | 13.353 | 15.556 | 17.744 |
| 12.50 | 0.184 | 0.701 | 1.438 | 2.176 | 4.387 | 6.594 | 8.796 | 10.991 | 13.176 | 15.350 | 17.511 |
| 12.75 | 0.182 | 0.692 | 1.419 | 2.147 | 4.329 | 6.508 | 8.681 | 10.848 | 13.005 | 15.152 | 17.286 |
| 13.00 | 0.180 | 0.683 | 1.401 | 2.120 | 4.274 | 6.424 | 8.570 | 10.709 | 12.839 | 14.960 | 17.067 |
| 13.25 | 0.177 | 0.674 | 1.384 | 2.093 | 4.220 | 6.343 | 8.462 | 10.575 | 12.679 | 14.773 | 16.856 |
| 13.50 | 0.175 | 0.666 | 1.366 | 2.067 | 4.168 | 6.265 | 8.358 | 10.445 | 12.523 | 14.592 | 16.650 |
| 13.75 | 0.173 | 0.658 | 1.350 | 2.042 | 4.117 | 6.189 | 8.257 | 10.318 | 12.372 | 14.417 | 16.451 |
| 14.00 | 0.171 | 0.650 | 1.334 | 2.018 | 4.068 | 6.115 | 8.158 | 10.196 | 12.226 | 14.247 | 16.258 |

*Note*: * $c_{el}$ (FEM-8) were derived from the free electron model with eight valence electrons (FEM-8) using the parameters for hexagonal close-packed (hcp) iron (details are in the Methods section, and the parameters of the equation of state for hcp-iron are in Table S2).

**Table S10.** Molar electronic heat capacity for hcp-iron by the LTD.

| $\rho$, density (g cm$^{-3}$) | $\Gamma_{el}T(\rho_0/\rho)$, molar electronic heat capacity by the linear temperature dependence model (J K$^{-1}$ mol$^{-1}$) * | | | | | | | | | | |
|---|---|---|---|---|---|---|---|---|---|---|---|
| | 300 K | 1000 K | 2000 K | 3000 K | 6000 K | 9000 K | 12000 K | 15000 K | 18000 K | 21000 K | 24000 K |
| 8.25 | 1.470 | 4.900 | 9.800 | 14.700 | 29.400 | 44.100 | 58.800 | 73.500 | 88.200 | 102.900 | 117.600 |
| 8.50 | 1.427 | 4.756 | 9.512 | 14.268 | 28.535 | 42.803 | 57.071 | 71.338 | 85.606 | 99.874 | 114.141 |
| 8.75 | 1.386 | 4.620 | 9.240 | 13.860 | 27.720 | 41.580 | 55.440 | 69.300 | 83.160 | 97.020 | 110.880 |
| 9.00 | 1.348 | 4.492 | 8.983 | 13.475 | 26.950 | 40.425 | 53.900 | 67.375 | 80.850 | 94.325 | 107.800 |
| 9.25 | 1.311 | 4.370 | 8.741 | 13.111 | 26.222 | 39.332 | 52.443 | 65.554 | 78.665 | 91.776 | 104.886 |
| 9.50 | 1.277 | 4.255 | 8.511 | 12.766 | 25.532 | 38.297 | 51.063 | 63.829 | 76.595 | 89.361 | 102.126 |
| 9.75 | 1.244 | 4.146 | 8.292 | 12.438 | 24.877 | 37.315 | 49.754 | 62.192 | 74.631 | 87.069 | 99.508 |
| 10.00 | 1.213 | 4.043 | 8.085 | 12.128 | 24.255 | 36.383 | 48.510 | 60.638 | 72.765 | 84.893 | 97.020 |
| 10.25 | 1.183 | 3.944 | 7.888 | 11.832 | 23.663 | 35.495 | 47.327 | 59.159 | 70.990 | 82.822 | 94.654 |
| 10.50 | 1.155 | 3.850 | 7.700 | 11.550 | 23.100 | 34.650 | 46.200 | 57.750 | 69.300 | 80.850 | 92.400 |
| 10.75 | 1.128 | 3.760 | 7.521 | 11.281 | 22.563 | 33.844 | 45.126 | 56.407 | 67.688 | 78.970 | 90.251 |
| 11.00 | 1.103 | 3.675 | 7.350 | 11.025 | 22.050 | 33.075 | 44.100 | 55.125 | 66.150 | 77.175 | 88.200 |
| 11.25 | 1.078 | 3.593 | 7.187 | 10.780 | 21.560 | 32.340 | 43.120 | 53.900 | 64.680 | 75.460 | 86.240 |
| 11.50 | 1.055 | 3.515 | 7.030 | 10.546 | 21.091 | 31.637 | 42.183 | 52.728 | 63.274 | 73.820 | 84.365 |
| 11.75 | 1.032 | 3.440 | 6.881 | 10.321 | 20.643 | 30.964 | 41.285 | 51.606 | 61.928 | 72.249 | 82.570 |
| 12.00 | 1.011 | 3.369 | 6.738 | 10.106 | 20.213 | 30.319 | 40.425 | 50.531 | 60.638 | 70.744 | 80.850 |
| 12.25 | 0.990 | 3.300 | 6.600 | 9.900 | 19.800 | 29.700 | 39.600 | 49.500 | 59.400 | 69.300 | 79.200 |
| 12.50 | 0.970 | 3.234 | 6.468 | 9.702 | 19.404 | 29.106 | 38.808 | 48.510 | 58.212 | 67.914 | 77.616 |
| 12.75 | 0.951 | 3.171 | 6.341 | 9.512 | 19.024 | 28.535 | 38.047 | 47.559 | 57.071 | 66.582 | 76.094 |
| 13.00 | 0.933 | 3.110 | 6.219 | 9.329 | 18.658 | 27.987 | 37.315 | 46.644 | 55.973 | 65.302 | 74.631 |
| 13.25 | 0.915 | 3.051 | 6.102 | 9.153 | 18.306 | 27.458 | 36.611 | 45.764 | 54.917 | 64.070 | 73.223 |
| 13.50 | 0.898 | 2.994 | 5.989 | 8.983 | 17.967 | 26.950 | 35.933 | 44.917 | 53.900 | 62.883 | 71.867 |
| 13.75 | 0.882 | 2.940 | 5.880 | 8.820 | 17.640 | 26.460 | 35.280 | 44.100 | 52.920 | 61.740 | 70.560 |
| 14.00 | 0.866 | 2.888 | 5.775 | 8.663 | 17.325 | 25.988 | 34.650 | 43.313 | 51.975 | 60.638 | 69.300 |

*Note*: * $\Gamma_{el}T(\rho_0/\rho)$ were derived from the linear temperature dependence with $\Gamma_{el}$ = 4.90 (mJ K$^{-2}$ mol$^{-1}$) (*ref. 59*) using the parameters for hexagonal close-packed (hcp) iron (details are in the Methods section, and the parameters of the equation of state for hcp-iron are in Table S2).

## *Supplementary References*

Supplementary Data

# Density deficit of the Earth's core revealed by a multi-megabar primary pressure scale

Daijo Ikuta*, Eiji Ohtani*, Hiroshi Fukui, Tatsuya Sakamaki,
Rolf Heid, Daisuke Ishikawa, and Alfred Q. R. Baron*

* Corresponding authors. Email: dikuta@okayama-u.ac.jp,
eohtani@tohoku.ac.jp, and baron@spring8.or.jp

***This PDF file includes:***

*All spectra and fitting results*

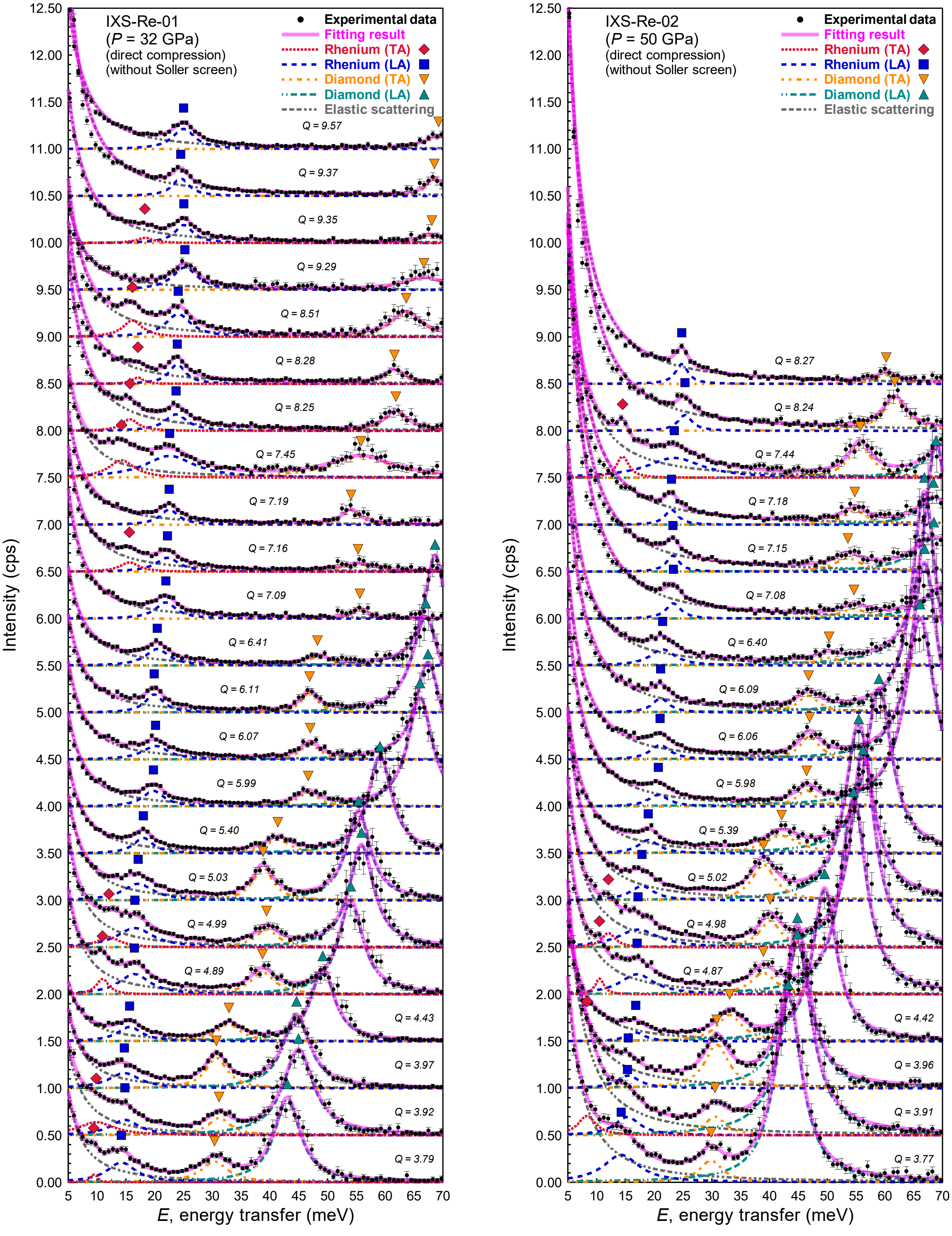

Inelastic x-ray scattering (IXS) spectra and fitting results in high pressure conditions (IXS-Re-01 and IXS-Re-02). The symbols and lines are the same as those in Fig. 1A in the main text.

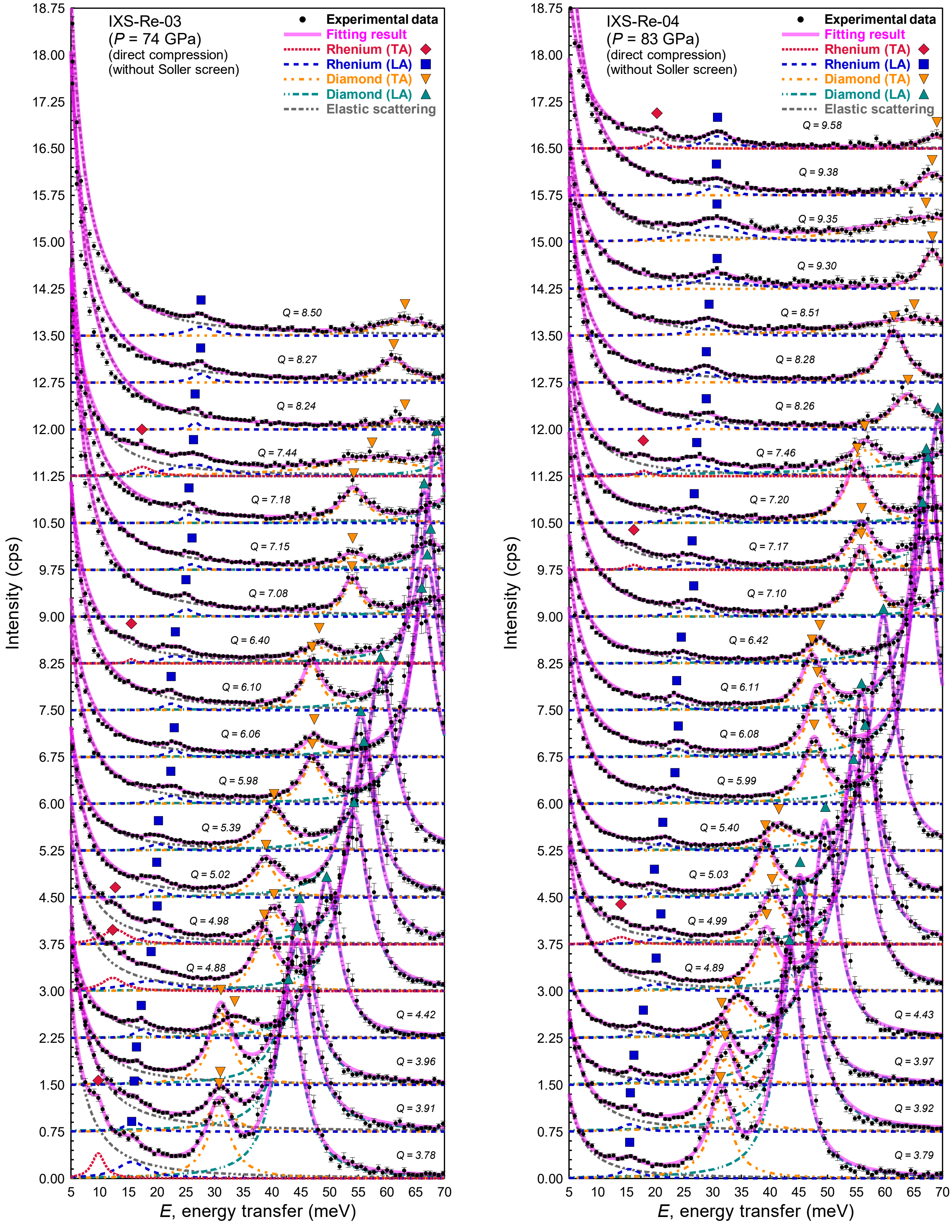

IXS spectra and fitting results in high pressure conditions (IXS-Re-03 and IXS-Re-04).
The symbols and lines are the same as those in Fig. 1A in the main text.

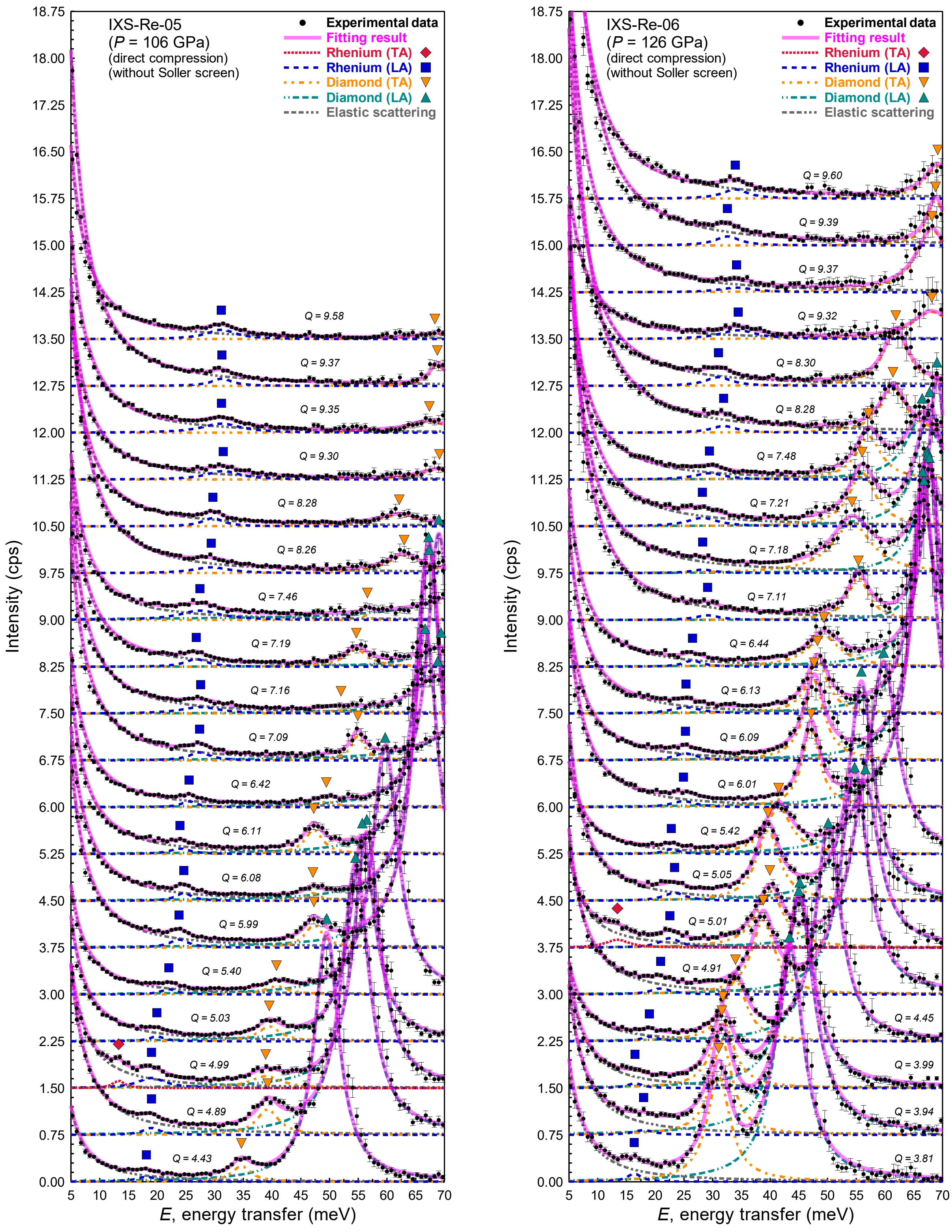

IXS spectra and fitting results in high pressure conditions (IXS-Re-05 and IXS-Re-06).
The symbols and lines are the same as those in Fig. 1A in the main text.

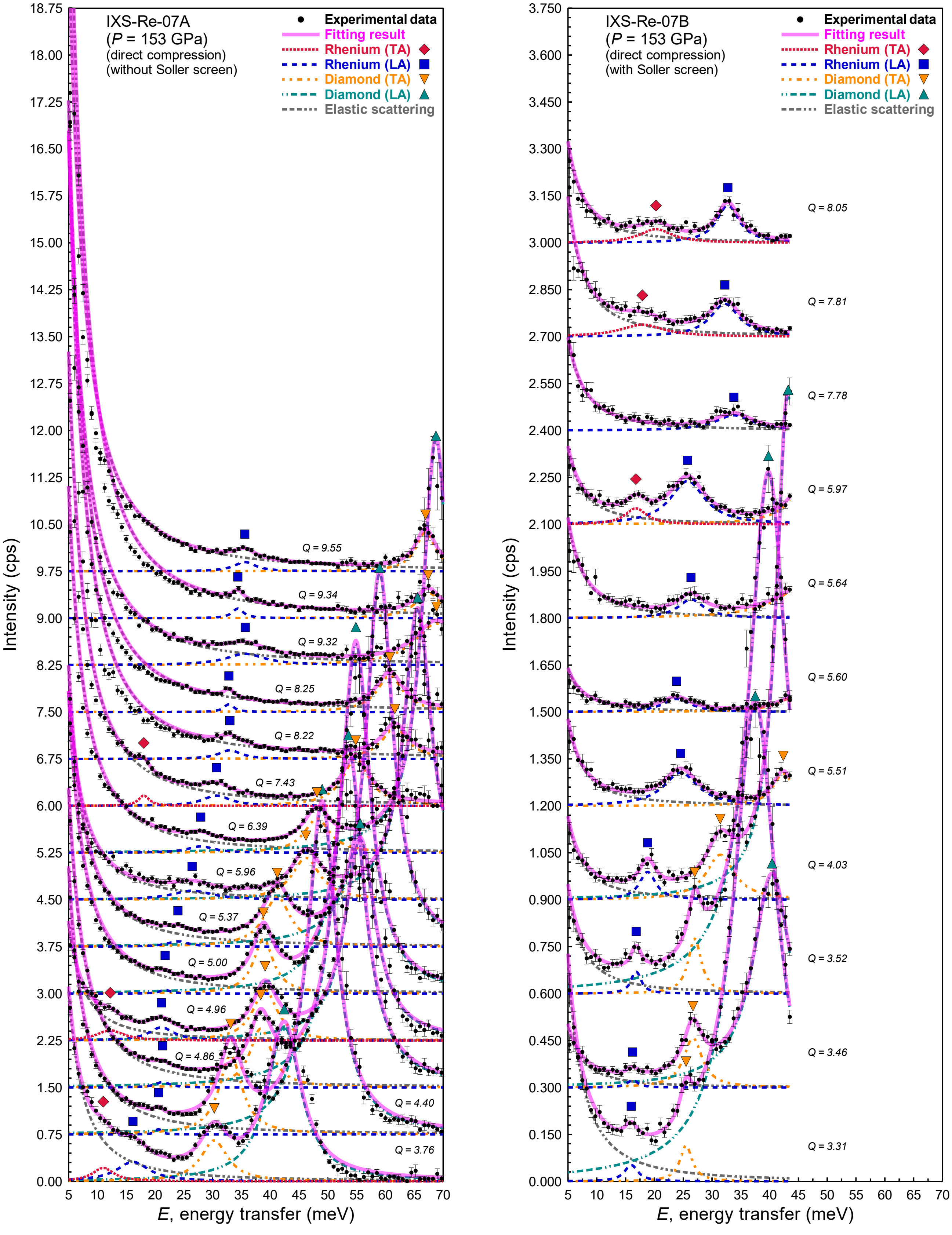

IXS spectra and fitting results in high pressure conditions (IXS-Re-07).
The symbols and lines are the same as those in Fig. 1A in the main text.

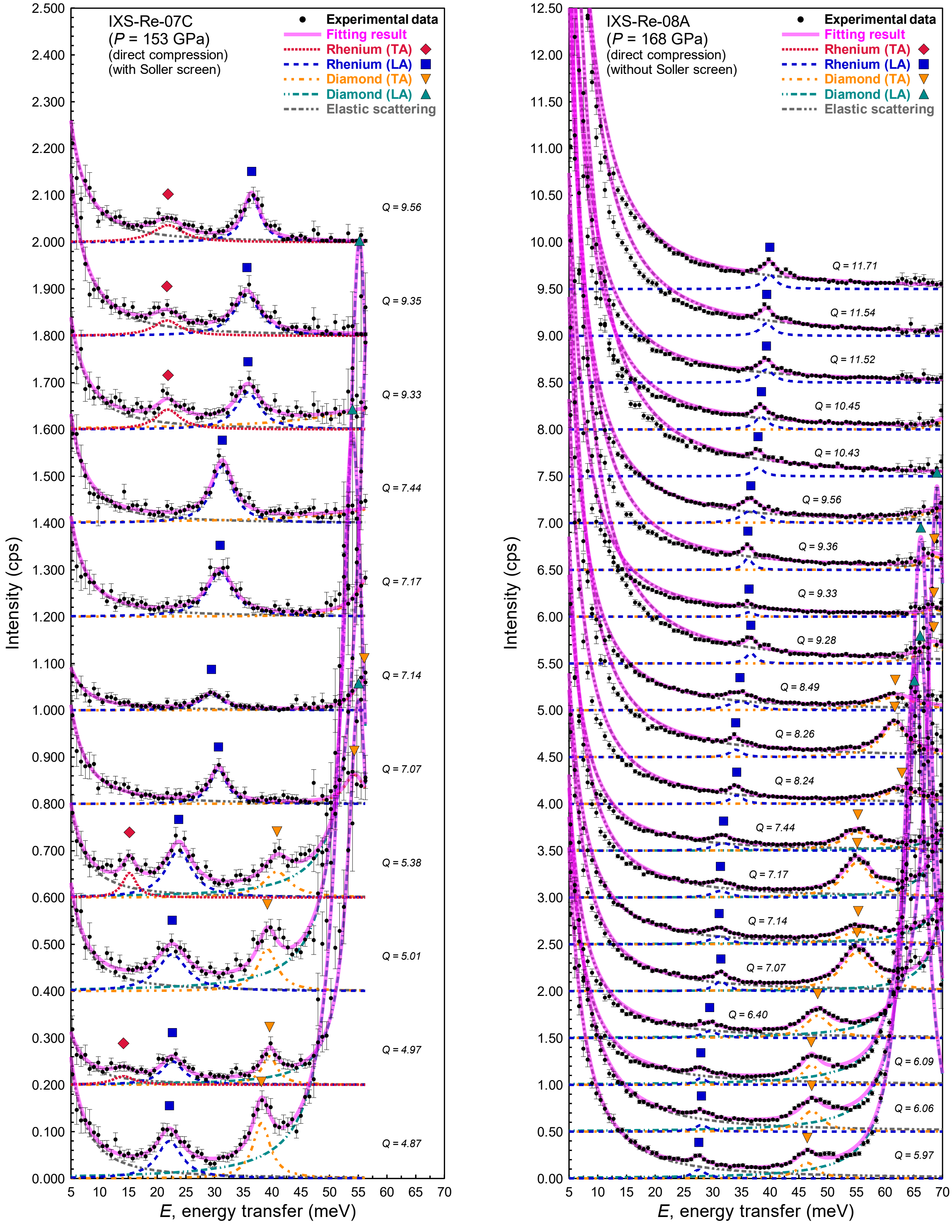

IXS spectra and fitting results in high pressure conditions (IXS-Re-07 and IXS-Re-08).
The symbols and lines are the same as those in Fig. 1A in the main text.

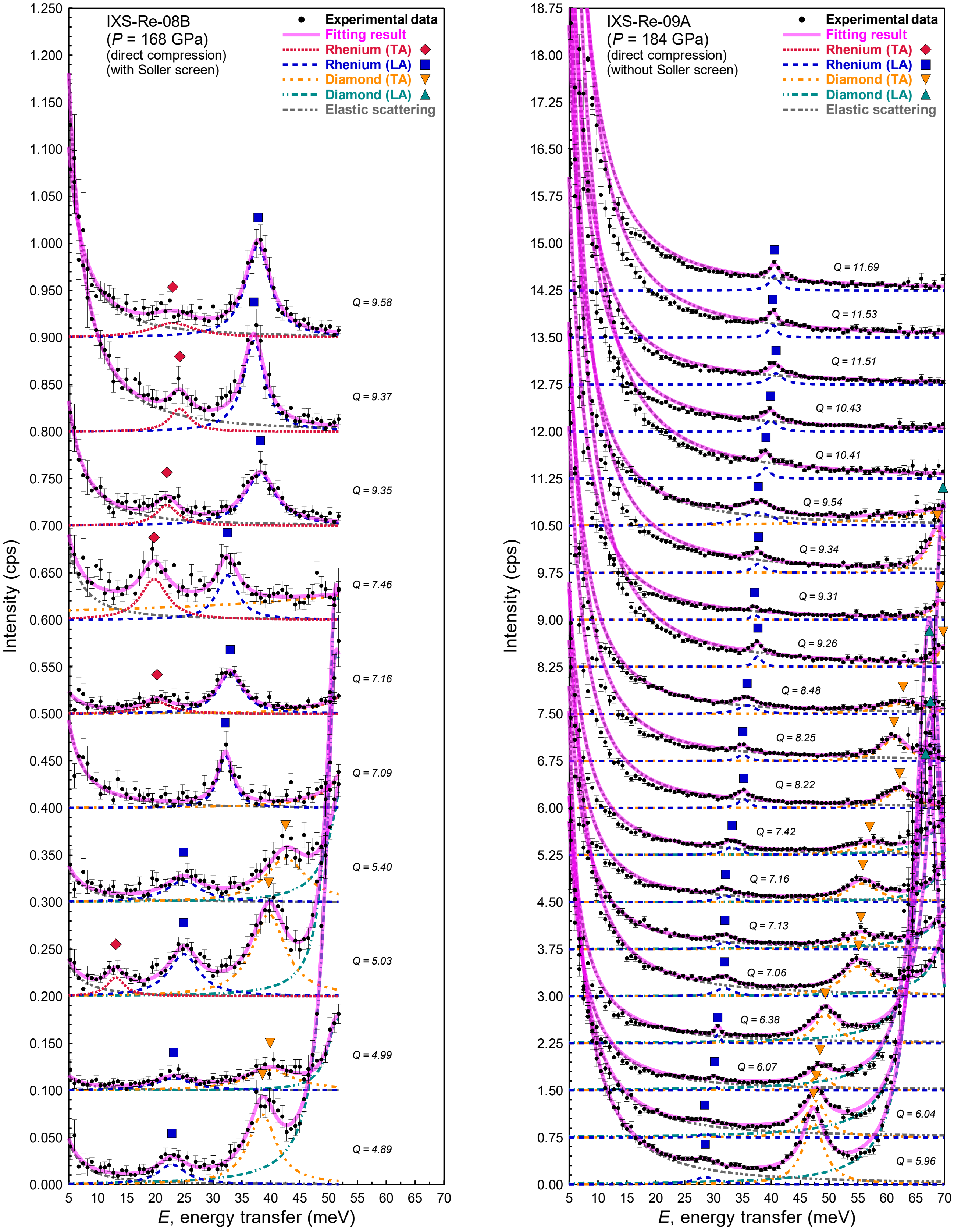

IXS spectra and fitting results in high pressure conditions (IXS-Re-08 and IXS-Re-09).
The symbols and lines are the same as those in Fig. 1A in the main text.

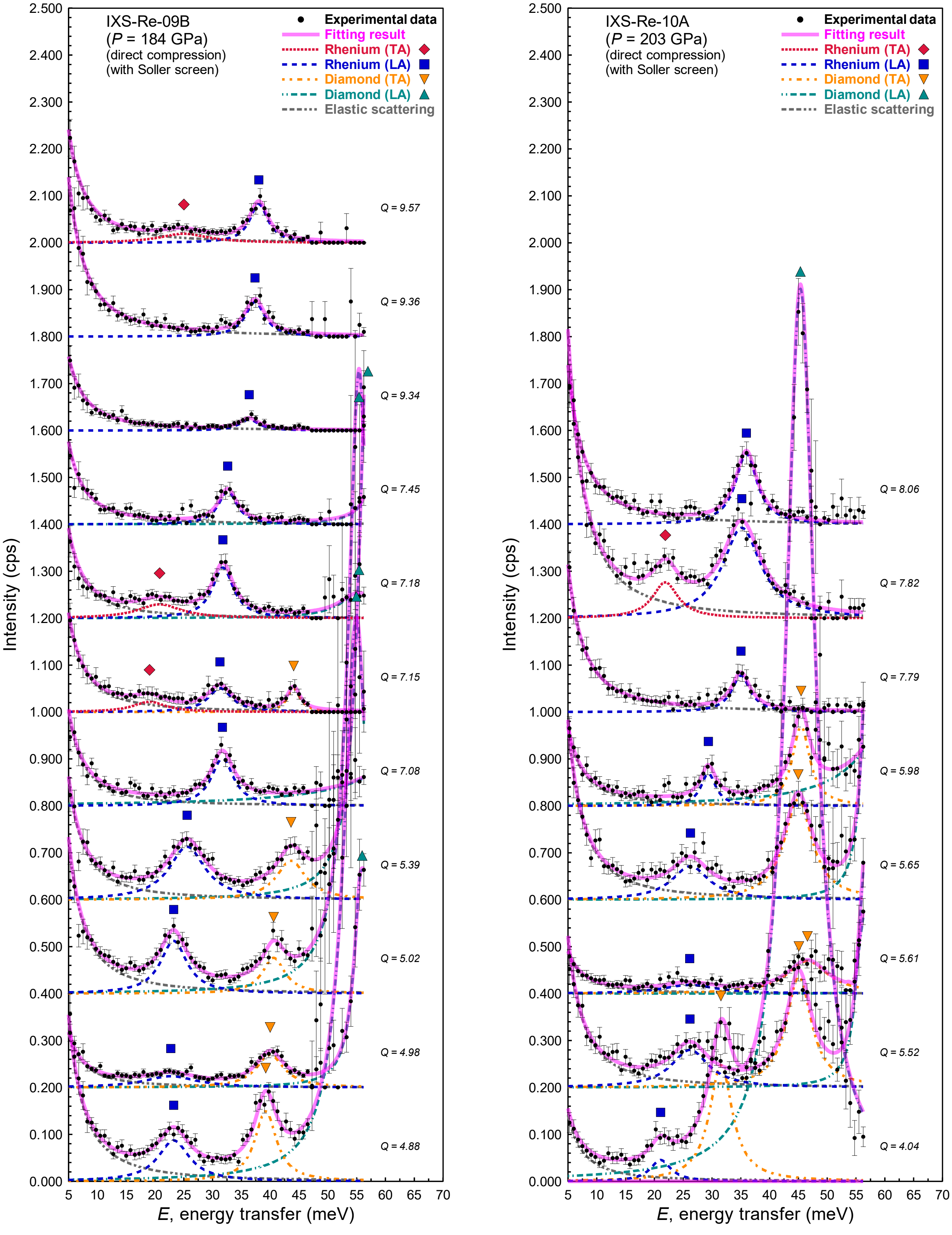

IXS spectra and fitting results in high pressure conditions (IXS-Re-09 and IXS-Re-10).
The symbols and lines are the same as those in Fig. 1A in the main text.

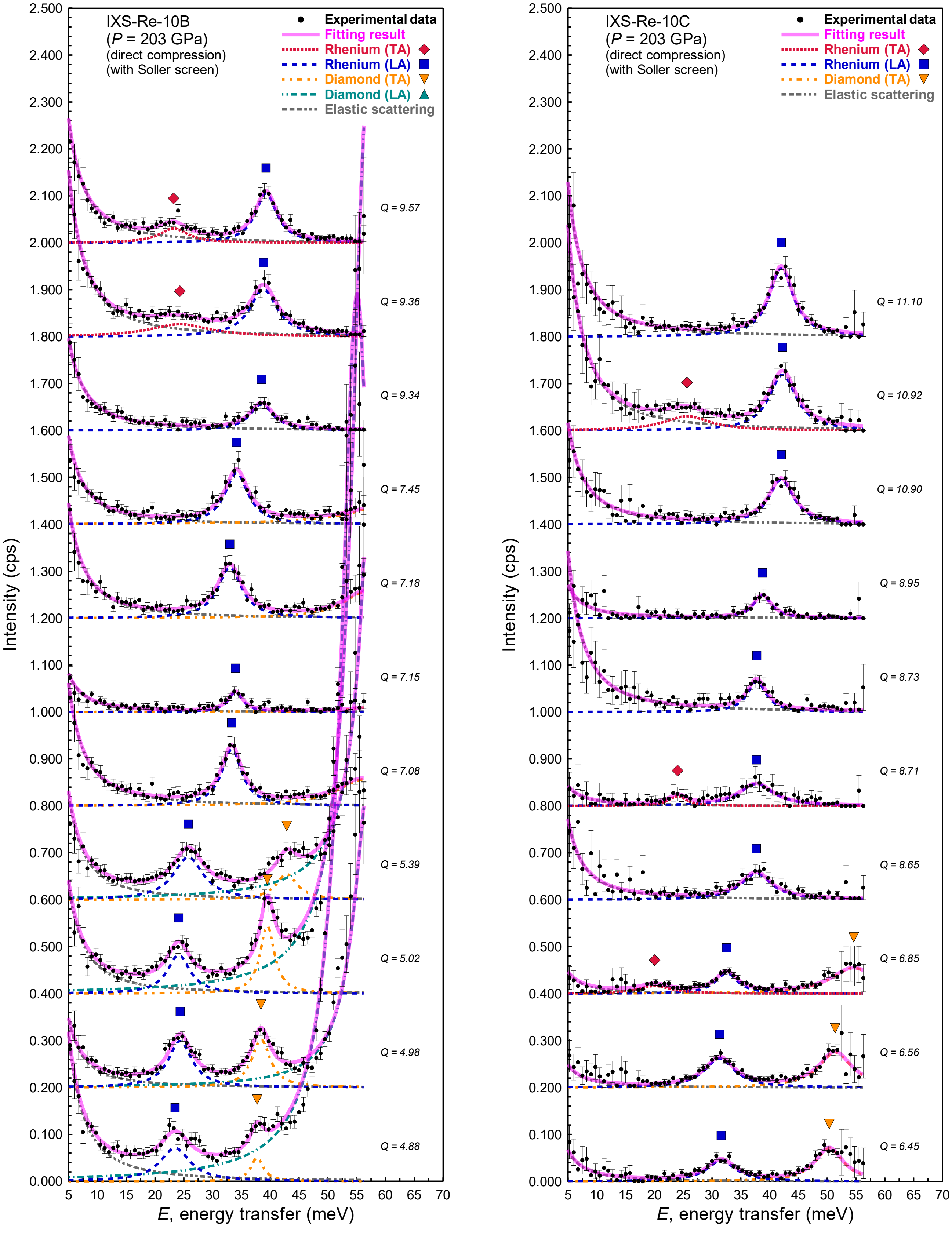

IXS spectra and fitting results in high pressure conditions (IXS-Re-10).
The symbols and lines are the same as those in Fig. 1A in the main text.

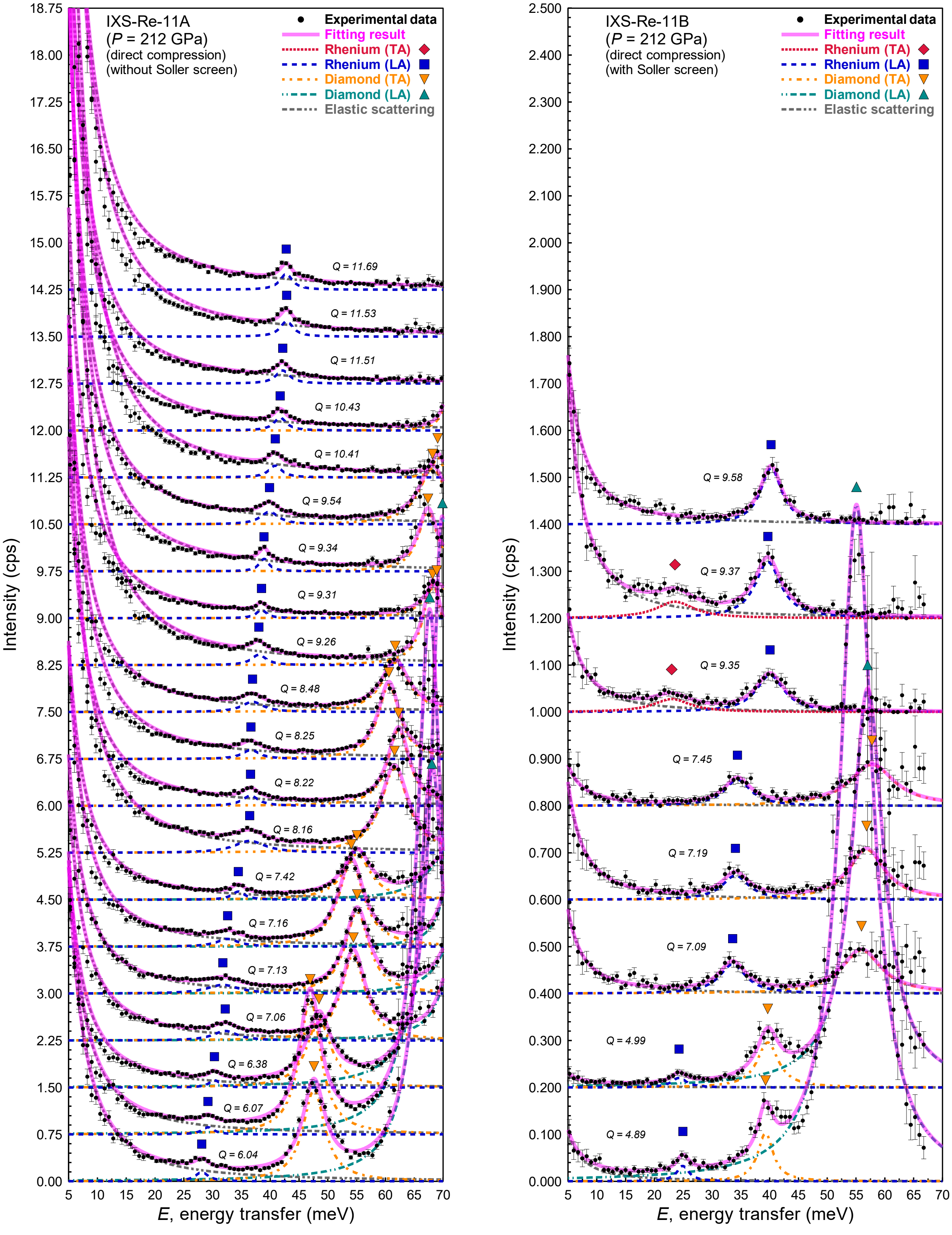

IXS spectra and fitting results in high pressure conditions (IXS-Re-11).
The symbols and lines are the same as those in Fig. 1A in the main text.

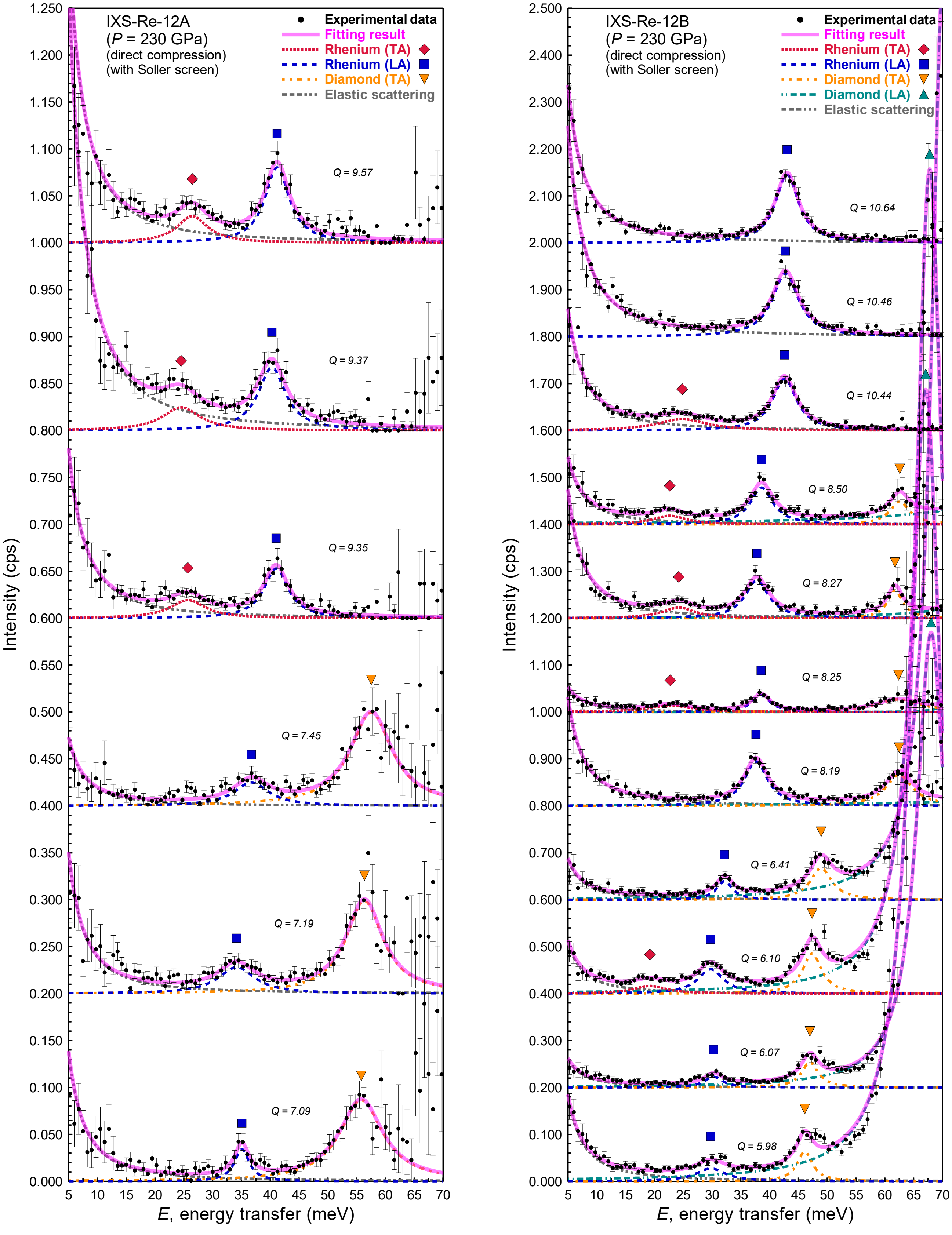

IXS spectra and fitting results in high pressure conditions (IXS-Re-12).
The symbols and lines are the same as those in Fig. 1A in the main text.

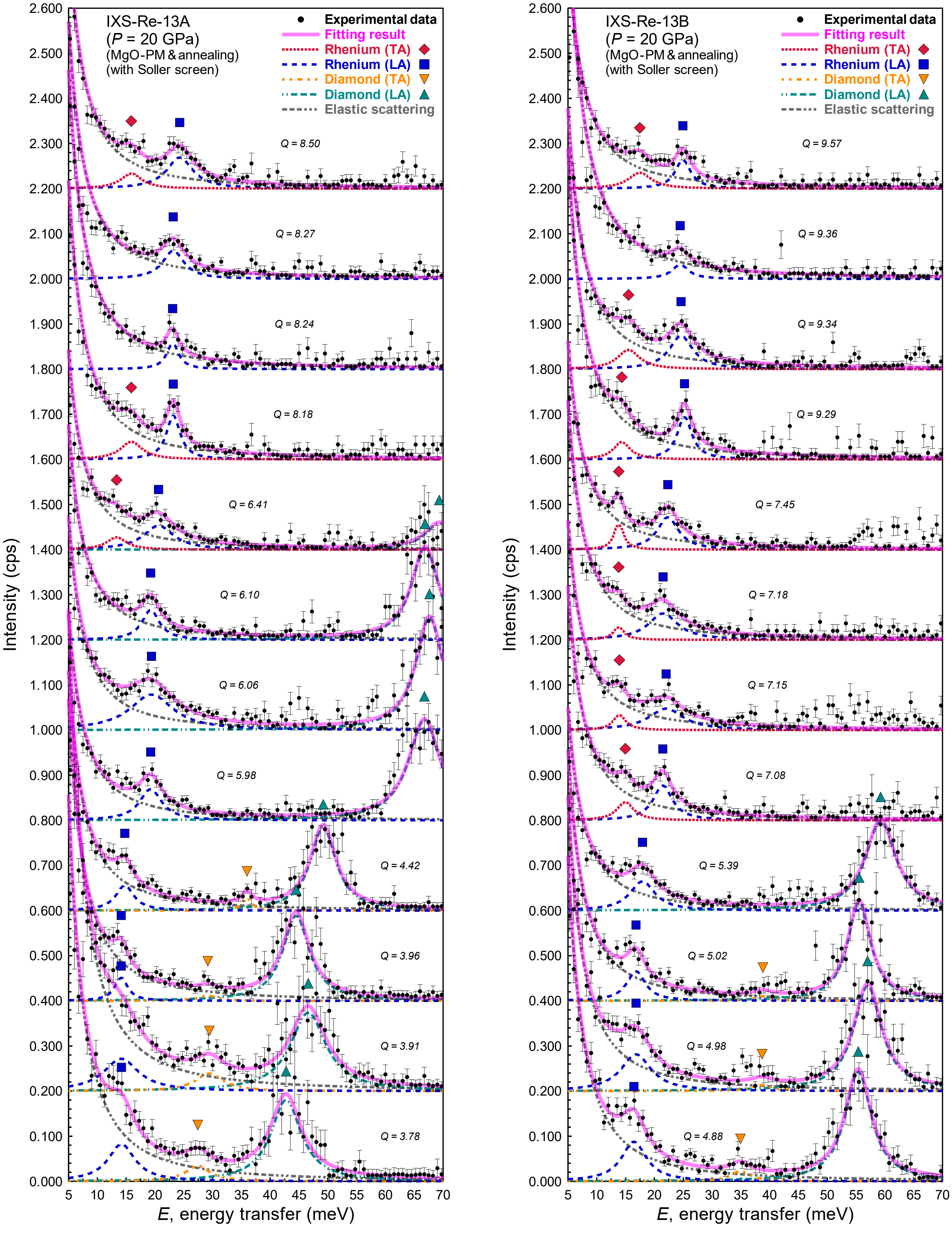

IXS spectra and fitting results in high pressure conditions (IXS-Re-13).
The symbols and lines are the same as those in Fig. 1A in the main text.

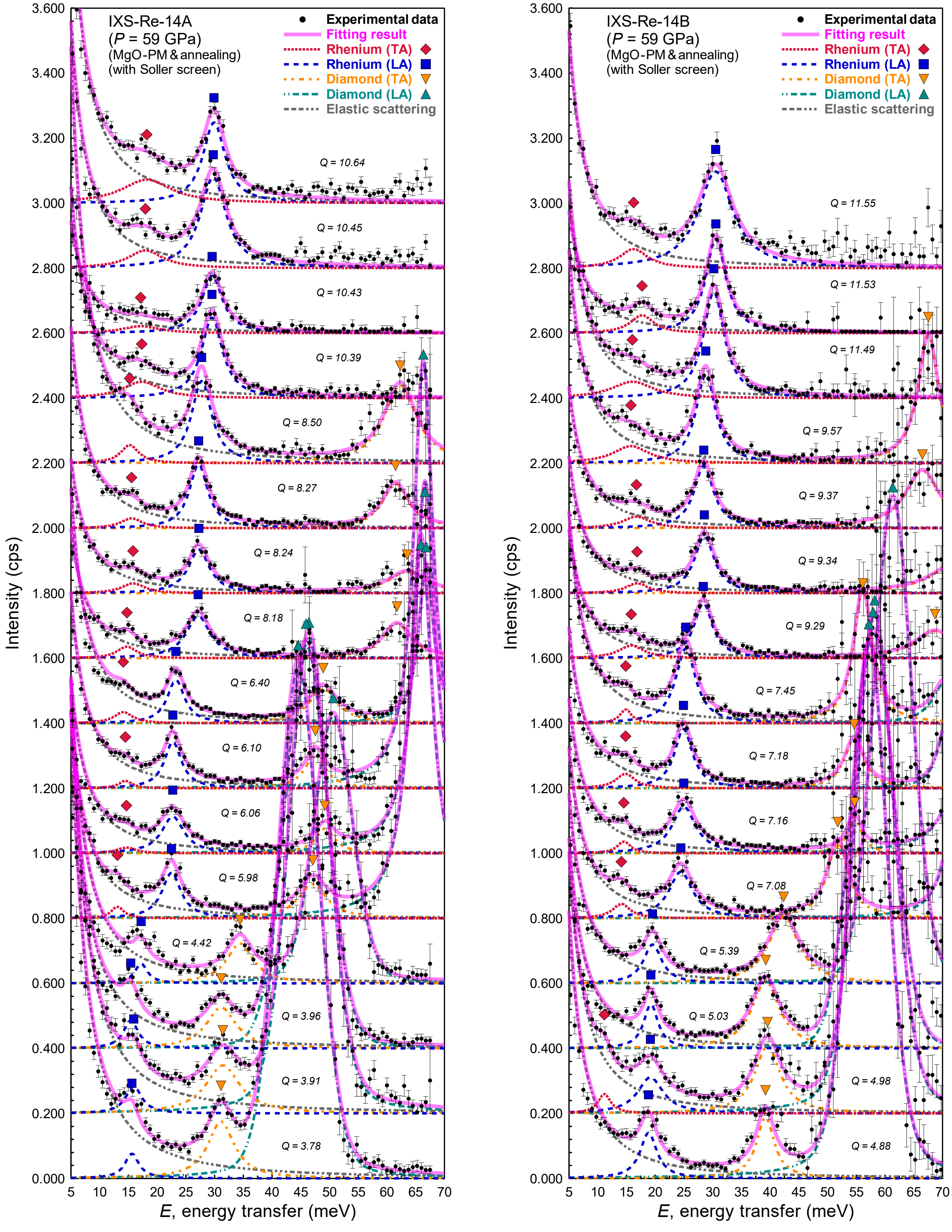

IXS spectra and fitting results in high pressure conditions (IXS-Re-14).
The symbols and lines are the same as those in Fig. 1A in the main text.

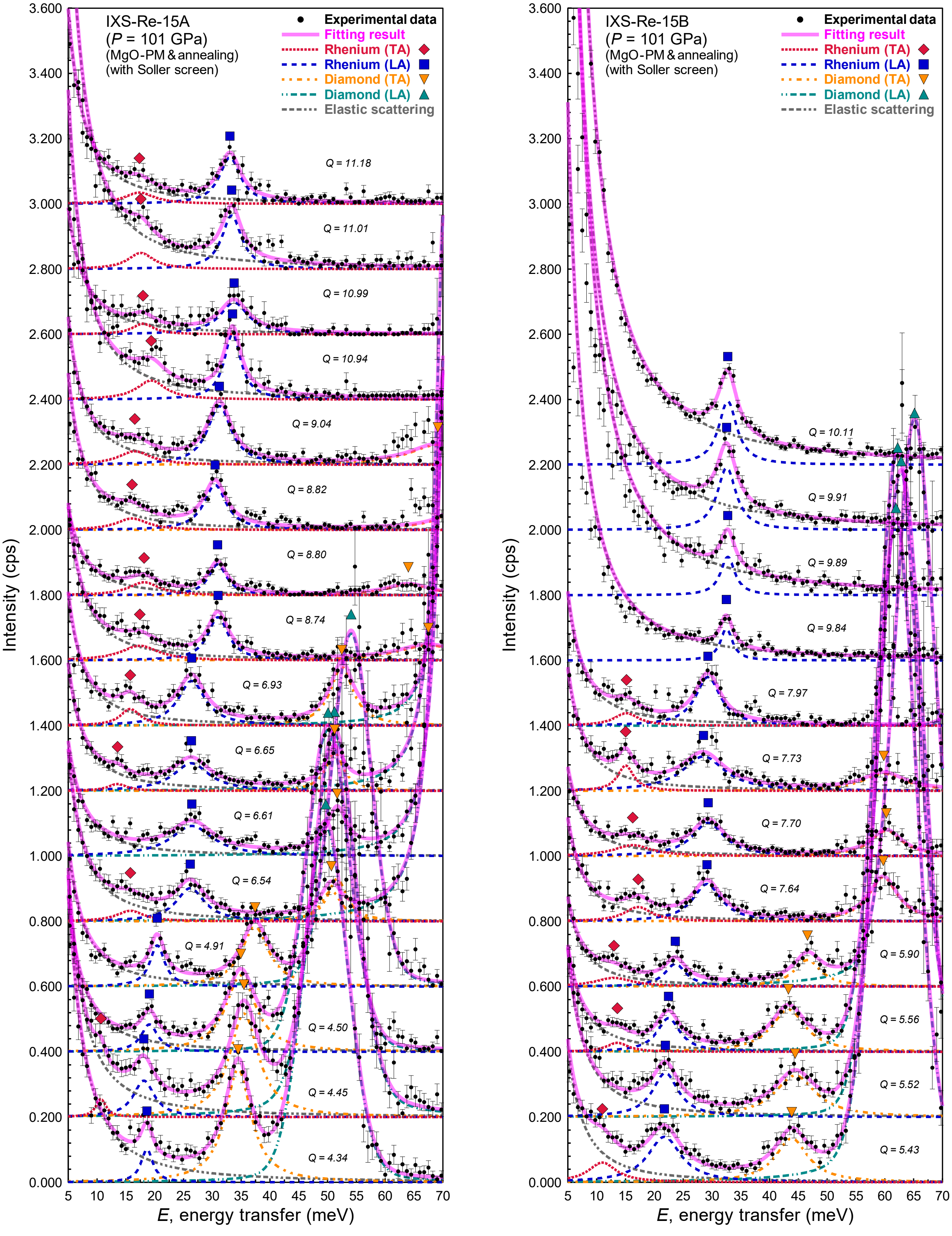

IXS spectra and fitting results in high pressure conditions (IXS-Re-15).
The symbols and lines are the same as those in Fig. 1A in the main text.

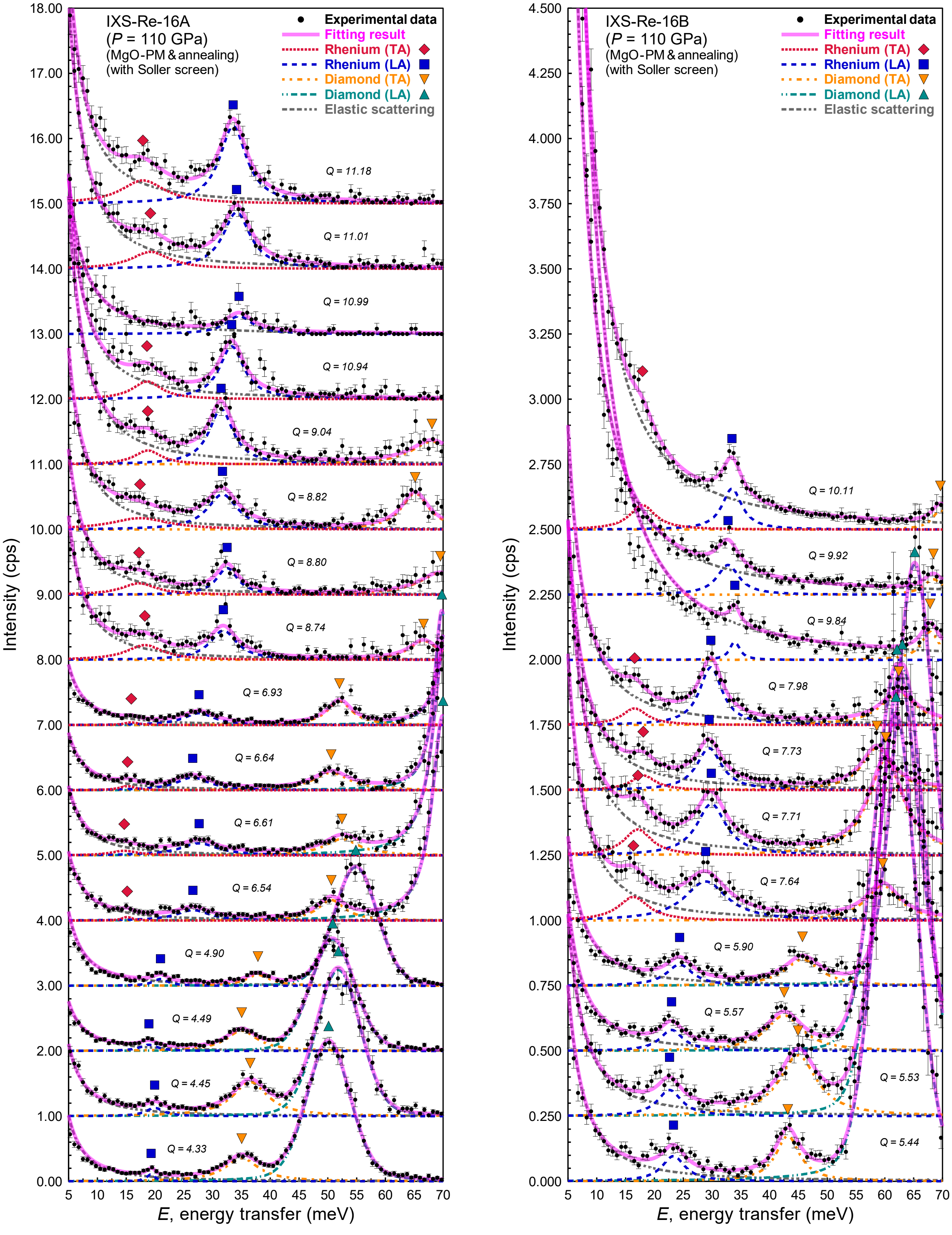

IXS spectra and fitting results in high pressure conditions (IXS-Re-16).
The symbols and lines are the same as those in Fig. 1A in the main text.

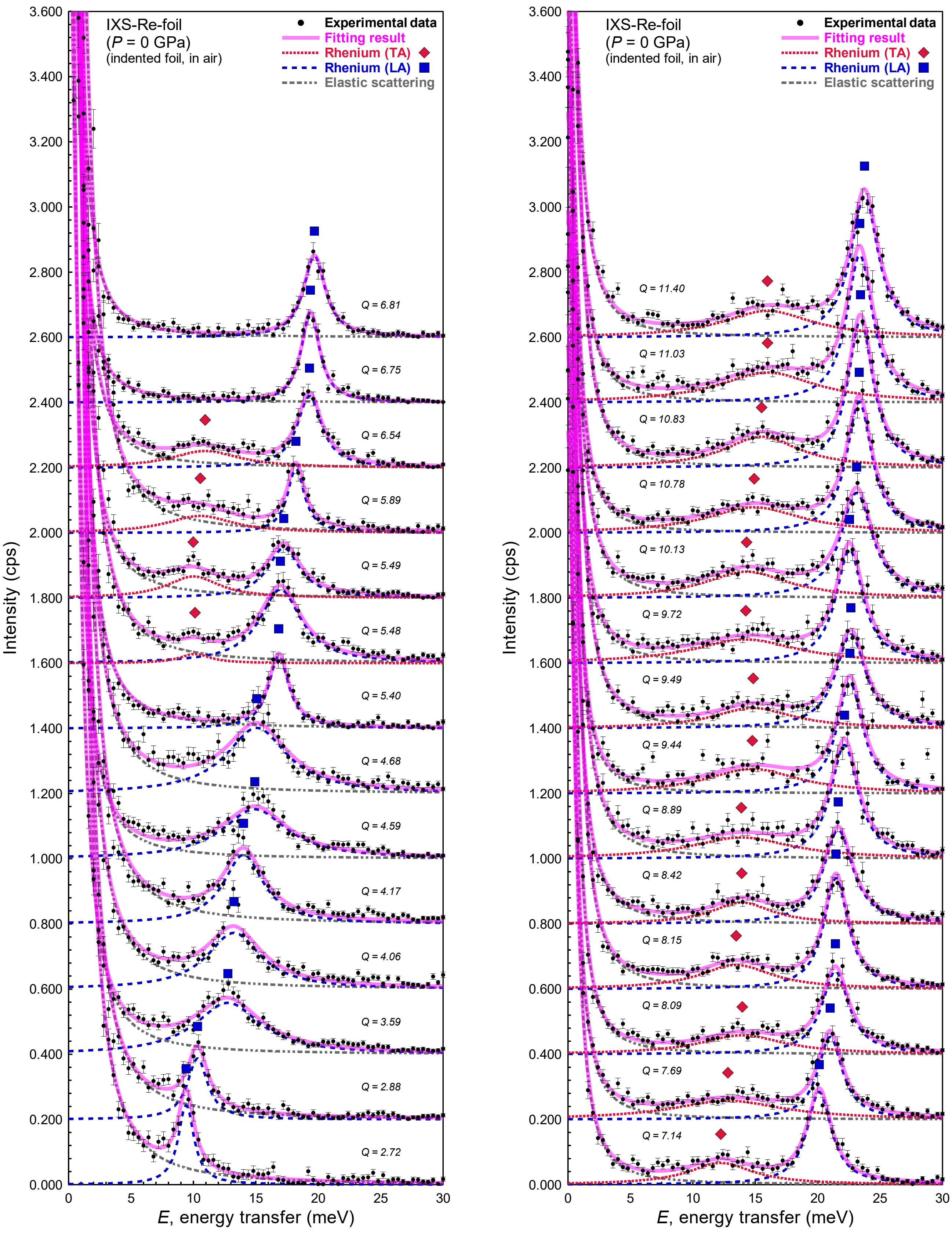

IXS spectra and fitting results in ambient conditions (IXS-Re-foil).
The symbols and lines are the same as those in Fig. S2A in the supplementary materials.